\pdfoutput=1
\documentclass[11pt,twoside,a4paper,cmspaper,final,collab]{cms-tdr}
\def\svnVersion{d2cefcc-D}\def\svnDate{2021/06/09}\def\cmsCernNoTag{CERN-EP-2021-038}\def\cmsCernDate{\today}\def\cmsMessage{"Published in the Journal of High Energy Physics as \href{http://dx.doi.org/10.1007/JHEP07(2021)027}{\doi{10.1007/JHEP07(2021)027}.}"}
\begin{document}\cmsNoteHeader{HIG-19-015}

\newcommand{\sqrts}{\ensuremath{\sqrt{s}=13\TeV}\xspace} 
\newcommand{\intlumi}{\ensuremath{137\fbinv}\xspace} 
\newcommand{\Hgg}{\ensuremath{\PH\!\to\!\PGg\PGg}\xspace}
\newcommand{\mgg}{\ensuremath{m_{\PGg\PGg}}\xspace}                                               
\newcommand{\mH}{\ensuremath{m_\PH}\xspace}  
\newcommand{\Hrap}{\ensuremath{\abs{y_\PH}}\xspace}
\newcommand{\mjj}{\ensuremath{m_{\text{jj}}}\xspace} 
\newcommand{\mll}{\ensuremath{m_{\ell\ell}}\xspace} 
\newcommand{\Zee}{\ensuremath{\PZ \to \Pe\Pe }\xspace}            
\newcommand{\Zmumu}{\ensuremath{\PZ \to \PGmp\PGmm}\xspace}                        
\newcommand{\Zmumug}{\ensuremath{\PZ \to \PGm\PGm\gamma}\xspace}                 
\newcommand{\gamplusjet}{\ensuremath{\PGg\!+\!\text{jet}}\xspace}                 
\newcommand{\gamplusjets}{\ensuremath{\PGg\!+\!\text{jets}}\xspace}                 
\newcommand{\gamgamplusjets}{\ensuremath{\PGg\PGg\!+\!\text{jets}}\xspace}                 
\newcommand{\jetplusjet}{\ensuremath{\text{jet}\!+\!\text{jet}}\xspace}                 
\newcommand{\ttH}{\ensuremath{\ttbar\PH}\xspace}                           
\newcommand{\bbH}{\ensuremath{\bbbar\PH}\xspace}                           
\newcommand{\ttZ}{\ensuremath{\ttbar\PZ}\xspace}                           
\newcommand{\srv}{\ensuremath{\sigma_{\text{rv}}}\xspace}                                                    
\newcommand{\swv}{\ensuremath{\sigma_{\text{wv}}}\xspace}                                                    
\newcommand{\seff}{\ensuremath{\sigma_{\text{eff}}}\xspace}                                           
\newcommand{\shm}{\ensuremath{\sigma_{\text{HM}}}\xspace}                                                    
\newcommand{\Like}{\ensuremath{\mathcal{L}}\xspace}                                                   
\newcommand{\NLL}{\ensuremath{2\text{NLL}}\xspace}                                                  
\newcommand{\dNLL}{\ensuremath{-2\Delta\ln{\text{L}}}\xspace}                                           
\newcommand{\sigmaSM}{\ensuremath{\sigma_{SM}}\xspace}                                                
\newcommand{\ptgg}{\ensuremath{\pt^{\PGg\PGg}}\xspace}                                            
\newcommand{\ptH}{\ensuremath{\pt^\PH}\xspace}                                                          
\newcommand{\ptHjj}{\ensuremath{\pt^{\PH\text{jj}}}\xspace}                                                    
\newcommand{\ptggjj}{\ensuremath{\pt^{\PGg\PGg\text{jj}}}\xspace}                                                    
\newcommand{\ptV}{\ensuremath{\pt^{\PV}}\xspace}                                                          
\newcommand{\kv}{\ensuremath{\kappa_\PV}\xspace}                                                          
\newcommand{\kf}{\ensuremath{\kappa_\text{F}}\xspace}                                                          
\newcommand{\kglu}{\ensuremath{\kappa_\Pg}\xspace}                                                          
\newcommand{\kgam}{\ensuremath{\kappa_\PGg}\xspace}                                                          
\newcommand{\ggH}{\ensuremath{\Pg\Pg\PH}\xspace}
\newcommand{\qqH}{\ensuremath{\cPq\cPq\PH}\xspace}
\newcommand{\VH}{\ensuremath{\PV\PH}\xspace}
\newcommand{\WH}{\ensuremath{\PW\PH}\xspace}
\newcommand{\ZH}{\ensuremath{\PZ\PH}\xspace}
\newcommand{\ggZH}{\ensuremath{\Pg\Pg\PZ\PH}\xspace}
\newcommand{\ggZHhad}{\ensuremath{\Pg\Pg\PZ(\qqbar)\PH}\xspace}
\newcommand{\tHq}{\ensuremath{\PQt\PH\PQq}\xspace}                           
\newcommand{\tHW}{\ensuremath{\PQt\PH\PW}\xspace}                           
\newcommand{\tH}{\ensuremath{\PQt\PH}\xspace}                           

\newlength\cmsTabSkip\setlength{\cmsTabSkip}{1ex}

\providecommand{\cmsTable}[1]{\resizebox{\textwidth}{!}{#1}}
\providecommand{\cmsTableAlt}[1]{\resizebox{!}{0.7\textwidth}{#1}}

\cmsNoteHeader{HIG-19-015}
\title{Measurements of Higgs boson production cross sections and couplings in the diphoton decay channel at \texorpdfstring{$\sqrt{s} = 13\TeV$}{sqrt(s) = 13 TeV}}

\date{\today}

\abstract{
 Measurements of Higgs boson production cross sections and couplings in events where the Higgs boson decays into a pair of photons are reported. Events are selected from a sample of proton-proton collisions at $\sqrt{s}=13\TeV$ collected by the CMS detector at the LHC from 2016 to 2018, corresponding to an integrated luminosity of 137\fbinv. Analysis categories enriched in Higgs boson events produced via gluon fusion, vector boson fusion, vector boson associated production, and production associated with top quarks are constructed.  The total Higgs boson signal strength, relative to the standard model (SM) prediction, is measured to be $1.12\pm0.09$.  Other properties of the Higgs boson are measured, including SM signal strength modifiers, production cross sections, and its couplings to other particles.  These include the most precise measurements of gluon fusion and vector boson fusion Higgs boson production in several different kinematic regions, the first measurement of Higgs boson production in association with a top quark pair in five regions of the Higgs boson transverse momentum, and an upper limit on the rate of Higgs boson production in association with a single top quark.  All results are found to be in agreement with the SM expectations.
}

\hypersetup{%
pdfauthor={CMS Collaboration},%
pdftitle={Measurements of Higgs boson production cross sections and couplings in the diphoton decay channel at sqrt{s} = 13 TeV},%
pdfsubject={CMS},%
pdfkeywords={CMS, full Run 2, Higgs, diphoton, simplified template cross sections}}

\maketitle

\section{Introduction}
\label{sec:intro}

Since the discovery of a Higgs boson (\PH) by the ATLAS and CMS Collaborations in 2012
\cite{Aad:2012tfa,Chatrchyan:2012xdj,Chatrchyan:2013lba}, an extensive programme of measurements focused
on characterising its properties and testing its compatibility with the standard model (SM) of particle physics has been performed.
Analysis of data collected during the second run of the CERN LHC at \sqrts has already resulted in the observation of 
Higgs boson production mechanisms and decay modes predicted by the SM~\cite{Sirunyan:2018hoz,Aaboud:2018urx,CMS_VH,ATLAS_VH}.
The most precise measurements are obtained by combining results from different Higgs boson decay channels. 
Such combinations have enabled the total Higgs boson production cross section to be measured with an uncertainty of less than 7\%~\cite{Sirunyan:2018koj,Aad:2019mbh}.
All reported results have so far been consistent with the corresponding SM predictions.

In the SM, the \Hgg decay has a small branching fraction of approximately 0.23\% for a Higgs boson mass (\mH) around 125\GeV~\cite{YR4}.
However, its clean final-state topology with
two well-reconstructed photons provides a narrow invariant mass (\mgg) peak 
that can be used to effectively distinguish it from background processes.
As a result, \Hgg is one of the most important channels for 
precision measurements of Higgs boson properties.
Furthermore, it is one of the few decay channels that is sensitive
to all principal Higgs boson production modes. 

The results reported in this paper build upon previous analyses performed by the CMS Collaboration~\cite{HggRun1,HIG-16-040}.
Here, the data collected by the CMS experiment between 2016 and 2018 are analysed together.
The resulting statistical power of the combined data set improves the precision on existing measurements 
and allows new measurements to be made. 
The structure of this analysis is designed to enable measurements 
within the simplified template cross section (STXS) framework~\cite{YR4}.
Using this structure, various measurements of Higgs boson properties can be performed.
These include SM signal strength modifiers, production cross sections, 
and the Higgs boson's couplings to other particles.
Measurements of all these quantities are reported in this paper. 

The STXS framework provides a coherent approach with which 
to perform precision Higgs boson measurements. 
Its goal is to minimise the theory dependence of Higgs boson measurements, 
lessening the direct impact of SM predictions on the results,
and to provide access to kinematic regions likely to be affected by physics beyond the SM (BSM). 
At the same time, this approach permits the use of advanced analysis techniques to optimise sensitivity.
Reducing theory-dependence is desirable because it makes the measurements 
both easier to reinterpret and means they are less affected by potential updates 
to theoretical predictions, making them useful over a longer period of time~\cite{Cepeda:2019klc}.
The results presented within the STXS framework nonetheless depend on the SM simulation 
used to model the experimental acceptance of the signal processes, 
which could be modified in BSM scenarios.

The strategy employed in this analysis is to construct analysis categories enriched in events from
as many different kinematic regions as possible, 
thereby providing sensitivity to the individual regions defined in the STXS framework.
This permits measurements to be performed across all the major Higgs boson production modes, 
including gluon fusion (\ggH), vector boson fusion (VBF), vector boson associated production (\VH), 
production associated with a top quark-antiquark pair (\ttH), 
and production in association with a single top quark (\tH).

In addition to measurements within the STXS framework, 
this paper contains several other interpretations of the data.
The event categorisation designed to target the individual STXS regions also provides sensitivity 
to signal strength modifiers, both for inclusive Higgs boson production and for individual production modes, 
as well as measurements within the $\kappa$-framework~\cite{YR3}.

The paper is structured as follows. 
The CMS detector is described in Section~\ref{sec:detector}.
An overview of the STXS framework is given in Section~\ref{sec:strategy},
together with a summary of the overall strategy of this analysis.
In Section~\ref{sec:samples}, details of the data and simulation used to design and perform the analysis are given.
The reconstruction of candidate \Hgg events is described in Section~\ref{sec:event_reco}, 
before the event categorisation procedure is explained in Section~\ref{sec:categorisation}.
The techniques used to model the signal and background are outlined in Section~\ref{sec:sig_bkg},
with the associated systematic uncertainties listed in Section~\ref{sec:systematics}.
The results are presented in Section~\ref{sec:results}, 
with tabulated versions provided in HEPDATA~\cite{hepdata}.
Finally, the paper is summarised in Section~\ref{sec:summary}.

\section{The CMS detector}
\label{sec:detector}

The central feature of the CMS apparatus is a superconducting solenoid of 6\unit{m} internal diameter, 
providing a magnetic field of 3.8\unit{T}. 
Within the solenoid volume are a silicon pixel and strip tracker, 
a lead tungstate crystal electromagnetic calorimeter (ECAL), 
and a brass and scintillator hadron calorimeter (HCAL), 
each composed of a barrel and two endcap sections. 
The ECAL consists of 75\,848 lead tungstate crystals, 
which provide coverage in pseudorapidity $\abs{\eta} < 1.48 $ in the barrel region 
and $1.48 < \abs{\eta} < 3.00$ in the two endcap regions. 
Preshower detectors consisting of two planes of silicon sensors interleaved with 
a total of 3 radiation lengths of lead are located in front of each EE detector.
Forward calorimeters extend the $\eta$ coverage provided by the barrel and endcap detectors. 
Muons are detected in gas-ionisation chambers embedded 
in the steel flux-return yoke outside the solenoid.

Events of interest are selected using a two-tiered trigger system~\cite{Khachatryan:2016bia}. 
The first level, composed of custom hardware processors, 
uses information from the calorimeters and muon detectors to select events 
at a rate of around 100\unit{kHz} within a fixed time interval of less than 4\mus. 
The second level, known as the high-level trigger, 
consists of a farm of processors running a version of the full event reconstruction software 
optimised for fast processing, and reduces the event rate to around 1\unit{kHz} before data storage~\cite{CMS_HLT}.

The particle-flow (PF) algorithm~\cite{ParticleFlow} aims to reconstruct 
and identify each individual particle (PF candidate) in an event, 
with an optimised combination of information from the various elements of the CMS detector. 
The energy of photons is obtained from the ECAL measurement. 
The energy of electrons is determined from a combination of the electron momentum 
at the primary interaction vertex as determined by the tracker, 
the energy of the corresponding ECAL cluster, 
and the energy sum of all bremsstrahlung photons spatially compatible 
with originating from the electron track. 
The energy of muons is obtained from the curvature of the corresponding track. 
The energy of charged hadrons is determined from a combination of their momentum 
measured in the tracker and the matching ECAL and HCAL energy deposits, 
corrected for zero-suppression effects 
and for the response function of the calorimeters to hadronic showers. 
Finally, the energy of neutral hadrons is obtained 
from the corresponding corrected ECAL and HCAL energies.

For each event, hadronic jets are clustered from these reconstructed particles 
using the infrared and collinear safe anti-\kt algorithm~\cite{AntiKt, Cacciari:2011ma} 
with a distance parameter of 0.4. 
Jet momentum is determined as the vectorial sum of all particle momenta in the jet, 
and is found from simulation to be, on average, within 5 to 10\% of the true momentum 
over the whole transverse momentum (\pt) spectrum and detector acceptance. 
Additional proton-proton interactions within the same or nearby bunch crossings (pileup) 
can contribute additional tracks and calorimetric energy depositions to the jet momentum. 
To mitigate this effect, charged particles identified to be originating from pileup vertices 
are discarded and an offset correction is applied to correct for remaining contributions. 
Jet energy corrections are derived from simulation to bring the measured response of jets 
to that of particle level jets on average. 
In situ measurements of the momentum balance in dijet, 
$\text{photon}+\text{jet}$, $\PZ+\text{jet}$, 
and multijet events are used to account for any residual differences 
in the jet energy scale between data and simulation~\cite{JECperformance}. 
The jet energy resolution amounts typically to 15--20\% at 30\GeV, 10\% at 100\GeV, 
and 5\% at 1\TeV~\cite{JECperformance}. 
Additional selection criteria are applied to each jet to remove jets potentially dominated 
by anomalous contributions from various subdetector components or reconstruction failures.

The missing transverse momentum vector \ptvecmiss is computed as the negative vector \pt sum 
of all the PF candidates in an event, and its magnitude is denoted as \ptmiss~\cite{METperformance}. 
The \ptvecmiss is modified to account for corrections to the energy scale of the reconstructed jets in the event.

A more detailed description of the CMS detector, 
together with a definition of the coordinate system used and the relevant kinematic variables, 
can be found in Ref.~\cite{CMSdetector}.

\section{Analysis strategy}
\label{sec:strategy}

\subsection{The STXS framework}
\label{sec:STXS}
In the STXS framework, kinematic regions based upon Higgs boson production modes are defined.
These regions, or bins, exist in varying degrees of granularity, following sequential ``stages".
At the so-called STXS stage 0, the bins correspond closely to the different Higgs boson production mechanisms.
Events where the absolute value of the Higgs boson rapidity, \Hrap, is greater than 2.5 are not included 
in the definition of the bins because they are typically outside of the experimental acceptance.
Measurements of stage-0 cross sections in the \Hgg decay channel were presented by the CMS Collaboration in Ref.~\cite{HIG-16-040}.
Additionally, an analysis probing the coupling between the top quark and Higgs boson in the diphoton decay channel was recently performed by the CMS Collaboration~\cite{HIG-19-013}.
Several other stage-0 measurements in different decay channels have also been made by both the ATLAS and CMS Collaborations 
\cite{ATLAS_Hgg36,ATLAS_ZZ,ATLASstage0_WW,ATLASstage0_tt,CMSstage0_ZZa,CMSstage0_WW,CMSstage0_tt}.
Each experiment has also presented results combining the various analyses \cite{Aad:2019mbh,Sirunyan:2018koj}.

At STXS stage 1, a further splitting of the bins using the events' kinematic properties is performed~\cite{Stage1p1}. 
This provides additional information for different theoretical interpretations of the measurements, 
and enhances the sensitivity to possible signatures of BSM physics. 
Furthermore, increasing the number of independent bins reduces the theory-dependence 
of the measurements; within each bin SM kinematic properties are assumed, 
and thus splitting the bins allows these assumptions to be partially lifted.

Measurements at stage 1 of the framework have already been reported 
by the ATLAS Collaboration \cite{ATLAS_Hgg36,ATLAS_ZZ,ATLAS_VHbb}. 
Following these results, adjustments to the framework and its definitions were made, 
such that the most recent definition of STXS bins is referred to as STXS stage 1.2. 
The first measurement of STXS stage-1.2 cross sections was recently performed by the CMS Collaboration~\cite{HIG-19-001}.

\begin{figure}
   \centering
   \includegraphics[width=1\textwidth]{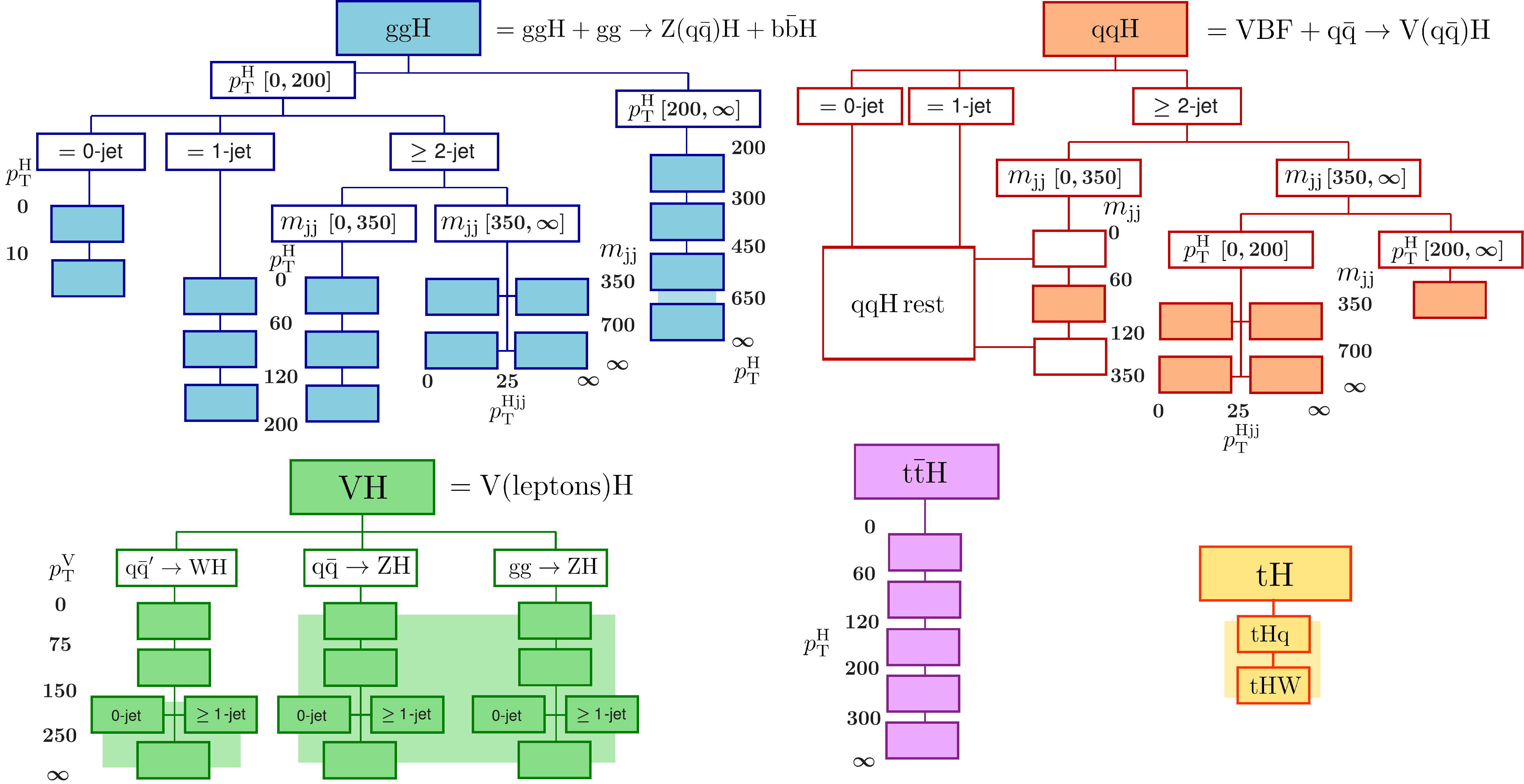}
   \caption{Diagram showing the full set of STXS stage-1.2 bins, adapted from Ref.~\cite{YR4},
            defined for events with $\Hrap < 2.5$.
            The solid boxes represent each STXS stage-1.2 bin. 
            The units of \ptH, \mjj, \ptHjj, and \ptV are in \GeVns. 
            The shaded regions indicate the STXS bins that are divided at stage 1.2, but are not measured
            independently in this analysis. 
           }
   \label{fig:allSTXSbins}
\end{figure}

The full set of STXS stage-1.2 bins is described below and an illustration is given in Fig.~\ref{fig:allSTXSbins}.
The \ggH region (blue) is split into STXS bins using the Higgs boson transverse momentum (\ptH), 
the number of jets, and additionally has a VBF-like region with high dijet mass (\mjj). 
This VBF-like region is split into four STXS bins according to \mjj and the transverse momentum of the Higgs
boson plus dijet system (\ptHjj).
Events originating from \bbH production are grouped with the \ggH production mode, 
as are those from gluon-initiated production in association with a vector boson (\ggZH) 
where the vector boson decays hadronically. 
The VBF and hadronic \VH modes are considered together as electroweak \qqH production (orange).
Here the STXS bins are defined using the number of jets, \ptH, \mjj and \ptHjj.
The four STXS bins which define the \qqH rest region are not explicitly probed in this analysis.
The leptonic \VH STXS bins (green) are split into three separate regions representing the 
\WH, \ZH, and \ggZH production modes, which are further divided according to
the number of jets and the transverse momentum of the vector boson (\ptV) that decays leptonically.
The \ttH production mode (pink) is split only by \ptH. 
Finally, the \tH STXS bin includes contributions from both the \tHq and \tHW production modes.
All references to STXS bins hereafter imply the STXS stage-1.2 bins.
Further details on the exact definitions are contained in Section~\ref{sec:categorisation}, describing the event categorisation.
All the production mechanisms shown in Fig.~\ref{fig:allSTXSbins} are measured independently in this analysis.

\subsection{Analysis categorisation}
\label{sec:analysis_categorisation}

To perform measurements of Higgs boson properties,
analysis categories must first be constructed where the narrow signal peak is distinguishable 
from the falling background \mgg spectrum.
The categorisation procedure uses properties of the reconstructed diphoton system 
and any additional final-state particles to improve the sensitivity of the analysis.
As part of the categorisation, dedicated selection criteria and classifiers are used to select events consistent with the
\tH, \ttH, \VH, VBF, and \ggH production modes.
This both increases the analysis sensitivity and enables measurements
of individual production mode cross sections to be performed. 

In order to measure cross sections of STXS bins individually, 
events deemed to be compatible with a given production mode 
are further divided into analysis categories that differentiate between the various STXS bins.
For most production modes, the divisions are made using the detector-level equivalents 
of the particle-level quantities used to define the STXS bins; 
an example is using \ptgg to construct analysis categories targeting STXS bins defined by \ptH values.
Increasing the total number of analysis categories to target individual STXS bins 
in this way does not degrade the analysis' sensitivity 
to the individual production mode and total Higgs boson cross sections.
For each production mode, the event categorisation is designed to target 
all of the STXS bins to which some sensitivity can be obtained in the diphoton decay channel with the available data.

Several different machine learning (ML) algorithms are used throughout this analysis 
for both regression and classification tasks.
Examples include regressions that improve the agreement between simulation and data, 
and classification to improve the discrimination between signal and background processes.
The usage of ML techniques for event categorisation is also found to improve the separation 
between different STXS bins, which further improves the sensitivity of STXS measurements. 
For the training of boosted decision trees (BDTs), either the \textsc{xgboost}~\cite{xgboost} 
or the TMVA package~\cite{TMVA} package is used.
The \textsc{TensorFlow}~\cite{TensorFlow} package is used to train deep neural networks (DNNs).

For the \ggH phase space, almost all of the STXS bins can be measured individually, without any bin merging (blue in Fig.~\ref{fig:allSTXSbins}). 
The exceptions are the high dijet mass (\mjj) STXS bins, which are difficult to distinguish from VBF events.
Furthermore, the sensitivity to STXS bins with particularly high \ptH is limited.
Analysis categories are constructed using a 
BDT to assign the most probable STXS bin for each event.
The amount of background is reduced using another BDT, referred to as the diphoton BDT.
The diphoton BDT is trained to discriminate between all Higgs boson signal events 
and all other modes of SM diphoton production.
Throughout the analysis, events originating from the \bbH production mode
are grouped together with \ggH events.

The VBF production mode and \VH production where the vector boson decays hadronically 
are considered together as (EW) \qqH production (orange in Fig.~\ref{fig:allSTXSbins}).
A set of analysis categories enriched in VBF-like events, where a dijet with high \mjj is present, is defined. 
These analysis categories make use of the same diphoton BDT used in the analysis categories targeting \ggH to reduce the number of background events.
Additionally, a BDT based on the kinematic properties of the characteristic VBF dijet system, known as the dijet BDT, is utilised.
The dijet BDT is trained to distinguish between three different classes of events with a VBF-like topology: 
VBF events, \ggH events, and events produced by all other SM processes.
This enables VBF events to be effectively separated from both VBF-like \ggH events and other SM backgrounds.
At least one analysis category is defined to target each VBF-like \qqH STXS bin.
Additional analysis categories enriched in \VH-like events, 
where the vector boson decays hadronically to give a dijet whose \mjj is consistent with a \PW or \PZ boson,
are defined.
These make use of a dedicated \VH hadronic BDT to reduce both the number of background events 
and contamination from \ggH events.

Analysis categories targeting \VH leptonic production (green in Fig.~\ref{fig:allSTXSbins}) are divided into three categorisation regions, 
containing either zero, one, or two reconstructed charged leptons (electrons or muons).
Each categorisation region uses a dedicated BDT to reduce the background contamination. 
It is not possible to measure STXS bins individually with the available data set. 
Nonetheless, where a sufficient number of events exists, analysis categories are constructed 
to provide sensitivity to merged groups of STXS bins.

In this analysis, \ttH and \tH production cross sections are measured independently 
(\ttH STXS bins are purple in Fig.~\ref{fig:allSTXSbins}, whilst \tH is yellow).
For this purpose, a dedicated DNN referred to as the top DNN is trained to discriminate between \tH and \ttH events. 
An analysis category enriched in \tH events is defined that uses the top DNN to reduce the contamination from \ttH events, 
with a BDT used to reject background events from other sources.

The analysis categories targeting \ttH production are based on those described in Ref.~\cite{HIG-19-013}, 
with separate channels for hadronic and leptonic top quark decays.
In each channel, a dedicated BDT is trained to reject background events. 
Furthermore, the top DNN is used to reduce the amount of contamination from \tH events.
The analysis categories are divided to provide the sensitivity to the STXS bins,
for which four \ptH ranges are defined.

It is possible for an event to pass the selection criteria for more than one analysis category.
To unambiguously assign each event to only one analysis category, a priority sequence is defined.
Events that could enter more than one analysis category are assigned to the analysis category with the highest priority. 
The priority sequence is based on the expected number of signal events, 
with a higher priority assigned to analysis categories with a lower expected signal yield.
This ordering enables the construction of analysis categories containing sufficiently high fractions of 
the Higgs boson production mechanisms with lower SM cross sections, 
which is necessary to perform independent measurements of these processes.

Events in data and the corresponding simulation for all three years of data-taking 
from 2016 to 2018 are grouped together in the final analysis categories.
This gives better performance than constructing analysis categories for each year individually, 
requiring fewer analysis categories in total for a comparable sensitivity. 
Separating the analysis categories by year would enable differences in the detector conditions ---
such as the variation in \mgg resolution --- to be exploited.
However this is found to be less important than the advantage of having a greater number of events
with which to train multivariate classifiers and optimise the analysis category definitions.
Furthermore, the variations in detector conditions are relatively modest, 
and in general not substantially greater than variations within a given year of data-taking, 
which allows all data collected in each of the three years to be analysed together.

Nonetheless, simulated events are generated for each year separately, 
with the corresponding detector conditions, before they are merged together. 
This accounts for the variation in the detector itself, in the event reconstruction procedure,
and in the LHC beam parameters.
Furthermore, corrections to the photon energy scale and other procedures relating to the event 
reconstruction are also performed for each year individually. 
Only when performing the final division of selected diphoton events into the analysis categories 
are the simulated and data events from different years processed together. 
The full description of all the analysis categories is given in Section~\ref{sec:categorisation}.

Once the selection criteria for each analysis category are defined, results are obtained 
by performing a simultaneous fit to the resulting \mgg distributions in all analysis categories.
The results of several different measurements with different observables are reported in Section~\ref{sec:results}.
For measurements within the STXS framework, it is not possible to measure each STXS bin individually. 
Therefore for each fit, a set of observables is defined by merging some STXS bins.
In this paper, the results of two scenarios with different parameterisations of the STXS bins are provided.
In addition, measurements of SM signal strength modifiers are reported, 
both for inclusive Higgs boson production and per production mode.
Finally, measurements of Higgs boson couplings within the $\kappa$-framework are also shown.

\section{Data samples and simulated events}
\label{sec:samples}

The analysis exploits proton-proton collision data at \sqrts, collected in 2016, 2017, and 2018 and corresponding to integrated luminosities of 35.9, 41.5, and 59.4\fbinv, respectively. 
The integrated luminosities of the 2016--2018 data-taking periods are individually known 
with uncertainties in the 2.3--2.5\% range~\cite{CMSlumi2016,CMSlumi2017,CMSlumi2018}, 
while the total (2016--2018) integrated luminosity has an uncertainty of 1.8\%, 
the improvement in precision reflecting the (uncorrelated) time evolution of some systematic effects.
In this section, the data sets and simulated event samples for all three years are described. 
Any differences between the years are highlighted in the text.

Events are selected using a diphoton high-level trigger with asymmetric photon \pt thresholds of 30 (30) and 18 (22)\GeV in 2016 (2017 and 2018) data. 
A calorimetric selection is applied at trigger level, based on the shape of the electromagnetic shower, the isolation of the photon candidate, 
and the ratio of the hadronic and electromagnetic energy deposits of the shower.  
The $\RNINE$ variable is defined as the energy sum of the $3{\times}3$ crystals centred on the most energetic crystal in the candidate electromagnetic cluster divided by the energy of the candidate. 
The value of $\RNINE$ is used to identify photons undergoing a conversion in the material upstream of the ECAL. 
Unconverted photons typically have narrower transverse shower profiles, resulting in higher values of the $\RNINE$ variable, compared to converted photons.
The trigger efficiency is measured from \Zee events using the ``tag-and-probe" technique \cite{CMS:2011aa}. 
The efficiency measured in data in bins of $\pt$, $\RNINE$, and $\eta$ is used to weight the simulated events to replicate the trigger efficiency observed in data.

A Monte Carlo (MC) simulated signal sample for each Higgs boson production mechanism is generated using \MGvATNLO (version 2.4.2) 
at next-to-leading order accuracy \cite{Alwall:2014hca} in perturbative quantum chromodynamics (QCD).
For each production mode, events are generated with $\mH=120$, 125, and 130\GeV.
Events produced via the gluon fusion mechanism are weighted as a function of \ptH
and the number of jets in the event, to match the prediction from the \textsc{nnlops} program~\cite{Hamilton:2013fea}.
All parton-level samples are interfaced with {\PYTHIA}8 version 8.226 (8.230) \cite{Sjostrand:2014zea} for parton showering and hadronization, 
with the CUETP8M1~\cite{Khachatryan:2015pea} (CP5~\cite{Sirunyan:2019dfx}) tune used for the simulation of 2016~(2017 and 2018)~data. 
Parton distribution functions (PDFs) are taken from the NNPDF~3.0~\cite{NNPDF3}~(3.1~\cite{NNPDF31})~set, when simulating 2016~(2017 and 2018)~data.
The production cross sections and branching fractions recommended by the LHC Higgs Working Group \cite{YR4} are used. 
The relative fraction of each STXS bin for each inclusive production mode at particle level is taken from simulation 
and used to compute the SM prediction for the production cross section in each STXS bin.
Additional signal samples generated with \POWHEG 2.0 \cite{Nason:2004rx,Frixione:2007vw,Alioli:2008tz,Nason:2009ai,Alioli:2010xd,Hartanto:2015uka}
at next-to-leading order accuracy in perturbative QCD are used to train some of the multivariate discriminants described in Section~\ref{sec:categorisation}.

The dominant source of background events in this analysis is due to SM diphoton production.
A smaller component comes from \gamplusjet or \jetplusjet events, in which jets are misidentified as photons.
In the final fits of the analysis, the background is estimated directly from the diphoton mass distribution in data. 
Simulated background events from different event generators are only used for the training of multivariate discriminants.
The diphoton background is generated with the \textsc{sherpa} (version 2.2.4) generator \cite{Gleisberg:2008ta}. 
It includes the Born processes with up to 3 additional jets, as well as the box processes at leading order accuracy. 
The \gamplusjet and \jetplusjet backgrounds are simulated at leading order with {\PYTHIA}8, after applying a filter at generator level to enrich the production of jets with a high electromagnetic activity. 
The filter requires a potential photon signal coming from photons, electrons, or neutral hadrons with $\PT>15\GeV$. 
In addition, the filter requires no more than two charged particles ($\PT>1.6\GeV$ and $\abs{\eta}<2.2$) in a cone of radius 
$R = \sqrt{\smash[b]{(\Delta\eta)^2+(\Delta\varphi)^2}}<0.2$ (where $\varphi$ is the azimuthal angle in radians) around the photon candidate, mimicking the tracker isolation described in Section~\ref{sec:event_reco}. 

A sample of Drell--Yan events is simulated with \MGvATNLO, 
and is used both to derive corrections for simulation and for validation purposes.

The response of the CMS detector is simulated using the \GEANTfour package~\cite{Agostinelli:2002hh}. 
This includes the simulation of the multiple proton-proton interactions taking place in each bunch crossing. 
These can occur at the nominal bunch crossing (in-time pileup) or at the crossing of previous and subsequent bunches (out-of-time pileup), and the simulation accounts for both. 
Simulated out-of-time pileup is limited to a window of $[ -12, +3 ]$ bunch crossings around the nominal, 
in which the effects on the observables reconstructed in the detector are most relevant. 
Simulated events are weighted to reproduce the distribution of the number of interaction vertices in data. 
The average number of interactions per bunch crossing in data in the 2016 (2017 and 2018) data sets is 23 (32).

\section{Event reconstruction}
\label{sec:event_reco}

\subsection{Photon reconstruction}
Efficiently reconstructing photons with an accurate and precise energy determination
plays a very important role in the sensitivity of this analysis.
This section describes in detail the procedures used to reconstruct the photon energy 
and the photon preselection criteria.

Photon candidates are reconstructed from energy clusters in the ECAL not linked to any charged-particle trajectories seeded in the pixel detector.
The clusters are built around a ``seed" crystal, identified as a local energy maximum above a given threshold. 
The clusters are grown with a so-called topological clustering, 
where crystals with at least one side in common with a crystal already in the cluster 
and with an energy above a given threshold are added to the existing cluster itself. 
Finally, the clusters are dynamically merged into ``superclusters" 
to ensure good containment of the shower, accounting for geometrical variation along $\eta$, 
and optimising the robustness of the energy resolution against pileup.
The energy of the photon is estimated by summing the energy of each crystal in the supercluster, 
calibrated and corrected for response variations in time \cite{Sirunyan:2020ycc}. 
The photon energy is corrected for the imperfect containment of the electromagnetic shower 
and the energy losses from converted photons. 
The correction is computed with a multivariate regression technique trained on simulated photons, 
which estimates simultaneously the energy of the photon and its uncertainty.

After the application of this simulation-derived correction, some differences 
remain between data and simulation.
A sequence of additional corrections are applied to improve the agreement between the two, 
using \Zee events where the electrons are reconstructed as photons.
First, any residual drift in the energy scale in data over time is corrected for in bins corresponding 
approximately to the duration of one LHC fill.
The second step involves modifying the energy scale in data and the energy resolution in simulation.
A set of corrections is derived to align the mean of the dielectron mass spectrum in data with the expected value from simulation, 
and to smear the resolution in simulation to match that observed in data.
These corrections are derived simultaneously in bins of $\abs{\eta}$ and \RNINE.
Further details on this procedure are contained in Ref.~\cite{HggMass}. 

\begin{figure}[htb!]
  \centering
  \includegraphics[width=0.49\textwidth]{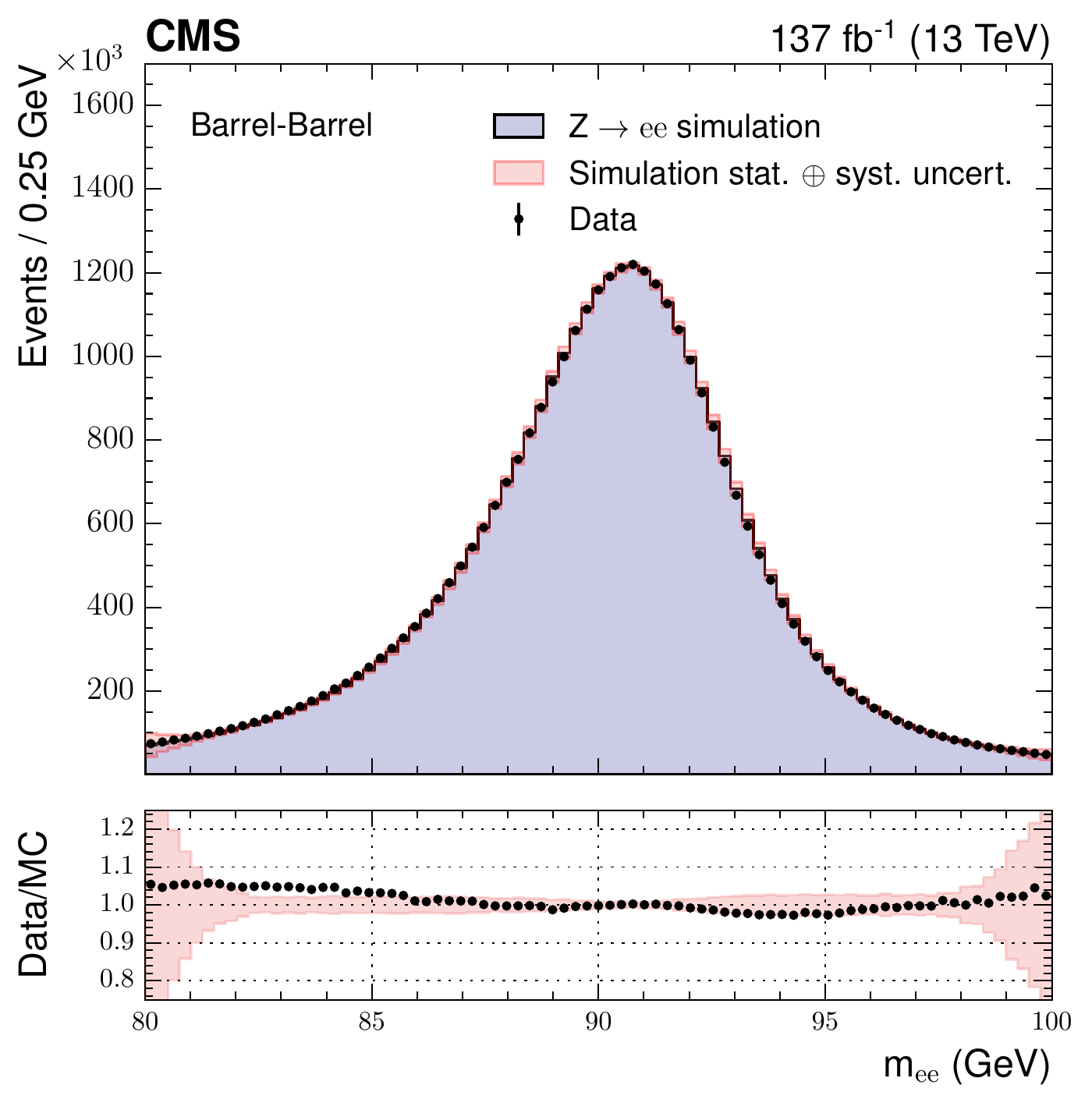}
  \includegraphics[width=0.49\textwidth]{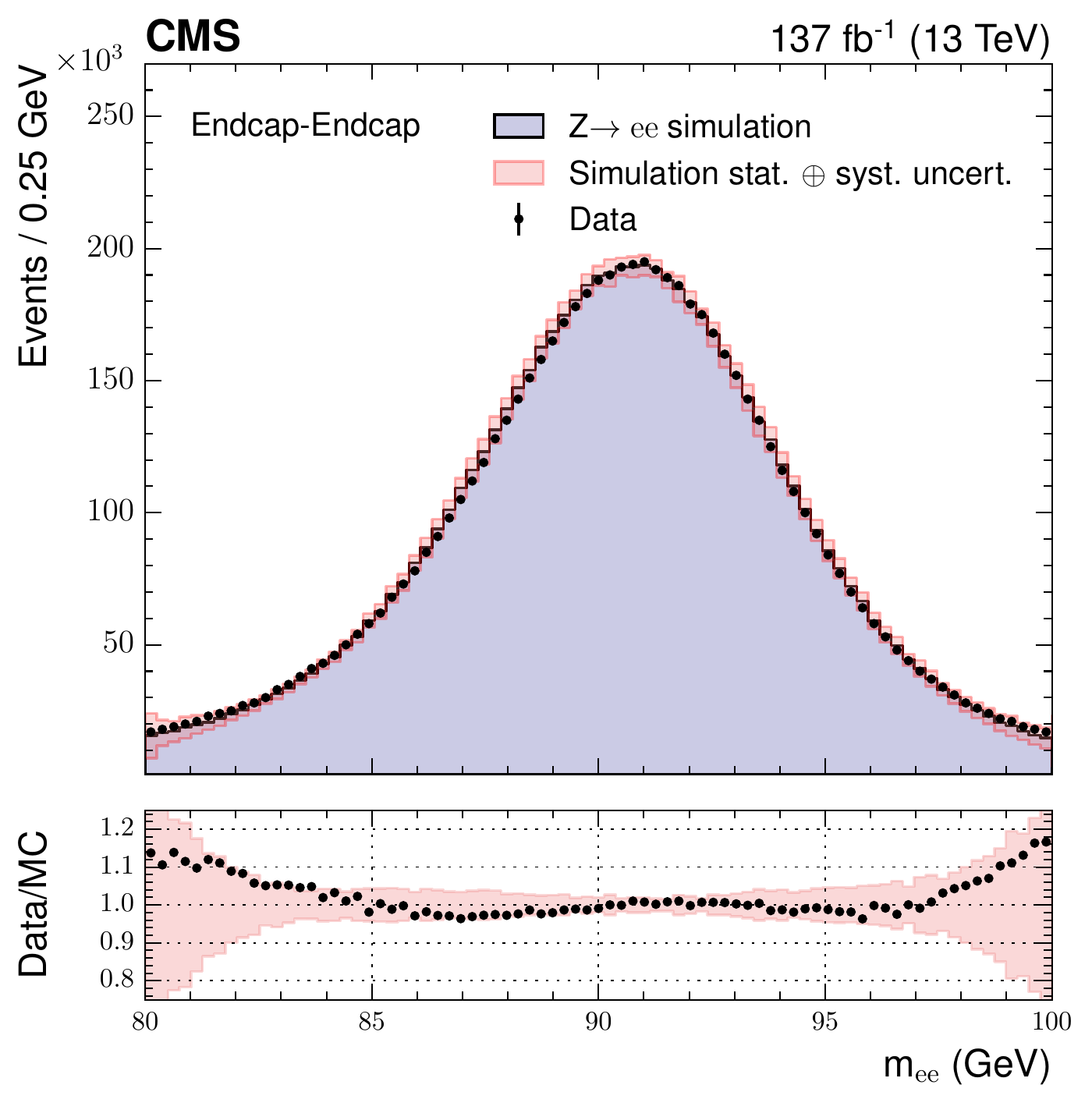}
  \caption{Comparison of the dielectron invariant mass spectra in data (black points) and simulation (blue histogram), 
  after applying energy scale corrections to data and energy smearing to the simulation, 
  for \Zee events with electrons reconstructed as photons. 
  The statistical and systematic uncertainty on the simulation is shown by the pink band.
  The comparison is shown for events where  both electrons are reconstructed in the ECAL barrel (left), and both in the ECAL endcaps (right).
  The lower panels show the ratio of the data to the MC simulation in black points, 
  with the uncertainty on the ratio represented by the pink band.
  The full data set collected in 2016--2018 and the corresponding simulation are shown.
  }
  \label{fig:ZeeMass}
\end{figure}

Figure~\ref{fig:ZeeMass} shows comparisons between data and simulation after all corrections are applied for two cases
where both electrons are reconstructed in the ECAL barrel and endcaps, respectively.
In both cases the dielectron invariant mass spectra for the data and simulation are compatible within the uncertainties.

Once the photon energy correction has been applied, photon candidates are preselected 
before being used to form diphoton candidates.
Requirements are placed on the photons' kinematic, 
shower shape, and isolation variables at values at least as stringent as those applied in the trigger.
The preselection criteria are as follows:

\begin{itemize}
        \item minimum \pt of the leading and subleading photons greater than 35 and 25\GeV, respectively;
        \item pseudorapidity of the photons $\abs{\eta}<2.5$ and not in the barrel-endcap transition of $1.44 < \abs{\eta} < 1.57$;
        \item preselection on the {$\RNINE$} variable and on
               $\sigma_{\eta \eta}$ --- the lateral extension of the
               shower, defined as the energy-weighted spread within the
               $5{\times}5$ crystal matrix centred on
               the crystal with the largest energy deposit in the
               supercluster --- to reject ECAL energy deposits
               incompatible with a single, isolated electromagnetic
               shower, such as those coming from neutral mesons;
       \item preselection on the ratio of the energy in the HCAL tower behind
               the supercluster's seed cluster to the energy in the supercluster (H/E),
               in order to reject hadronic showers;
       \item electron veto, which rejects the photon candidate if its 
         supercluster in the ECAL is near to the extrapolated path of a track compatible with an electron. 
         Tracks compatible with a reconstructed photon conversion vertex are not considered when applying this veto.
       \item requirement on the photon isolation ($\mathcal{I}_{\text{ph}}$),
	       defined as the \pt sum of the particles
	       identified as photons inside a cone of size $R=0.3$
	       around the photon direction;
       \item requirement on the track isolation in a hollow cone
	       ($\mathcal{I}_{\text{tk}}$), the \pt sum 
	       of all tracks in a cone of size $R=0.3$
	       around the photon candidate direction, 
         excluding tracks in an inner cone of size $R=0.04$ to avoid counting tracks arising 
         from photon conversion into electron-positron pairs;
       \item loose requirement on charged-hadron isolation
	       ($\mathcal{I}_{\text{ch}}$), the \pt sum of
	       charged hadrons inside a cone of size $R=0.3$ around the
	       photon candidate.
\end{itemize}

\begin{table}
 \centering
 \topcaption{Schema of the photon preselection requirements. 
             The requirements depend both on whether a photon is in the barrel or endcap, 
             and on its \RNINE value.}
 \begin{tabular}{cccccc}
                            & \RNINE      & H/E     & $\sigma_{\eta \eta}$ & $\mathcal{I}_{\text{ph}}$ (GeV) & $\mathcal{I}_{\text{tk}}$ (GeV) \\
   \hline
   \multirow{2}{*}{Barrel}  & [0.50, 0.85] & $<$0.08 & $<$0.015               & $<$4.0                    &  $<$6.0                   \\
                            & $>$0.85      & $<$0.08 &          \NA           & \NA                       &\NA                        \\[\cmsTabSkip]
   \multirow{2}{*}{Endcaps} & [0.80, 0.90] & $<$0.08 & $<$0.035               & $<$4.0                    &  $<$6.0                   \\
                            & $>$0.90      & $<$0.08 &          \NA           & \NA                       &\NA                        \\
 \end{tabular}
 \label{table:preselcuts}
\end{table}

The geometrical acceptance requirement is applied to the supercluster position in the ECAL. 
The requirement on the photon \pt is applied after the vertex assignment, 
which is described in further detail in Section~\ref{sec:event_reco_vertex}.
The preselection thresholds are shown in Table~\ref{table:preselcuts}. 
Additionally, photons are required 
to satisfy at least one of $\RNINE>0.8$, $\mathcal{I}_{\text{ch}}/\PT^{\PGg}<0.3$, and $\mathcal{I}_{\text{ch}}<20\GeV$.

The preselection efficiency is measured with the tag-and-probe technique using \Zee events in data, 
while the efficiency of the electron veto is measured in \Zmumug events in data.

\subsection{Photon identification}

Photons in events passing the preselection criteria are further required 
to satisfy a photon identification criterion based on a BDT
trained to separate genuine (``prompt") photons from jets mimicking a photon signature. 
This ID BDT is trained on a simulated sample of \gamplusjet events, 
where prompt photons are used as the signal, while jets are used as the background. 
Input variables to the ID BDT include shower shape variables, isolation variables, 
the photon energy and $\eta$, and global event variables sensitive to pileup, 
such as the median energy density per unit area $\rho$~\cite{HIG-16-040}.

Simulated inputs for the photon ID BDT, both shower shape and isolation variables,  
are corrected to agree with data using a chained quantile regression (CQR) method~\cite{CQR}. 
This method was developed to improve the agreement in the photon ID BDT output between data and simulation, 
thus reducing the size of the associated systematic uncertainty relative to previous analyses.  
Corrections are derived using an unbiased set of electrons 
from \Zee events selected with a tag-and-probe method. 
The CQR comprises a set of BDTs that predict the cumulative distribution function (CDF) 
of a given input variable.
Its prediction is conditional upon three electron kinematic variables 
(\pt, $\abs{\eta}$, $\phi$) and $\rho$. 
The CDFs extracted in this way from data and simulated events
are then used to derive a correction factor to be applied to any given simulated electron. 
These correction factors morph the CDF of the simulated shower shape onto the one observed in data.

The CQR method accounts for correlations among the shower shape variables and adjusts the correlation 
in the simulation to match the one observed in data.
To achieve this, an ordered chain  of the shower shape variables is constructed. 
The CDF of the first shower shape variable is predicted solely from the electron kinematic variables 
and event $\rho$ values, while the corrected values 
of the previously processed shower shape variables are also added as inputs for subsequent predictions.
The order of the different shower shape variables in the chain is optimised 
to minimise the final discrepancy of the ID BDT score between data and simulation.

The isolation variables are not included in the chain since their correlation 
with the shower shape variables is negligible. 
Furthermore, there is a \pt threshold on the particle candidates included 
in the computation of the isolation variables. 
This causes these variables to follow a disjoint distribution, 
with a gap present between the peak at zero and a tail at positive values.
The CDF of the isolation variables are therefore constant over the range of values between zero
and the start of the tail, which prevents the use of the same technique used for the shower shape variables.
The CQR method is thus extended with additional BDTs that are used to match, 
again based on the electron kinematic variables and the event $\rho$ value, the relative population 
of the peak and tail between data and simulation.
The tails of the isolation variable distributions themselves are then morphed 
using the same technique for the shower shape variables.

A systematic uncertainty associated with the corrections is also included in the analysis. 
This is estimated by rederiving the corrections with equally sized subsets 
of the \Zee events used for training.
Its magnitude corresponds to the standard deviation of the event-by-event differences 
in the corrected ID BDT output score obtained with the two training subsets. 
This uncertainty reflects the limited capacity of the network 
arising from the finite size of the training set.
The size of the resulting experimental uncertainty is smaller than that required to
cover discrepancies between data and simulation in previous versions of this analysis.

\begin{figure}[htb!]
  \centering
  \includegraphics[width=0.49\textwidth]{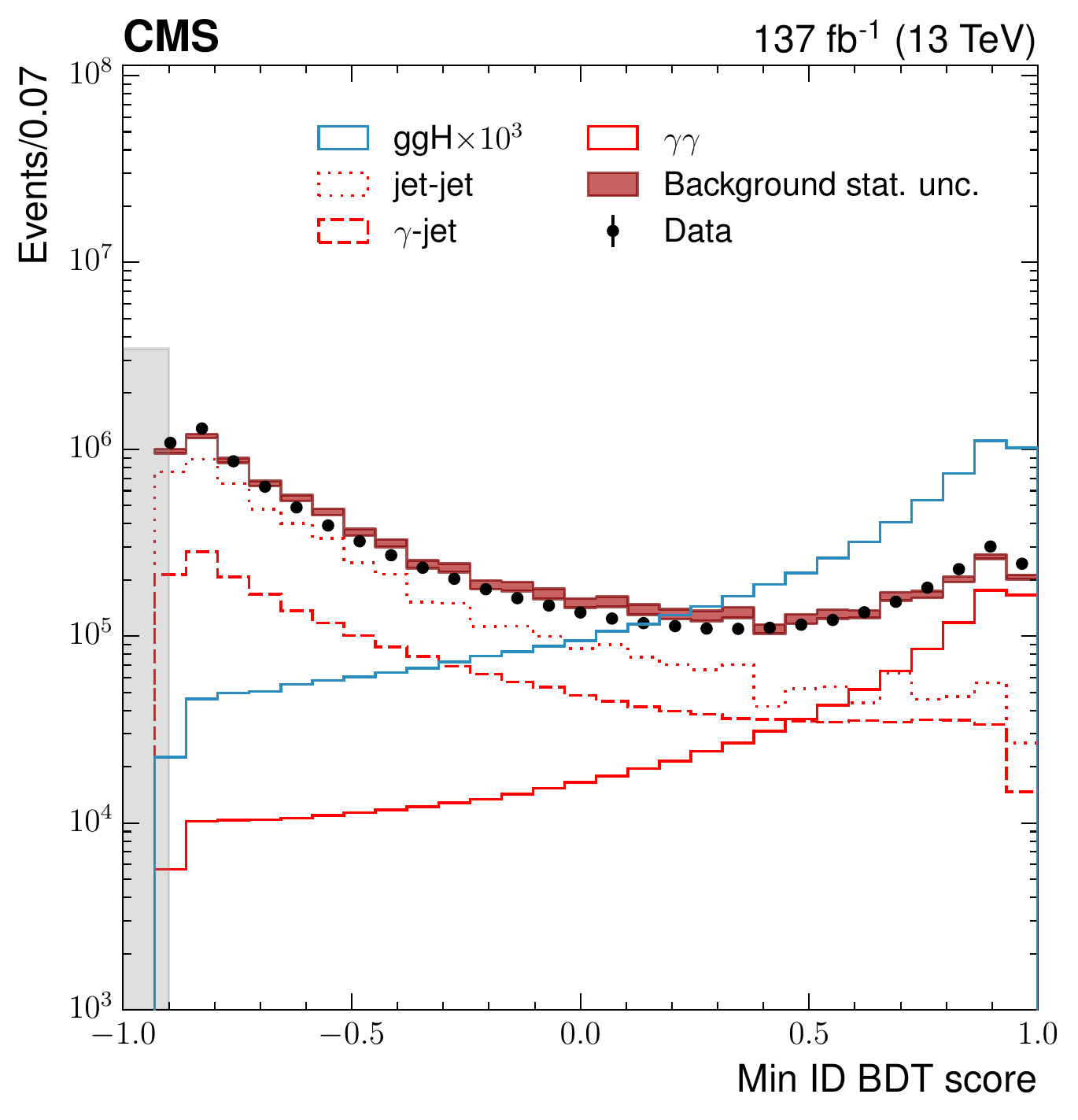}
  \includegraphics[width=0.49\textwidth]{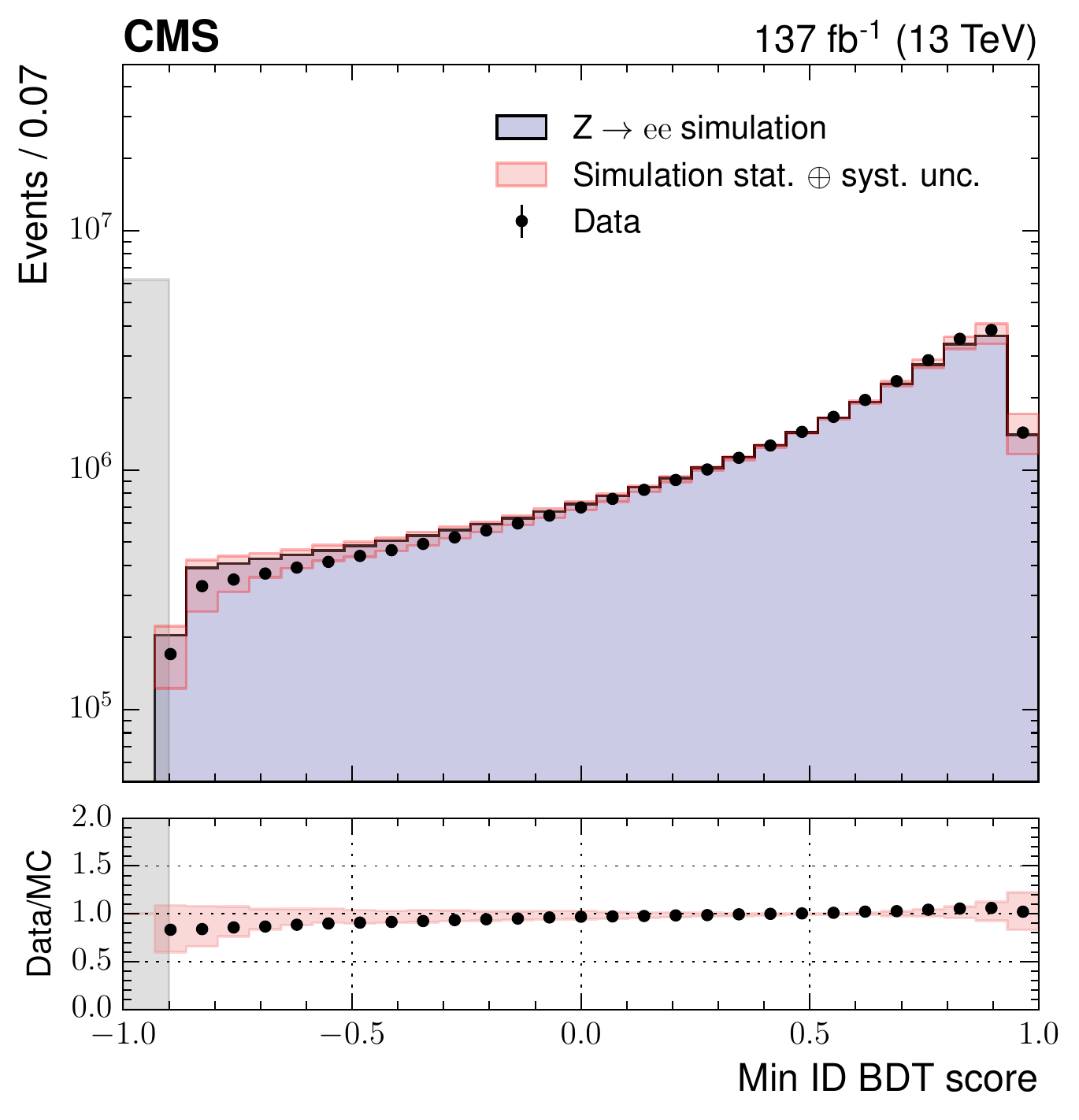}
  \caption{
  The left plot shows the distribution of the photon identification BDT score 
  of the lowest scoring photon in diphoton pairs with 
  $100 < \mgg < 180\GeV$, for data events passing the preselection (black points), 
  and for simulated background events (red band). 
  Histograms are also shown for different components of the simulated background.
  The blue histogram corresponds to simulated Higgs boson signal events.
  The right plot shows the same distribution 
  for \Zee events in data and simulation, where the electrons are reconstructed as photons. 
  The statistical and systematic uncertainty in simulation is also shown (pink band).
  Photons with an identification BDT score in the grey shaded region (below $-0.9$) are not considered in the analysis.
  The full data set collected in 2016--2018 and the corresponding simulation are shown.}
  \label{fig:IDMVA}
\end{figure}

The distribution of the photon ID BDT for the lowest scoring photon for signal events
and the different background components is shown in Fig.~\ref{fig:IDMVA}, 
together with a comparison of data and simulation using $\Zee$ events 
where the electrons are reconstructed as photons.
These \Zee events are chosen because of the similarity in the detector signature and reconstruction 
procedures for electrons and photons.
Here, the electrons being reconstructed as photons means that the track information is not used, 
and the energy is determined using the algorithm and corrections corresponding to photons rather than electrons.
The photon ID BDT distribution is also checked with photons using \Zmumug events, 
where data and simulation are found to agree within uncertainties. 

As an additional preselection criterion, 
photons are required to have a photon identification BDT score of at least $-0.9$.
Both photons pass this additional requirement in more than 99\% of simulated signal events.
The efficiency of the requirement in simulation is corrected to match that in data using \Zee events, 
and a corresponding systematic uncertainty is introduced.
 
\subsection{Diphoton vertex identification}
\label{sec:event_reco_vertex}

The determination of the primary vertex from which the two photons
originate has a direct impact on the \mgg resolution.
If the position along the beam axis ($z$) of the interaction producing
the diphoton is known to better than around 1\cm, the \mgg
resolution is dominated by the photon energy resolution.

The RMS of the distribution in $z$ of the reconstructed vertices in data in 2016--2018 varies in the range 
3.4--3.6\cm.
The corresponding distribution in each year's simulation is reweighted to match that in data.

The diphoton vertex assignment is performed using a BDT (the vertex identification BDT)
whose inputs are observables related to tracks
recoiling against the diphoton system \cite{HIG-16-040}.
It is trained on simulated \ggH events and identifies a single vertex in each event.

\begin{figure}
  \centering
  \includegraphics[width=0.47\textwidth]{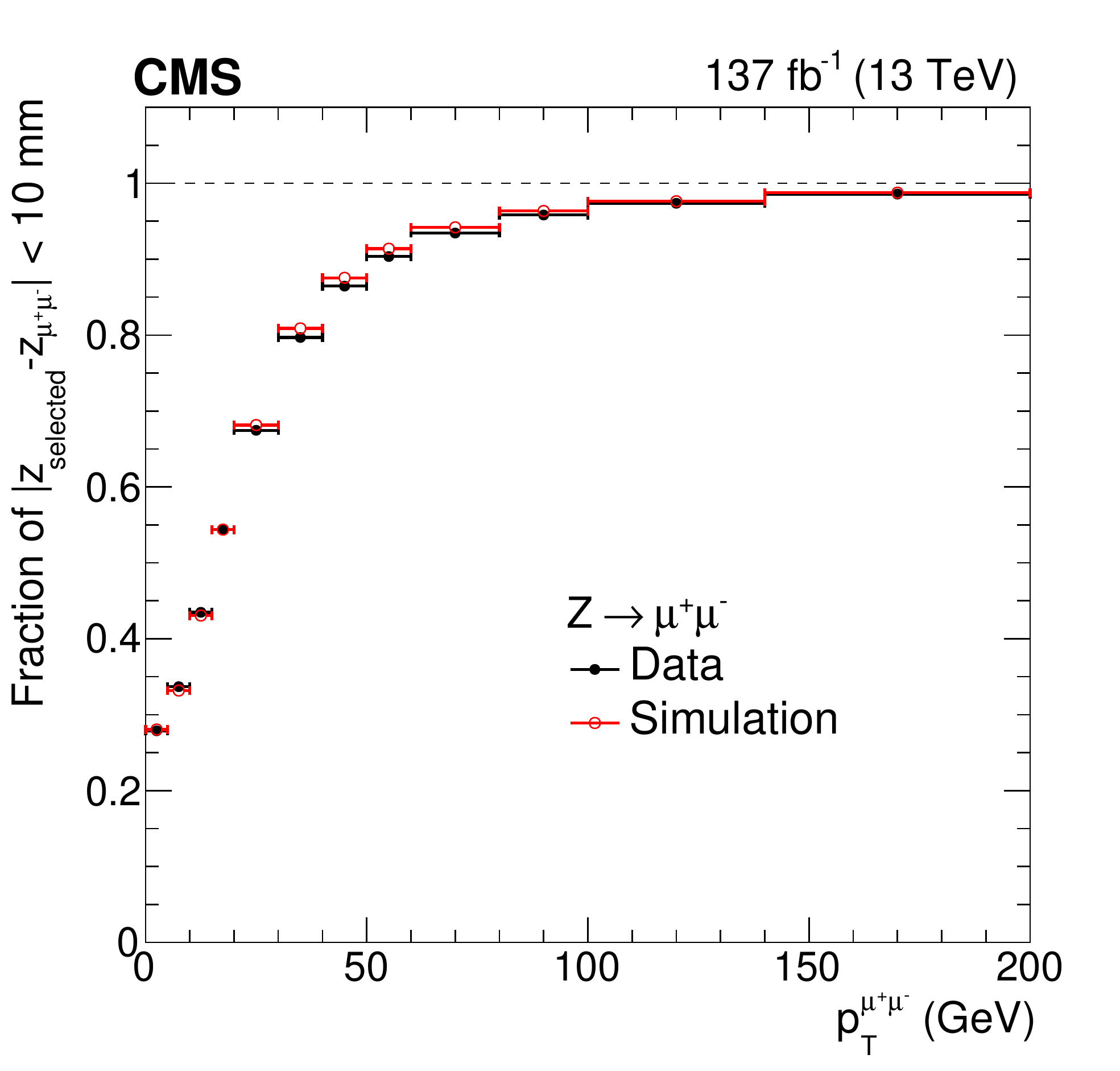}
  \includegraphics[width=0.49\textwidth]{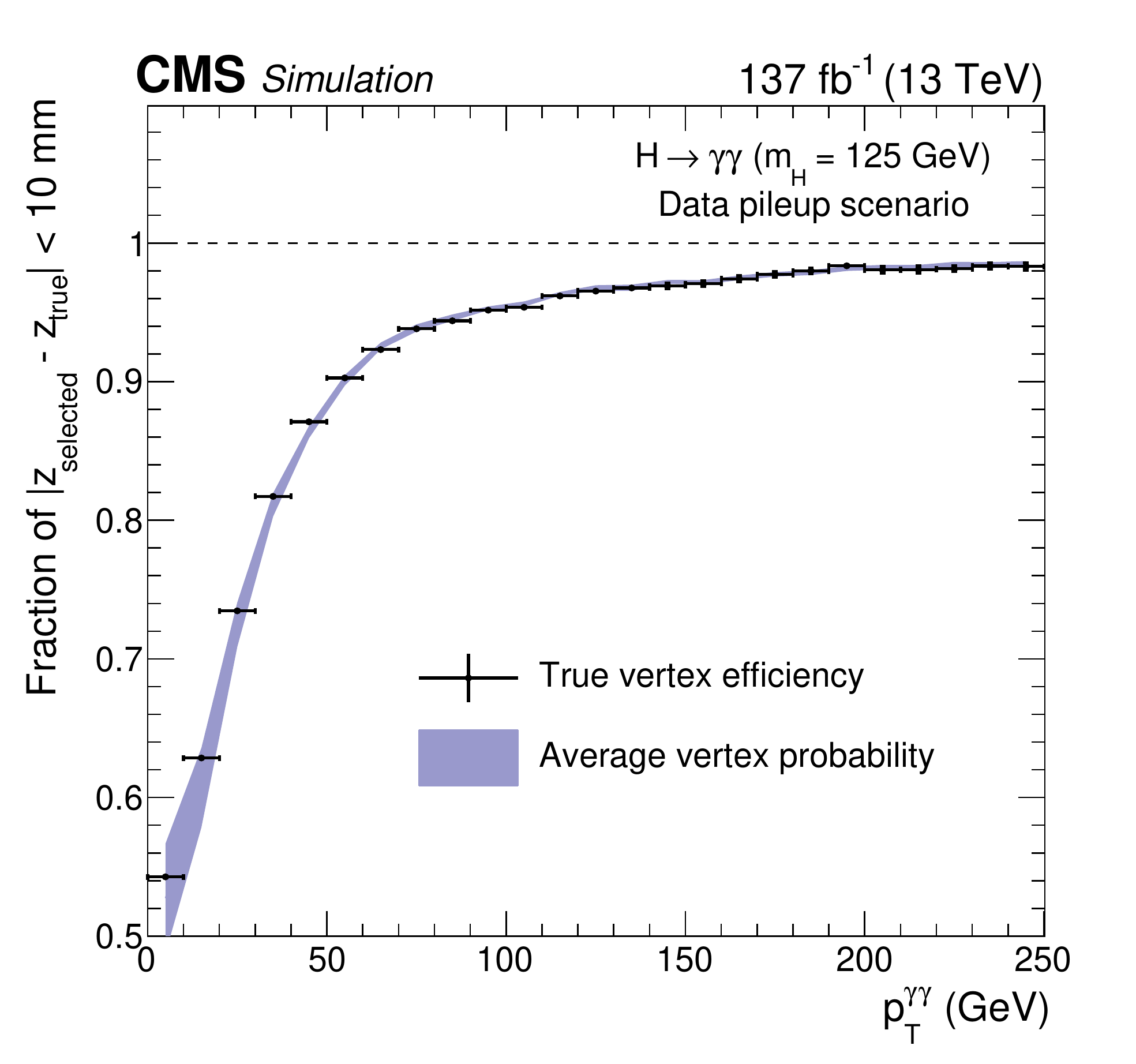}
  \caption{
  The left plot shows the validation of the \Hgg vertex identification algorithm on \Zmumu events, 
  where the muon tracks are omitted when performing the event reconstruction.
  This allows the fraction of events with the correctly assigned vertex estimated with simulation
  to be compared with data, as a function of the \pt of the dimuon system, 
  serving as a validation of the vertex identification BDT.
  Simulated events are weighted to match the distributions of pileup and distribution of vertices along the beam axis in data.
  The right plot demonstrates that the average vertex probability to be within 1\cm of the true vertex agrees with the true vertex efficiency
  in simulated events. 
  The full data set collected in 2016--2018 and the corresponding simulation are shown.
  }
  \label{fig:Zmumu}
\end{figure}

The performance of the vertex identification BDT is validated using \Zmumu events.
The vertices are refitted with the muon tracks omitted from the fit, to mimic a diphoton system.
Figure~\ref{fig:Zmumu} (left plot) shows the efficiency of correctly assigning the vertex, 
as a function of the dimuon \pt.
The data and simulation agree to within approximately 2\% across the entire \pt range. 
Nonetheless, the simulation is subsequently corrected to match the efficiencies measured in data, 
whilst preserving the total number of events.
A systematic uncertainty is introduced with a magnitude equal to the size of this correction.

The efficiency of assigning the diphoton vertex to be within
1\cm of the true vertex in simulated \Hgg events is approximately 79\%.
The events with an incorrectly-assigned vertex are primarily \ggH events with zero additional jets, 
and the associated systematic uncertainty affects \ggH events only.

A second vertex-related multivariate discriminant, the vertex probability BDT,
estimates the probability that the vertex, chosen by the vertex identification BDT, 
is within 1\cm of the vertex from which the diphoton originated.
The vertex probability BDT is trained on simulated \Hgg events using
input variables relating to the vertices in the event, 
their vertex identification BDT scores, the number of photons with 
associated conversion tracks, and the \pt of the diphoton system.
Agreement is observed between the average vertex probability 
and the vertex efficiency in simulation, 
as shown in Fig.~\ref{fig:Zmumu} (right plot).

\subsection{Additional objects}

Objects in the event other than the two photons are reconstructed as described in Section~\ref{sec:detector}.
Charged hadrons originating from interaction vertices other than the one chosen
by the vertex identification BDT are removed from the analysis.
In addition, all jets are required to have $\pt > 25\GeV$, be within $\abs{\eta} < 4.7$, 
and be separated from both photons by $\Delta R(\text{jet},\PGg) > 0.4$.
Depending on the analysis category, more stringent constraints on the jet \pt and $\abs{\eta}$ may be imposed;
this is described in the text where relevant.
In addition, some analysis categories require that jets also pass an identification criterion 
designed to reduce the number of selected jets originating from pileup collisions~\cite{PUJID}.
Jets from the hadronization of bottom quarks are tagged using a DNN that takes secondary vertices
and PF candidates as inputs~\cite{CMSbtagging}.

Electrons and muons are used in the analysis categories targeting \ttH and leptonic \VH production.
Electrons are required to have $\pt > 10\GeV$ and be within $\abs{\eta} < 2.4$, 
excluding the barrel-endcap transition region.
Muons must have $\pt > 5\GeV$ and fall within $\abs{\eta} < 2.4$.
In addition, isolation and identification requirements are imposed on both~\cite{ElectronPerformance,MuonPerformance}.

\section{Event categorisation}
\label{sec:categorisation}

The event selection in all analysis categories requires the two leading preselected photon candidates to have $\pt^{\PGg 1} > \mgg/3$ 
and $\pt^{\PGg 2} > \mgg/4$, respectively, with an invariant mass in the range $100 < \mgg < 180\GeV$.
The requirements on the scaled photon \pt
prevent distortions at the lower end of the \mgg spectrum.
As described in Section~\ref{sec:strategy}, events are divided into analysis categories 
to provide sensitivity to different production mechanisms and STXS bins.
Each analysis category is designed to select as many events as possible from a given STXS bin, 
or set of bins, referred to here as the target bin or bins.
The requirements for each analysis category should also select as few events from other, non-targeted
STXS bins as possible, to enable simultaneous measurements of different cross sections.
Finally, the selection should also reject as many background events as possible, 
to maximise the measurements' eventual sensitivity.
This section describes the several different categorisation schemes used 
for different event topologies, and the relevant STXS bins for each.

The STXS bins themselves are defined using particle-level quantities.
In all targeted bins, \Hrap is required to be less than 2.5. 
Jets are clustered using the anti-\kt algorithm~\cite{AntiKt} with a distance parameter of 0.4. 
All stable particles, except for those arising from the decay of the Higgs boson or the leptonic decay 
of an associated vector boson, are included in the clustering.
Jets are also required to have $\pt>30\GeV$. 
The definition of leptons includes electrons, muons, and tau leptons. 
Further details of the objects used to define the STXS bins can be found in Ref.~\cite{YR4}.

In many of the categorisation schemes, ML algorithms are used to classify signal events 
or discriminate between signal and background processes.
The output scores of the algorithms can then form part of the selection criteria used to define analysis categories.
Where these ML techniques are used to classify events, two types of validation are performed.
Firstly, in the typical case where simulated signal and background events are used to train the algorithm, 
a comparison of the simulated background to the corresponding data is performed.
Good agreement between the two gives confidence that the background processes are accurately modelled
and therefore that the ML algorithm performs well in its classification task.
Since the background model used in the final maximum likelihood fit is derived directly from data, 
poor agreement in background-like regions cannot induce any biases, 
but only result in sub-optimal performance of the classifier.
The second form of validation involves finding a signal-like region in which to compare 
the classifier output scores in simulation and data.
Here the aim of the comparison is to instil confidence that simulated Higgs boson signal events,
which do enter the final measurement, are sufficiently well-modelled.
Therefore simulation and data should be expected 
to agree within statistical and systematic uncertainties in these cases.
Furthermore, for all of the classifiers, the input variables are chosen such that \mgg
cannot be inferred.
This prevents distortion of the \mgg spectrum when applying selection thresholds on
the output scores.

A summary of all the analysis categories, together with the STXS bin or bins each analysis category targets, 
is given in Section~\ref{sec:categorisation_summary}.

\subsection{Event categories for \texorpdfstring{\ggH}{ggH} production}

The definitions of the \ggH STXS bins are given in Table~\ref{tab:ggH_definitions}, 
corresponding to the blue entries in Fig.~\ref{fig:allSTXSbins}.
The bins are defined using \ptH, the number of jets, and \mjj.
Those bins with $\ptH>200\GeV$ are referred to as ``BSM" bins 
because they have a cross section that is predicted to be low in the SM, 
but which could be enhanced by the presence of additional BSM particles.
Events originating from \ggZH production in which the \PZ boson decays hadronically are included in the definition of \ggH.
Analysis categories are defined to target each \ggH STXS bin independently, except for those in the VBF-like phase space. 
Events from the VBF-like bins are categorised separately, as described in Section~\ref{sec:vbf_categorisation}.

\begin{table}
    \topcaption{
    Definition of the \ggH STXS bins. 
    The product of the cross section and branching fraction ($\mathcal{B}$), 
    evaluated at \sqrts and $\mH = 125\GeV$, is given for each bin in the last column.
    The fraction of the total production mode cross section from each STXS bin is also shown.
    Events originating from \ggZH production, in which the \PZ decays hadronically, 
    are grouped together with the corresponding \ggH STXS bin 
    in the STXS measurements and are shown as a separate column in the table. 
    The \bbH production mode, whose $\sigma_{\text{SM}}\mathcal{B}=1.054\,\text{fb}$, 
    is grouped together with the \ggH 0J high \ptH bin.
    Unless stated otherwise, the STXS bins are defined for $\Hrap < 2.5$.
    Events with $\Hrap > 2.5$ are mostly outside of the experimental acceptance
    and therefore have a negligible contribution to all analysis categories.
    }
    \label{tab:ggH_definitions}
    \centering
    \cmsTable{
      \begin{tabular}{lcccc}
         \multirow{2}{*}{STXS bin} & \multirow{2}{*}{\begin{tabular}[c]{@{}c@{}}Definition\\ units of $\ptH$, $\mjj$ and $\ptHjj$ in \GeVns\end{tabular}} & \multicolumn{2}{c}{Fraction of cross section} & \multirow{2}{*}{$\sigma_{\text{SM}}\mathcal{B}$~(fb)} \\ 
          &  & \ggH & $\Pg\Pg\to\PZ(\qqbar)\PH$ &  \\ [\cmsTabSkip] \hline
         \ggH forward & $\Hrap > 2.5$ & 8.09\% & 2.73\% & 8.93 \\ [\cmsTabSkip]
         \ggH 0J low $\ptH$ & Exactly 0 jets, $\ptH<10$ & 13.87\% & 0.01\% & 15.30 \\ 
         \ggH 0J high $\ptH$ & Exactly 0 jets, $10<\ptH<200$ & 39.40\% & 0.29\% & 43.45 \\ [\cmsTabSkip]
         \ggH 1J low $\ptH$ & Exactly 1 jet, $\ptH<60$ & 14.77\% & 2.00\% & 16.29 \\ 
         \ggH 1J med $\ptH$ & Exactly 1 jet, $60<\ptH<120$ & 10.23\% & 5.34\% & 11.29 \\ 
         \ggH 1J high $\ptH$ & Exactly 1 jet, $120<\ptH<200$ & 1.82\% & 3.53\% & 2.01 \\  [\cmsTabSkip]
         \ggH $\geq$2J low $\ptH$ & At least 2 jets, $\ptH<60$, $\mjj<350$ & 2.56\% & 5.74\% & 2.83 \\ 
         \ggH $\geq$2J med $\ptH$ & At least 2 jets, $60<\ptH<120$, $\mjj<350$ & 4.10\% & 19.63\% & 4.56 \\ 
         \ggH $\geq$2J high $\ptH$ & At least 2 jets, $120<\ptH<200$, $\mjj<350$ & 1.88\% & 29.55\% & 2.13 \\  [\cmsTabSkip]
         \ggH BSM $200<\ptH<300$ & No jet requirements, $200<\ptH<300$ & 0.98\% & 13.93\% & 1.11 \\ 
         \ggH BSM $300<\ptH<450$ & No jet requirements, $300<\ptH<450$ & 0.25\% & 3.86\% & 0.28 \\ 
         \ggH BSM $450<\ptH<650$ & No jet requirements, $450<\ptH<650$ & 0.03\% & 0.77\% & 0.03 \\
         \ggH BSM $\ptH>650$ & No jet requirements, $\ptH>650$ & 0.01\% & 0.20\% & 0.01 \\ [\cmsTabSkip]
         \ggH VBF-like low \mjj low \ptHjj & \begin{tabular}[c]{@{}c@{}}At least 2 jets,   $\ptH<200$,\\ $350<\mjj<700$, $\ptHjj<25$\end{tabular} & 0.63\% & 1.14\% & 0.70 \\ 
         \ggH VBF-like low \mjj high \ptHjj & \begin{tabular}[c]{@{}c@{}}At least 2 jets,  $\ptH<200$,\\ $350<\mjj<700$, $\ptHjj>25$\end{tabular} & 0.77\% & 8.06\% & 0.86 \\ 
         \ggH VBF-like high \mjj low \ptHjj & \begin{tabular}[c]{@{}c@{}}At least 2 jets,  $\ptH<200$,\\ $\mjj>700$, $\ptHjj<25$\end{tabular} & 0.28\% & 0.36\% & 0.31 \\ 
         \ggH VBF-like high \mjj high \ptHjj & \begin{tabular}[c]{@{}c@{}}At least 2 jets, $\ptH<200$,\\ $\mjj>700$, $\ptHjj>25$\end{tabular} & 0.32\% & 2.85\% & 0.36 \\ 
      \end{tabular}
    }
\end{table}

The \ggH categorisation procedure can be summarised as follows.
First, events are classified using the so-called \ggH BDT. 
The \ggH BDT predicts the probability that a diphoton event belongs to a given \ggH STXS class. 
Each class corresponds either to an individual STXS bin or to a set of multiple STXS bins.
The first eight classes considered by the \ggH BDT are individual STXS bins. 
These comprise the zero, one, and two jet bins with $\ptH<200\GeV$ and $\mjj<350\GeV$, 
corresponding to the eight leftmost \ggH STXS bins in Fig.~\ref{fig:allSTXSbins}.
To minimise model-dependence, the \ggH BDT is not trained to distinguish between 
the STXS bins with $\ptH > 200\GeV$.
Instead, all events with $\ptH > 200\GeV$ are treated as a single class, 
consisting of a set of four STXS bins.
Hence, the task of the \ggH BDT amounts to predicting one of nine \ggH classes, 
which are uniquely defined by \ptH and the number of jets. 
Each event is then assigned to an analysis category based upon its most probable STXS bin,
as determined by the \ggH BDT.
Events for which the maximum probability corresponds to the $\ptH > 200\GeV$ class 
are assigned into an analysis category targeting one of the four STXS bins with $\ptH > 200\GeV$. 
This assignment is performed using the event's reconstructed \ptgg value. 
Finally, the analysis' sensitivity is maximised by further dividing the analysis categories using the diphoton BDT, 
which is trained to discriminate between signal and background processes and described in further detail below.

The \ggH BDT is trained using simulated \ggH events only.
Input features to the \ggH BDT are properties of the photons 
and quantities related to the kinematic properties of up to three $\pt > 20\GeV$ jets.
The photon features used are the photon kinematic variables, ID BDT scores, 
\mgg resolution estimates, and the vertex probability estimate.
The \ptgg value is also included as an input.
As previously mentioned, the set of variables is chosen such that \mgg cannot be inferred from the inputs;
for this reason, the \pt/\mgg values of each photon, rather than \pt, are used. 
The variables related to jets include the kinematic variables and pileup ID scores of the three leading jets in the event. 

The resulting STXS bin assignment performs better than simply using the reconstructed \ptgg and number of jets. 
The fraction of selected \ggH events in simulation that are assigned to the correct STXS bin 
increases from 77 to 82\% when using the \ggH BDT rather than the reconstructed \ptgg and number of jets.
This improvement can be explained by the fact that the \ggH BDT is able to 
exploit the correlations between the photon and jet kinematic properties.
In this way, the well-measured photon quantities can be used to infer information about the less well-measured jets. 
As a result, the contamination of analysis categories due to migration across jet bins is reduced; 
the migration across \ptgg boundaries is much smaller and essentially unchanged by the \ggH BDT. 
The \ggH BDT therefore slightly improves the analysis sensitivity, most noticeably in the zero- and one-jet bins. 
Furthermore, the correlations between cross section parameters in the final fits are reduced. 

\begin{figure}[htb!]
  \centering
  \includegraphics[width=0.6\textwidth]{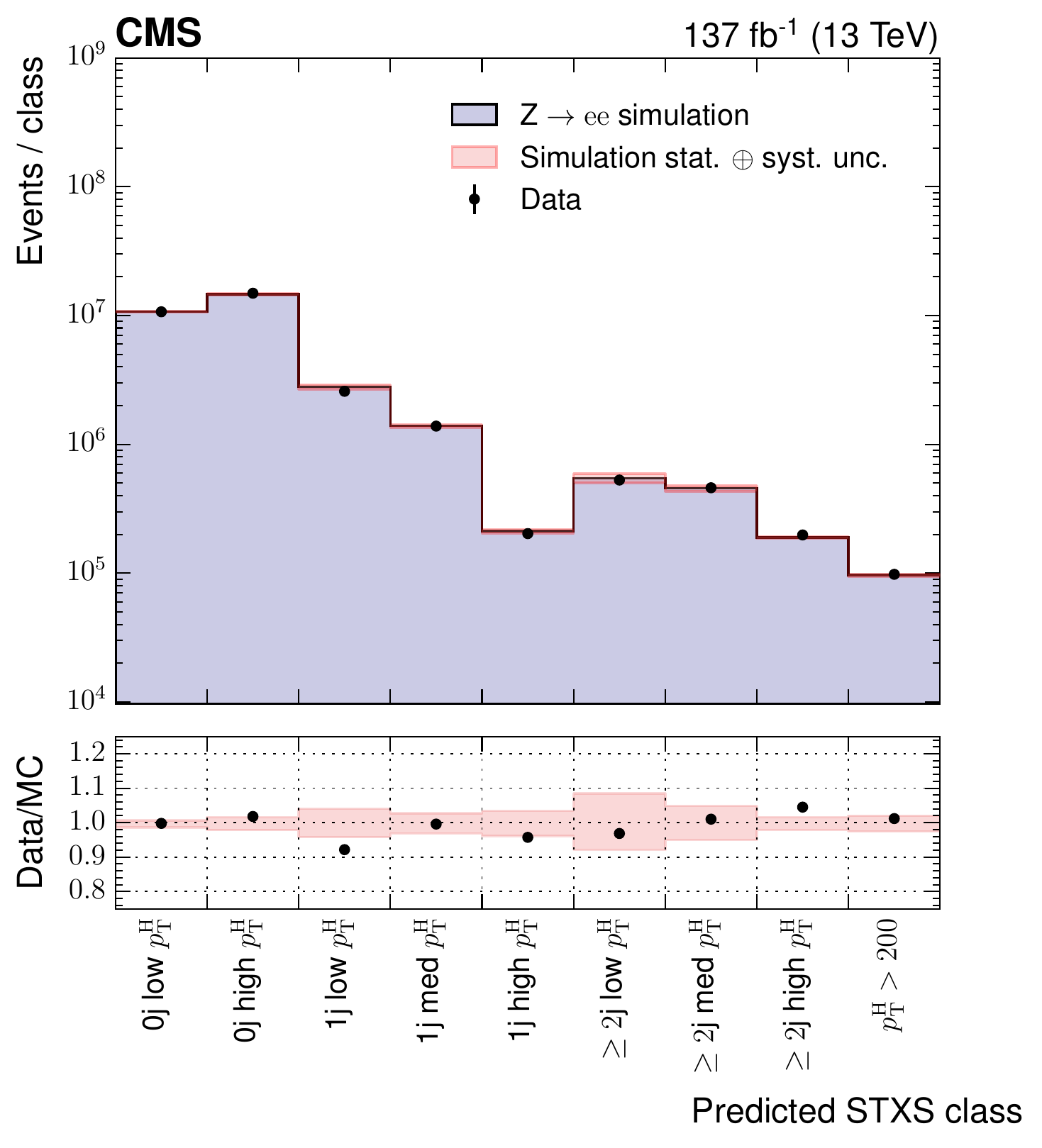} 
  \caption{The most probable STXS class from the \ggH BDT in \Zee events where 
           the electrons are reconstructed as photons is shown.
           The points show the predicted class for data, 
           whilst the histogram shows predicted score for simulated Drell--Yan events, 
           including statistical and systematic uncertainties (pink band). 
           The full data set collected in 2016--2018 and the corresponding simulation are shown.}
  \label{fig:multiclass_validation}
\end{figure}

To validate the modelling of the \ggH BDT and its input variables, 
the agreement in the STXS class prediction between data and simulation in \Zee events, with electrons reconstructed as photons, is checked. 
Figure~\ref{fig:multiclass_validation} shows the number of events predicted to belong to each event class.
The uncertainties in the photon ID BDT, the photon energy resolution, 
and the jet energy scale and resolution are included.
There is good agreement between data and simulation in this signal-like region.

The diphoton BDT is used, after events are classified by the \ggH BDT, to reduce the background 
from SM diphoton production, thereby maximising the analysis sensitivity.
The diphoton BDT is trained with all Higgs boson signal events against 
SM diphoton production as background.
A high score is assigned to events with photons showing signal-like kinematic properties, 
good \mgg resolution, and high photon identification BDT score. 
The input variables to the classifier are the photon kinematic variables, ID BDT scores, 
\mgg resolution estimates and the vertex probability estimate.

\begin{figure}[htb!]
  \centering
  \includegraphics[width=0.49\textwidth]{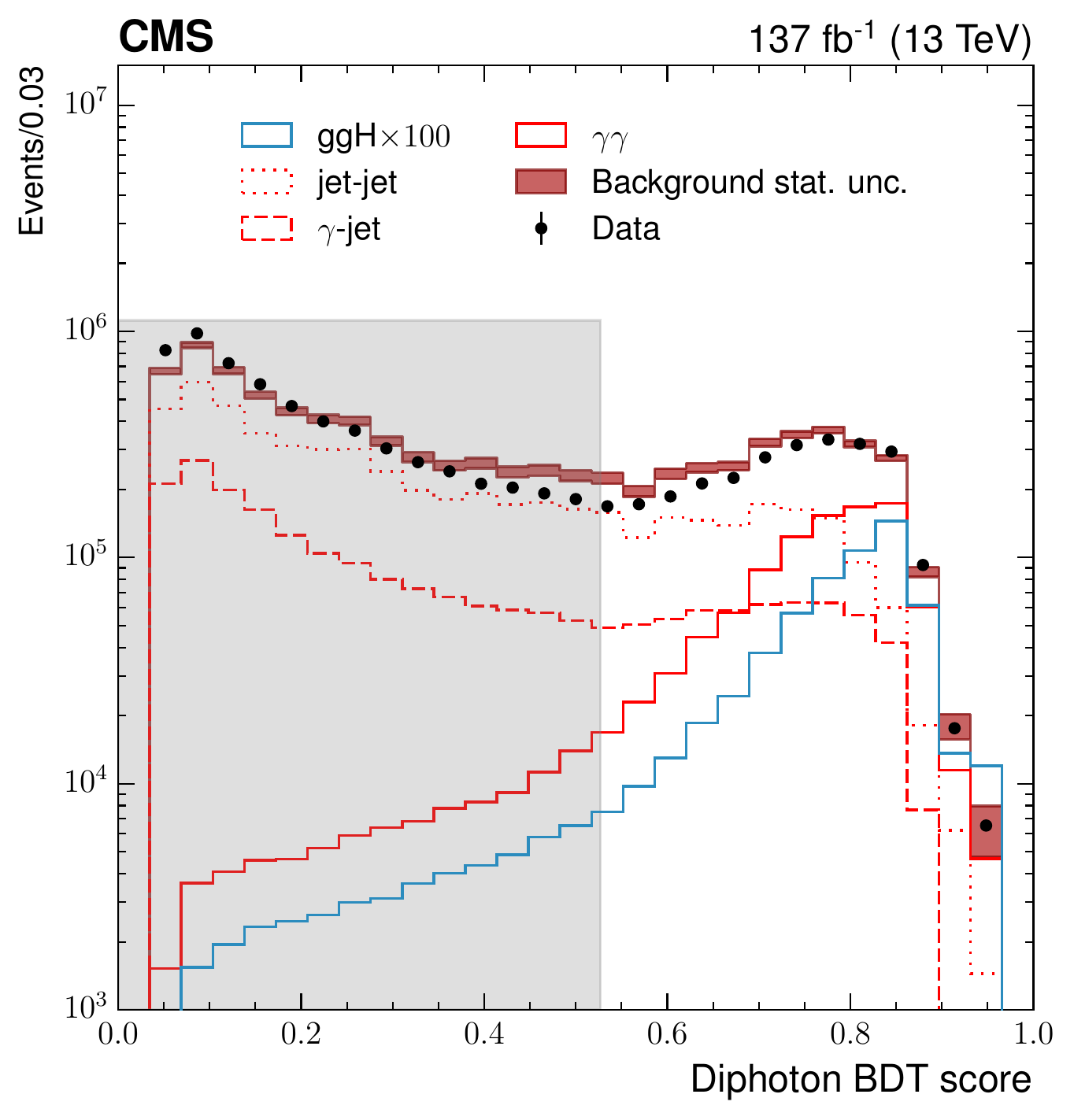} 
  \includegraphics[width=0.49\textwidth]{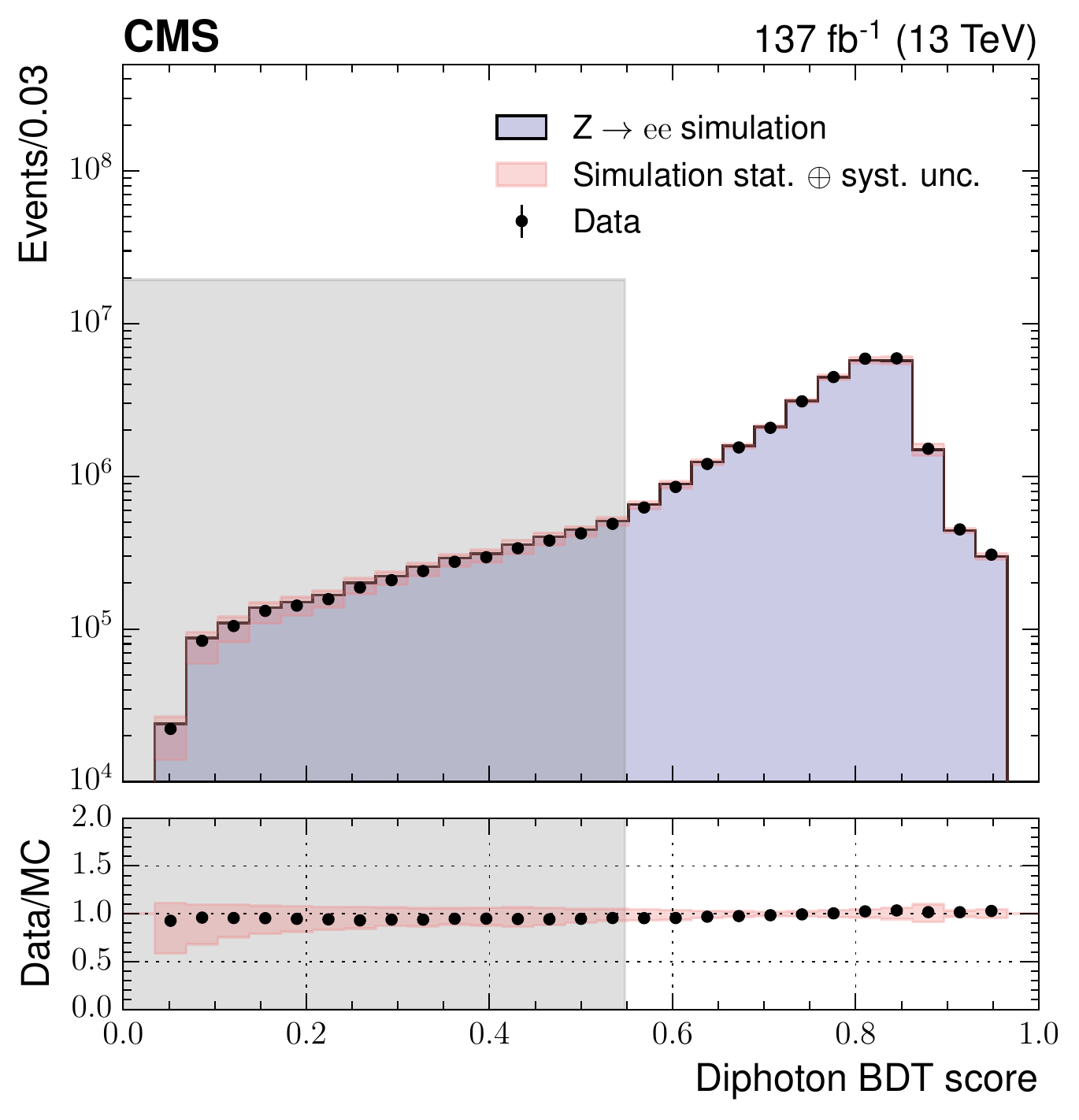} 
  \caption{
  The left plot shows the distribution of the diphoton BDT score 
  in events with \mgg in the range 100--120 or 130--180\GeV, 
  for data events passing the preselection (black points), 
  and for simulated background events (red band). 
  Histograms are also shown for different components of the simulated background in red.
  The blue histogram corresponds to simulated Higgs boson signal events ($\times 100$).
  The right plot shows the same distribution in \Zee events where 
  the electrons are reconstructed as photons. 
  The points show the score for data, 
  the histogram shows the score for simulated Drell--Yan events, 
  including statistical and systematic uncertainties (pink band). 
  The regions shaded grey contain diphoton BDT scores 
  below the lowest threshold used to define an analysis category.
  The full data set collected in 2016--2018 and the corresponding simulation are shown.}
  \label{fig:diphoBDT}
\end{figure}

Figure~\ref{fig:diphoBDT} shows the output score of the diphoton BDT for signal and background
events, together with corresponding data from the \mgg sidebands, 
meaning $100 < \mgg < 120\GeV$ or $130 < \mgg < 180\GeV$.
A validation of the diphoton BDT obtained in \Zee events,
where the electrons are reconstructed as photons, is also shown in Fig.~\ref{fig:diphoBDT}.
Here the data and simulation agree within the statistical and systematic uncertainties.

After being classified by the \ggH BDT, events are divided into analysis categories using the diphoton BDT, 
with the boundaries chosen to maximise the expected sensitivity.
The resulting analysis categories are referred to as "tags".
For \ggH production, there is at least one tag targeting each individual STXS bin, 
except for the VBF-like bins.
The tag names are given in decreasing order of the expected ratio of signal-to-background events (S/B). 
For example, the tag with the highest S/B targeting the \ggH zero jet bin 
with $\ptH < 10\GeV$ is denoted 0J low \ptgg Tag 0.

The expected signal and background yields in each \ggH analysis category are shown in Table~\ref{tab:ggH_yields}.
The yields shown in this and subsequent tables correspond to those in the final analysis categories, 
meaning that events selected by analysis categories with higher priority are not considered.

\begin{table}
    \topcaption{The expected number of signal events for $\mH = 125\GeV$
    in analysis categories targeting \ggH production, excluding those targeting the VBF-like phase space,
    shown for an integrated luminosity of 137\fbinv. 
    The fraction of the total number of events arising from each production mode in each analysis category is provided, 
    as is the fraction of events originating from the targeted STXS bin or bins. 
    Entries with values less than 0.05\% are not shown. 
    Here \qqH includes contributions from both VBF and hadronic \VH production, 
    whilst ``Top" includes \ttH and \tH together. 
    The $\seff$, defined as the smallest interval containing 68.3\% of the \mgg distribution, 
    is listed for each analysis category. 
    The final column shows the expected ratio of signal to signal-plus-background, S/(S+B), 
    where S and B are the numbers of expected signal and background events 
    in a $\pm1\seff$ window centred on $\mH$.}
    \label{tab:ggH_yields}
    \centering
    \cmsTable{
      \begin{tabular}{lccccccccc}
          \multirow{3}{*}{Analysis categories} & \multicolumn{8}{c}{SM 125\GeV Higgs boson expected signal} & \multirow{3}{*}{S/(S+B)} \\
           & \multirow{2}{*}{Total} & \multirow{2}{*}{\begin{tabular}[c]{@{}c@{}}Target\\STXS bin(s)\end{tabular}} & \multicolumn{5}{c}{Fraction of total events} & \multirow{2}{*}{\begin{tabular}[c]{@{}c@{}}$\seff$\\(GeV)\end{tabular}} & \\
           & & & \ggH & \bbH & \qqH & \VH lep & Top & & \\ \hline
           0J low $\ptgg$ Tag0 & 296.2 & 86.6\% & 97.9\% & 1.1\% & 0.8\% & 0.1\% &\NA& 1.89 & 0.06 \\
           0J low $\ptgg$ Tag1 & 340.0 & 88.5\% & 98.0\% & 1.0\% & 0.8\% & 0.1\% &\NA& 2.31 & 0.03 \\
           0J low $\ptgg$ Tag2 & 279.6 & 89.3\% & 98.1\% & 1.0\% & 0.8\% & 0.1\% &\NA& 2.53 & 0.02 \\
           [\cmsTabSkip]
           0J high $\ptgg$ Tag0 & 612.4 & 81.9\% & 95.6\% & 1.4\% & 2.6\% & 0.4\% &\NA& 1.64 & 0.09 \\
           0J high $\ptgg$ Tag1 & 1114.6 & 79.4\% & 95.4\% & 1.3\% & 2.8\% & 0.4\% &\NA& 2.19 & 0.05 \\
           0J high $\ptgg$ Tag2 & 1162.6 & 78.3\% & 95.3\% & 1.4\% & 2.7\% & 0.5\% &\NA& 2.56 & 0.02 \\
           [\cmsTabSkip]
           1J low $\ptgg$ Tag0 & 132.0 & 66.2\% & 88.8\% & 0.8\% & 9.4\% & 0.8\% & 0.1\% & 1.53 & 0.11 \\
           1J low $\ptgg$ Tag1 & 340.0 & 66.3\% & 88.6\% & 0.8\% & 9.6\% & 0.9\% & 0.1\% & 1.95 & 0.05 \\
           1J low $\ptgg$ Tag2 & 260.6 & 66.2\% & 88.3\% & 0.8\% & 9.7\% & 1.0\% & 0.1\% & 2.37 & 0.02 \\
           [\cmsTabSkip]
           1J med $\ptgg$ Tag0 & 184.1 & 65.2\% & 81.7\% & 0.5\% & 16.3\% & 1.4\% & 0.2\% & 1.65 & 0.15 \\
           1J med $\ptgg$ Tag1 & 310.2 & 66.3\% & 83.6\% & 0.4\% & 14.3\% & 1.6\% & 0.1\% & 1.91 & 0.08 \\
           1J med $\ptgg$ Tag2 & 291.4 & 65.0\% & 83.7\% & 0.5\% & 13.8\% & 1.8\% & 0.2\% & 2.13 & 0.03 \\
           [\cmsTabSkip]
           1J high $\ptgg$ Tag0 & 37.3 & 61.9\% & 75.7\% & 0.2\% & 22.8\% & 1.0\% & 0.2\% & 1.55 & 0.30 \\
           1J high $\ptgg$ Tag1 & 31.2 & 61.7\% & 75.0\% & 0.3\% & 23.4\% & 1.1\% & 0.2\% & 1.73 & 0.16 \\
           1J high $\ptgg$ Tag2 & 80.9 & 62.2\% & 76.5\% & 0.2\% & 21.5\% & 1.6\% & 0.2\% & 1.97 & 0.07 \\
           [\cmsTabSkip]
           $\geq$2J low $\ptgg$ Tag0 & 17.7 & 52.7\% & 76.7\% & 0.6\% & 19.0\% & 1.3\% & 2.4\% & 1.56 & 0.06 \\
           $\geq$2J low $\ptgg$ Tag1 & 57.6 & 54.0\% & 74.4\% & 0.6\% & 20.5\% & 1.4\% & 3.0\% & 1.88 & 0.03 \\
           $\geq$2J low $\ptgg$ Tag2 & 43.9 & 50.5\% & 72.7\% & 0.6\% & 20.8\% & 1.7\% & 4.2\% & 2.46 & 0.01 \\
           [\cmsTabSkip]
           $\geq$2J med $\ptgg$ Tag0 & 21.2 & 64.9\% & 80.6\% & 0.3\% & 16.3\% & 1.0\% & 1.8\% & 1.42 & 0.17 \\
           $\geq$2J med $\ptgg$ Tag1 & 70.1 & 61.4\% & 77.9\% & 0.3\% & 18.1\% & 1.1\% & 2.6\% & 1.82 & 0.07 \\
           $\geq$2J med $\ptgg$ Tag2 & 135.4 & 57.5\% & 74.8\% & 0.4\% & 19.7\% & 1.4\% & 3.8\% & 2.08 & 0.03 \\
           [\cmsTabSkip]
           $\geq$2J high $\ptgg$ Tag0 & 29.0 & 65.5\% & 77.8\% & 0.2\% & 18.7\% & 1.3\% & 2.1\% & 1.48 & 0.23 \\
           $\geq$2J high $\ptgg$ Tag1 & 52.5 & 62.3\% & 76.1\% & 0.2\% & 19.6\% & 1.5\% & 2.6\% & 1.76 & 0.11 \\
           $\geq$2J high $\ptgg$ Tag2 & 45.5 & 58.4\% & 73.8\% & 0.2\% & 20.4\% & 1.9\% & 3.7\% & 1.92 & 0.05 \\
           [\cmsTabSkip]
           BSM $200<\ptgg<300$ Tag0 & 30.7 & 75.8\% & 77.5\% & 0.2\% & 19.4\% & 1.2\% & 1.6\% & 1.41 & 0.39 \\
           BSM $200<\ptgg<300$ Tag1 & 39.6 & 69.9\% & 73.8\% & 0.1\% & 21.5\% & 1.7\% & 2.8\% & 1.90 & 0.11 \\
           [\cmsTabSkip]
           BSM $300<\ptgg<450$ Tag0 & 15.5 & 74.8\% & 76.3\% & 0.1\% & 19.7\% & 1.7\% & 2.2\% & 1.53 & 0.34 \\
           BSM $300<\ptgg<450$ Tag1 & 2.6 & 66.3\% & 67.9\% & 0.1\% & 22.5\% & 2.6\% & 7.0\% & 1.42 & 0.09 \\
           [\cmsTabSkip]
           BSM $450<\ptgg<650$ & 3.1 & 58.1\% & 61.8\% & 0.1\% & 30.0\% & 2.4\% & 5.6\% & 1.55 & 0.20 \\
           [\cmsTabSkip]
           BSM $\ptgg>650$ & 0.9 & 72.5\% & 72.3\% & 0.1\% & 21.0\% & 2.9\% & 3.8\% & 1.21 & 0.36 \\
           [\cmsTabSkip]
\end{tabular}
    }
\end{table}

\subsection{Event categories for VBF production}
\label{sec:vbf_categorisation}

In the STXS framework, the \qqH production mode includes both VBF events 
and \VH events where the vector boson decays hadronically.
Within \qqH production, there are five STXS bins that correspond to typical VBF-like events, 
with a single bin for \VH-like events.
The precise definitions of the \qqH STXS bins are given in Table~\ref{tab:vbf_definitions}. 
These correspond to the orange entries in Fig.~\ref{fig:allSTXSbins}.

\begin{table}
    \topcaption{
    Definition of the \qqH STXS bins. 
    The product of the cross section and branching fraction ($\mathcal{B}$),
    evaluated at \sqrts and $\mH = 125\GeV$, is given for each bin in the last column.
    The fraction of the total production mode cross section from each STXS bin is also shown.
    Unless stated otherwise, the STXS bins are defined for $\Hrap < 2.5$.
    Events with $\Hrap > 2.5$ are mostly outside of the experimental acceptance
    and therefore have a negligible contribution to all analysis categories.
    }
    \label{tab:vbf_definitions}
    \centering
    \cmsTable{
      \begin{tabular}{lccccc}
         \multirow{2}{*}{STXS bin} & \multirow{2}{*}{\begin{tabular}[c]{@{}c@{}}Definition\\ units of $\ptH$, $\mjj$ and $\ptHjj$ in \GeVns\end{tabular}} & \multicolumn{3}{c}{Fraction of cross section} & \multirow{2}{*}{$\sigma_{\text{SM}}\mathcal{B}$~(fb)} \\ 
          &  & VBF & $\qqbar'\to\PW(\qqbar')\PH$ & $\qqbar\to\PZ(\qqbar)\PH$ &  \\ [\cmsTabSkip] \hline
         \qqH forward & $\Hrap > 2.5$ & 6.69\% & 12.57\% & 9.84\% & 0.98 \\ [\cmsTabSkip]
         \qqH 0J & Exactly 0 jets & 6.95\% & 5.70\% & 3.73\% & 0.77 \\ 
         \qqH 1J & Exactly 1 jet & 32.83\% & 31.13\% & 25.03\% & 3.82 \\ 
         \qqH $\mjj<60$ & At least 2 jets, $\mjj<60$ & 1.36\% & 3.58\% & 2.72\% & 0.23 \\ 
         \qqH \VH-like & At least 2 jets, $60<\mjj<120$ & 2.40\% & 29.43\% & 28.94\% & 1.23 \\ 
         \qqH $120<\mjj<350$ & At least 2 jets, $120<\mjj<350$ & 12.34\% & 13.92\% & 12.59\% & 1.53 \\ [\cmsTabSkip]
         \qqH VBF-like low \mjj low \ptHjj & \begin{tabular}[c]{@{}c@{}}At least 2 jets,   $\ptH<200$,\\ $350<\mjj<700$, $\ptHjj<25$\end{tabular} & 10.26\% & 0.44\% & 0.35\% & 0.90 \\ 
         \qqH VBF-like low \mjj high \ptHjj & \begin{tabular}[c]{@{}c@{}}At least 2 jets,  $\ptH<200$,\\ $350<\mjj<700$, $\ptHjj>25$\end{tabular} & 3.85\% & 1.86\% & 1.74\% & 0.39 \\ 
         \qqH VBF-like high \mjj low \ptHjj & \begin{tabular}[c]{@{}c@{}}At least 2 jets,  $\ptH<200$,\\ $\mjj>700$, $\ptHjj<25$\end{tabular} & 15.09\% & 0.09\% & 0.08\% & 1.30 \\ 
         \qqH VBF-like high \mjj high \ptHjj & \begin{tabular}[c]{@{}c@{}}At least 2 jets, $\ptH<200$,\\ $\mjj>700$, $\ptHjj>25$\end{tabular} & 4.25\% & 0.40\% & 0.39\% & 0.38 \\ 
         \qqH BSM & At least 2 jets, $\mjj>350$, $\ptH>200$ & 3.98\% & 0.88\% & 0.71\% & 0.37 \\ 
      \end{tabular}
    }
\end{table}

Events with a dijet system characteristic of the VBF production mode 
have a dedicated categorisation scheme in this analysis, described in this section.
Those events where the dijet is instead consistent with the decay of a vector boson 
are categorised separately, as described in Section~\ref{sec:vh_had_categorisation}.
No analysis categories are constructed to target the zero or one jet \qqH STXS bins, 
nor those with $\mjj<60\GeV$ or $120<\mjj<350\GeV$.

Following the STXS binning scheme, the particle-level definition of the VBF-like dijet system
requires two jets with $\pt > 30\GeV$, and whose $\mjj > 350\GeV$. 
These bins are defined analogously for EW \qqH production as well as from \ggH production.
When constructing the corresponding analysis categories at reconstruction level, 
a dijet preselection is applied that requires two jets within $\abs{\eta} < 4.7$, 
with $\pt > 40 (30)\GeV$ for the leading (subleading) jet, in addition to $\mjj > 350\GeV$.
Jets are also required to pass a threshold on a pileup identification score. 

The so-called dijet BDT is trained to estimate the probability that an event passing the VBF preselection
originated from VBF, \ggH, or non-Higgs boson SM diphoton production.
The inputs to the dijet BDT include various jet kinematic and angular variables, 
as well as the $\pt/\mgg$ of each photon and angular variables involving both jets and photons.
These inputs for VBF, \ggH, and non-Higgs boson SM production of two prompt photons 
are taken from simulation.
However, the modelling of backgrounds, where at least one of the two photons is a misreconstructed jet, is poor, 
predominantly due to the fact that very few simulated events pass the selection criteria. 
In this analysis, an approach is adopted whereby the simulated background events with
nonprompt photons are replaced with data from a dedicated control sample. 

The control sample is defined using the sideband of the photon ID BDT distribution, 
by requiring at least one photon ID BDT score to be below $-0.5$. 
The events in this control sample can potentially have both a different normalisation and 
different kinematic properties from those in the signal region.
To correct this, the events are reweighted in bins of the \pt and $\abs{\eta}$ of the photon passing the ID BDT requirement.
The required weights are derived from simulation, 
by estimating both the fraction of background events that contain nonprompt photons
and the ratio of the expected number of events in the signal region to the control sample.
The product of these two factors is applied as a weight to each data event in the control sample, 
and these reweighted events are subsequently used to train the dijet BDT. 

The resulting distributions of the dijet BDT input variables are compared to the 
\mgg sideband data and are found to be in reasonable agreement.
Furthermore, the increase in the number of events available for the training of the dijet BDT
leads to an improvement in its discrimination power. 

\begin{figure}[htb!]
  \centering
  \includegraphics[width=0.49\textwidth]{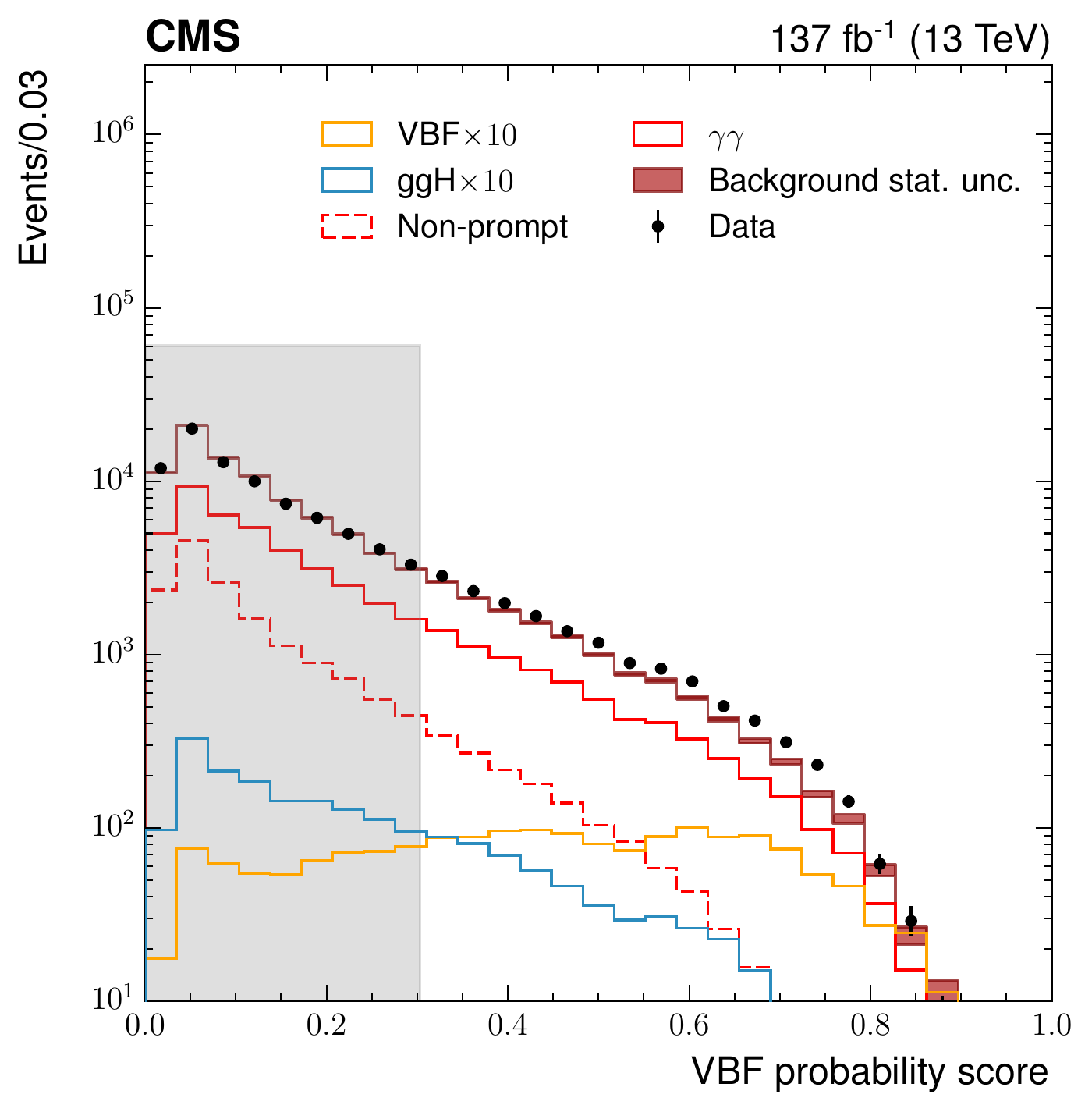} 
  \includegraphics[width=0.49\textwidth]{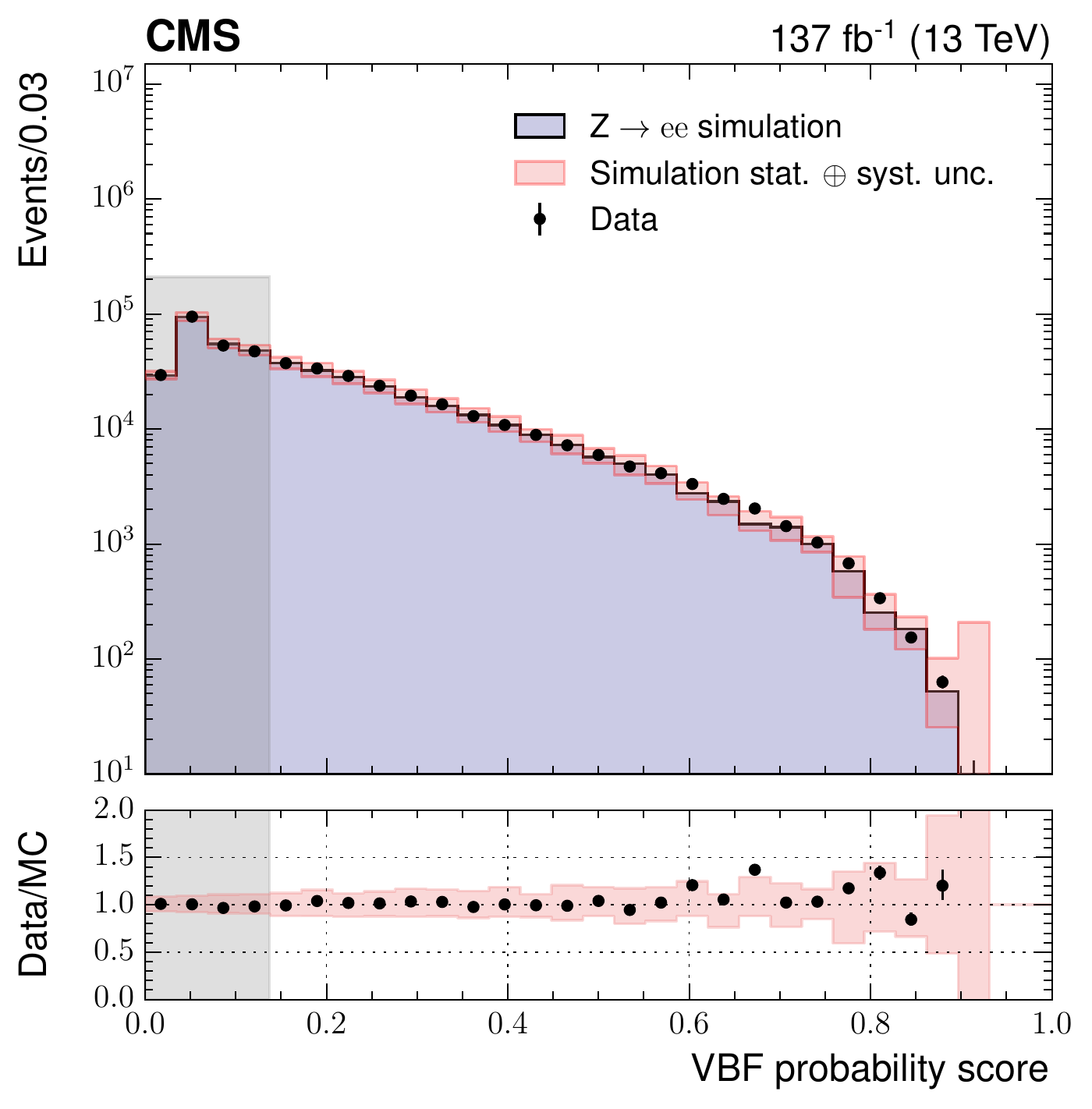} 
  \caption{
  The left plot shows the distribution of the dijet BDT output VBF probability
  in events with \mgg in the range 100--120 or 130--180\GeV, 
  for data events passing the dijet preselection (black points), 
  and for simulated background events (red band). 
  Histograms are also shown for different components of the simulated background in red.
  The orange histogram corresponds to simulated VBF signal events, 
  with the \ggH events shown in blue (both $\times 100$).
  The right plot shows the same distribution in \Zee events where 
  the electrons are reconstructed as photons. 
  The points show the score for data, 
  the histogram shows the score for simulated Drell--Yan events, 
  including statistical and systematic uncertainties (pink band). 
  The regions shaded grey contain VBF probability values
  below the lowest threshold used to define an analysis category.
  The full data set collected in 2016--2018 and the corresponding simulation are shown.}
  \label{fig:dijet_validation}
\end{figure}

The two independent output probabilities of the dijet BDT, taken to be the VBF probability and the \ggH probability, 
are validated in $\Zee+\text{jets}$ events with the electrons reconstructed as photons.
The dijet preselection criteria required to enter the VBF-like analysis categories are also 
applied to the $\Zee+\text{jets}$ events. 
The VBF probability distribution in simulation and data sidebands is shown in the left plot of Fig.~\ref{fig:dijet_validation}, 
while the right plot demonstrates good agreement between data and simulation in $\Zee+\text{jets}$ events. 
A similar level of agreement is observed in the \ggH probability distribution. 

Due to the use of the data control sample with photon ID BDT score below $-0.5$ in the dijet BDT training, 
an additional requirement that the two photons have a photon ID BDT score of larger than $-0.2$ 
is placed on events entering the VBF-like analysis categories. 
The final analysis categories are constructed following the structure of the STXS binning scheme.
Events can be assigned to analysis categories targeting one of five VBF-like STXS bins, 
as shown in Fig.~\ref{fig:allSTXSbins}.
The first is defined as having a high \ptH, with a threshold set at $200\GeV$.
The remaining four bins have $\ptH < 200\GeV$. 
They are defined by boundaries on \ptH and \ptHjj at $25\GeV$
and \mjj at $700\GeV$. 
The \ptHjj threshold is chosen to separate events containing two jets from those containing three or more, 
which are referred to as two-jet-like ($\ptHjj<25\GeV$) and three-jet-like ($\ptHjj>25\GeV$) bins, respectively.
The analysis categories are defined using the reconstructed observables corresponding to each 
particle-level quantity; these are the \ptgg, the reconstructed \mjj, and the reconstructed \ptHjj.

Events are further divided into analysis categories using both the dijet BDT output probabilities 
and the diphoton BDT score. 
For each of the five STXS bins, a set of analysis categories is constructed with events originating 
from VBF production considered as signal.
An optimisation is performed defining lower bounds on the dijet VBF probability and diphoton BDT score, 
with an upper bound on the dijet \ggH probability. 
Two analysis categories are constructed to target each STXS bin, 
the expected composition of which is given in Table~\ref{tab:vbf_yields}.

An additional set of analysis categories is defined covering the four STXS bins with $\ptH < 200\GeV$, 
but considering \ggH events as signal instead of VBF.
Two analysis categories targeting the set of four STXS bins together are constructed. 
Here lower bounds are set on the dijet \ggH probability and diphoton BDT score, 
with an upper bound placed on the dijet VBF probability.
The expected composition of these is also given in Table~\ref{tab:vbf_yields}.

\subsection{Event categories for hadronic VH production}
\label{sec:vh_had_categorisation}

In the EW \qqH STXS binning scheme, there is a bin representing hadronic \VH production, 
defined at the particle level by $60 < \mjj < 120\GeV$. 
Analysis categories targeting this bin are constructed in a similar way to those targeting VBF-like dijet events. 
The principal difference is in the selection of the two jets.
The hadronic \VH preselection requires two jets within $\abs{\eta} < 2.4$ and with $\pt > 30\GeV$, 
and satisfying a pileup jet identification criterion.
In addition, the reconstructed \mjj is required to be consistent with a decay of a vector boson, $60 < \mjj < 120\GeV$.

A BDT referred to as the \VH hadronic BDT is trained with
\VH hadronic events as signal, 
against \ggH and non-Higgs boson SM diphoton production together as background.
The training events of \VH, \ggH, and SM production of two prompt photons 
are taken from simulation.
The remaining background containing nonprompt photons is derived from a control sample 
in the same way as that employed for the dijet BDT training. 
The control sample is defined by requiring that at least one photon has 
a photon ID BDT score of less than $-0.5$, but otherwise passes the \VH hadronic preselection.
The resulting events are weighted to reproduce the expected number of background events 
and used in the BDT training of the \VH hadronic BDT.

The input variables for the \VH hadronic BDT are similar to those for the dijet BDT. 
Variables that aid in identifying events consistent with the vector boson decay are added, 
including the cosine of the difference of two angles: that of the diphoton system 
in the diphoton-dijet centre-of-mass frame, and that of the diphoton-dijet system in the lab frame.

The final two analysis categories use the output scores of both the \VH hadronic BDT and the diphoton BDT
to increase sensitivity, in addition to requiring a photon ID BDT score of greater than $-0.2$.

\begin{figure}[htb!]
  \centering
  \includegraphics[width=0.49\textwidth]{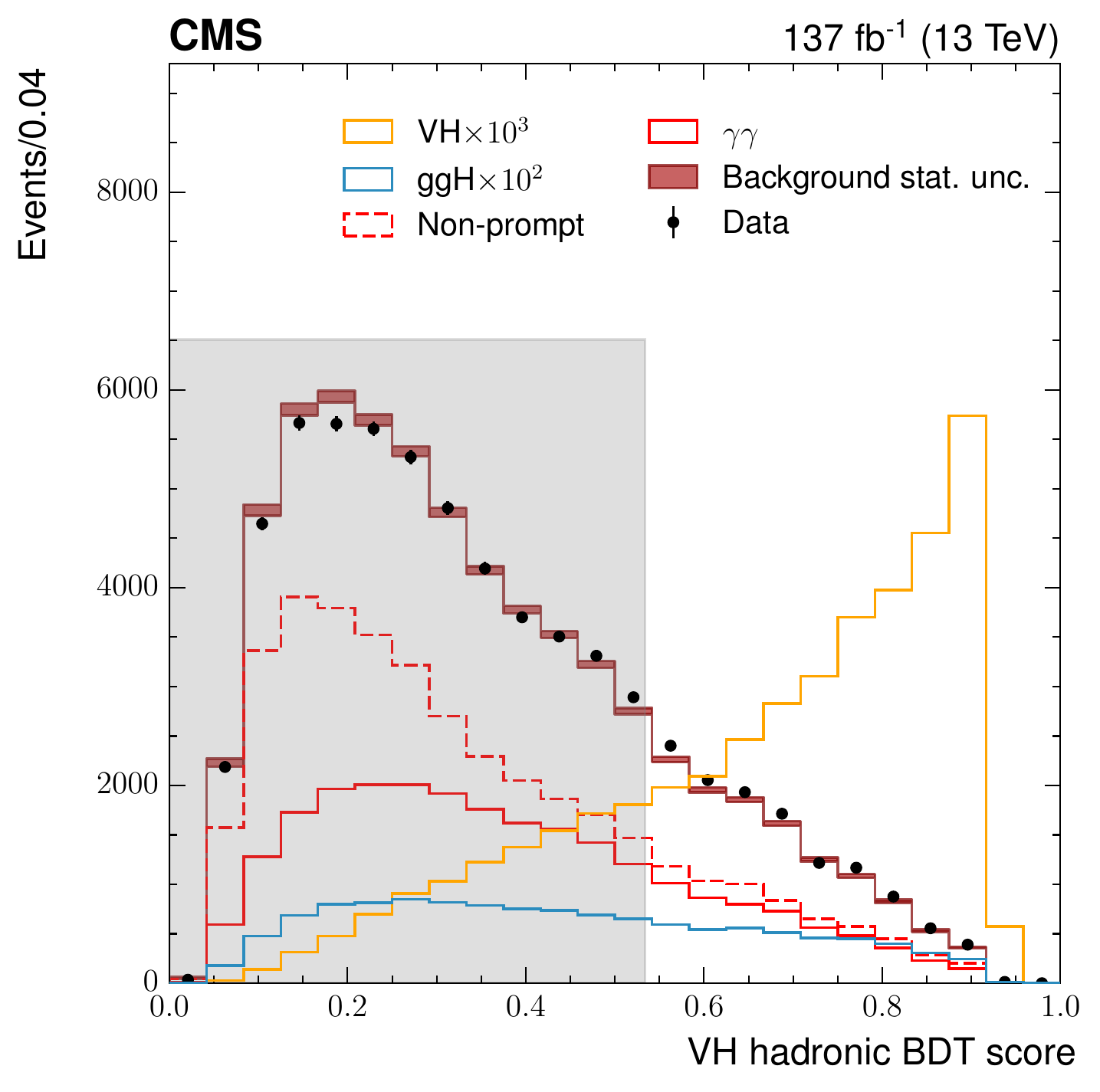} 
  \includegraphics[width=0.49\textwidth]{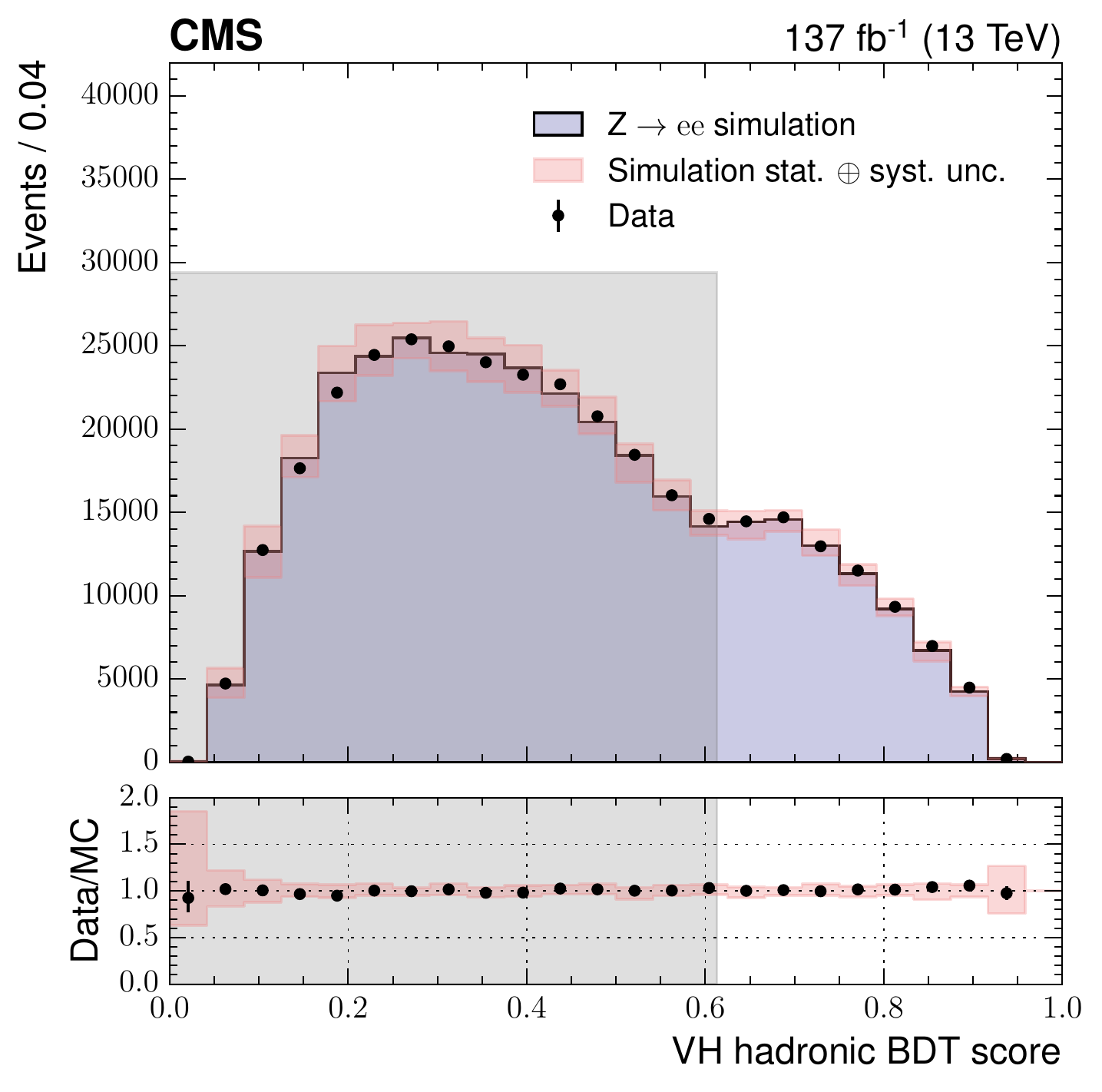} 
  \caption{
  The left plot shows the distribution of the \VH hadronic BDT output score 
  in events with \mgg in the range 100--120 or 130--180\GeV, 
  for data events passing the preselection (black points), 
  and for simulated background events (red band). 
  Histograms are also shown for different components of the simulated background in red.
  The sum of all background distributions is scaled to the data. 
  The orange histogram corresponds to simulated \VH hadronic signal events.
  The right plot shows the same distribution in $\Zee+\text{jets}$ events where 
  the electrons are reconstructed as photons. 
  The points show the score for data, 
  the histogram shows the score for simulated Drell--Yan events, 
  including statistical and systematic uncertainties (pink band). 
  The regions shaded grey contain \VH hadronic BDT scores 
  below the lowest threshold used to define an analysis category.
  The full data set collected in 2016--2018 and the corresponding simulation are shown.}
  \label{fig:vhhad_validation}
\end{figure}

The output score of the \VH hadronic BDT in simulation and data sidebands is shown in the left plot of Fig.~\ref{fig:vhhad_validation}.
The \VH hadronic BDT is also validated in $\Zee+\text{jets}$ events with electrons reconstructed as photons,
after the \VH hadronic preselection is applied.
The two distributions in simulation and data are shown in the right plot Fig.~\ref{fig:vhhad_validation} and exhibit good agreement.

The expected signal and background yields in each VBF and hadronic \VH analysis category 
are shown in Table~\ref{tab:vbf_yields}.

\begin{table}[htb!]
    \topcaption{
    The expected number of signal events for $\mH = 125\GeV$
    in analysis categories targeting VBF-like phase space and \VH production in which the vector boson decays hadronically,
    shown for an integrated luminosity of 137\fbinv.                
    The fraction of the total number of events arising from each production mode in each analysis category is provided, 
    as is the fraction of events originating from the targeted STXS bin or bins. 
    Entries with values less than 0.05\% are not shown. 
    Here \ggH includes contributions from the \ggZHhad and \bbH production modes, 
    whilst ``Top" represents both \ttH and \tH production together.
    The $\seff$, defined as the smallest interval containing 68.3\% of the \mgg distribution, 
    is listed for each analysis category. 
    The final column shows the expected ratio of signal to signal-plus-background, S/(S+B), 
    where S and B are the numbers of expected signal and background events 
    in a $\pm1\seff$ window centred on $\mH$.}
    \label{tab:vbf_yields}
    \centering
    \cmsTable{
      \begin{tabular}{lccccccccc}
          \multirow{3}{*}{Analysis categories} & \multicolumn{8}{c}{SM 125\GeV Higgs boson expected signal} & \multirow{3}{*}{S/(S+B)} \\
           & \multirow{2}{*}{Total} & \multirow{2}{*}{\begin{tabular}[c]{@{}c@{}}Target\\STXS bin(s)\end{tabular}} & \multicolumn{5}{c}{Fraction of total events} & \multirow{2}{*}{\begin{tabular}[c]{@{}c@{}}$\seff$\\(GeV)\end{tabular}} & \\
           & & & \ggH & VBF & \VH had & \VH lep & Top & & \\ \hline
           \ggH VBF-like Tag0 & 14.1 & 37.7\% & 65.9\% & 27.3\% & 3.8\% & 0.8\% & 2.3\% & 1.85 & 0.14 \\
           \ggH VBF-like Tag1 & 32.5 & 30.2\% & 61.3\% & 29.8\% & 4.1\% & 1.1\% & 3.7\% & 1.83 & 0.10 \\
           [\cmsTabSkip]
           \qqH low $\mjj$ low $\ptHjj$ Tag0 & 17.2 & 48.2\% & 36.6\% & 62.6\% & 0.4\% & 0.1\% & 0.3\% & 1.89 & 0.20 \\
           \qqH low $\mjj$ low $\ptHjj$ Tag1 & 13.5 & 48.5\% & 35.5\% & 63.4\% & 0.6\% & 0.1\% & 0.3\% & 1.74 & 0.19 \\
           [\cmsTabSkip]
           \qqH high $\mjj$ low $\ptHjj$ Tag0 & 27.0 & 70.4\% & 17.1\% & 82.7\% & 0.2\% &\NA& 0.1\% & 1.78 & 0.49 \\
           \qqH high $\mjj$ low $\ptHjj$ Tag1 & 12.9 & 58.2\% & 20.8\% & 78.7\% & 0.3\% & 0.1\% & 0.2\% & 1.99 & 0.27 \\
           [\cmsTabSkip]
           \qqH low $\mjj$ high $\ptHjj$ Tag0 & 10.4 & 15.0\% & 56.0\% & 41.3\% & 1.3\% & 0.4\% & 1.0\% & 1.92 & 0.12 \\
           \qqH low $\mjj$ high $\ptHjj$ Tag1 & 20.2 & 17.0\% & 57.9\% & 36.9\% & 2.4\% & 0.7\% & 2.1\% & 1.74 & 0.08 \\
           [\cmsTabSkip]
           \qqH high $\mjj$ high $\ptHjj$ Tag0 & 18.1 & 25.6\% & 28.1\% & 70.8\% & 0.4\% & 0.1\% & 0.5\% & 1.88 & 0.29 \\
           \qqH high $\mjj$ high $\ptHjj$ Tag1 & 17.5 & 23.8\% & 39.5\% & 57.8\% & 0.9\% & 0.3\% & 1.5\% & 1.98 & 0.13 \\
           [\cmsTabSkip]
           \qqH BSM Tag0 & 11.2 & 71.2\% & 24.4\% & 74.8\% & 0.1\% & 0.1\% & 0.6\% & 1.62 & 0.56 \\
           \qqH BSM Tag1 & 6.8 & 56.4\% & 36.9\% & 59.9\% & 1.1\% & 0.4\% & 1.7\% & 1.67 & 0.39 \\
           [\cmsTabSkip]
           \qqH \VH-like Tag0 & 16.3 & 55.8\% & 36.5\% & 2.8\% & 55.0\% & 1.4\% & 4.2\% & 1.72 & 0.25 \\
           \qqH \VH-like Tag1 & 47.1 & 26.8\% & 64.9\% & 4.7\% & 26.4\% & 1.2\% & 2.9\% & 1.66 & 0.13 \\
           [\cmsTabSkip]
      \end{tabular}
    }
\end{table}

\subsection{Event categories for leptonic VH production}
\label{sec:categorisation_vh}

The analysis categories described here target events in which the Higgs boson is produced in association 
with a \PW or \PZ vector boson that subsequently decays leptonically. 
Depending on the particular leptonic decay mode of the vector boson,  
the possible final states include zero, one, or two charged leptons.
The full definitions of each \VH leptonic STXS bin are given in Table~\ref{tab:vh_definitions}, 
corresponding to the green entries in Fig.~\ref{fig:allSTXSbins}.
The bins are defined using \ptV
and the number of jets in the event.

\begin{table}[htb!]
    \topcaption{
    Definition of the \VH leptonic STXS bins. 
    The product of the cross section and branching fraction ($\mathcal{B}$), 
    evaluated at \sqrts and $\mH = 125\GeV$, is given for each bin in the last column.
    The fraction of the total production mode cross section from each STXS bin is also shown.
    Unless stated otherwise, the STXS bins are defined for $\Hrap < 2.5$.
    Events with $\Hrap > 2.5$ are mostly outside of the experimental acceptance
    and therefore have a negligible contribution to all analysis categories.
    Only leptonic decays of the $\PW$ and $\PZ$ bosons are included in these definitions.
    }
    \label{tab:vh_definitions}
    \centering
    \cmsTable{
      \begin{tabular}{lccccc}
         \multirow{2}{*}{STXS bin} & \multirow{2}{*}{\begin{tabular}[c]{@{}c@{}}Definition\\ units of $\ptV$ in \GeVns\end{tabular}} & \multicolumn{3}{c}{Fraction of cross section} & \multirow{2}{*}{$\sigma_{\text{SM}}\mathcal{B}$~(fb)} \\ 
          &  & $\qqbar' \to$\WH & $\qqbar\to\ZH$ & $\Pg\Pg\to\ZH$ &  \\ [\cmsTabSkip] \hline
         \WH lep forward& \multirow{3}{*}{$\Hrap > 2.5$}& 12.13\%&\NA&\NA& 0.123 \\ 
         \ZH lep forward& &\NA& 11.21\%&\NA& 0.058 \\  
         \ggZH lep forward& &\NA&\NA& 2.71\%& 0.002 \\ [\cmsTabSkip]
         \WH lep $\ptV<75$ & No jet requirements, $\ptV<7$5 & 46.55\% &\NA&\NA& 0.473 \\ 
         \WH lep $75<\ptV<150$ & No jet requirements, $75<\ptV<150$ & 29.30\% &\NA&\NA& 0.298 \\ 
         \WH lep 0J $150<\ptV<250$ & Exactly 0 jets, $150<\ptV<250$ & 5.10\% &\NA&\NA& 0.052 \\ 
         \WH lep $\geq$1J $150<\ptV<250$ & At least 1 jet, $150<\ptV<250$ & 3.97\% &\NA&\NA& 0.040 \\
         \WH lep $\ptV>250$ & No jet requirements, $\ptV>250$ & 2.95\% &\NA&\NA& 0.030 \\ [\cmsTabSkip]
         \ZH lep $\ptV<75$ & No jet requirements, $\ptV<75$ &\NA& 45.65\% &\NA& 0.237 \\ 
         \ZH lep $75<\ptV<150$ & No jet requirements, $75<\ptV<150$ &\NA& 30.70\% &\NA& 0.160 \\ 
         \ZH lep 0J $150<\ptV<250$ & Exactly 0 jets, $150<\ptV<250$ &\NA& 5.16\% &\NA& 0.027 \\ 
         \ZH lep $\geq$1J $150<\ptV<250$ & At least 1 jet, $150<\ptV<250$ &\NA& 4.27\% &\NA& 0.022 \\ 
         \ZH lep $\ptV>250$ & No jet requirements, $\ptV>250$ &\NA& 3.01\% &\NA& 0.016 \\ [\cmsTabSkip]
         \ggZH lep $\ptV<75$ & No jet requirements, $\ptV<75$ &\NA&\NA& 15.96\% & 0.013 \\ 
         \ggZH lep $75<\ptV<150$ & No jet requirements, $75<\ptV<150$ &\NA&\NA& 43.32\% & 0.036 \\ 
         \ggZH lep 0J $150<\ptV<250$ & Exactly 0 jets, $150<\ptV<250$ &\NA&\NA& 9.08\% & 0.008 \\ 
         \ggZH lep $\geq$1J $150<\ptV<250$ & At least 1 jet, $150<\ptV<250$ &\NA&\NA& 20.49\% & 0.017 \\ 
         \ggZH lep $\ptV>250$ & No jet requirements, $\ptV>250$ &\NA&\NA& 8.45\% & 0.007 \\ 
      \end{tabular}
    }
\end{table}

For each of the three channels, a dedicated BDT classifier is used to discriminate 
between the \VH signal and background events. 
Each of these three BDTs are trained on simulated signal and background events. 
The exception is the zero-lepton final state, 
for which some simulated backgrounds are replaced by events derived from data, as described below.
The simulated SM background processes include photons plus jets, Drell--Yan, diboson production, and top quark pair production. 
The production modes of the Higgs boson other than \VH are also treated as backgrounds. 
Where there are a sufficient number of expected signal events, 
the categorisation regions are further split into analysis categories sensitive to merged groups of STXS bins.

The categorisation region with two same-flavour reconstructed leptons in the final state focuses on the $\PZ(\ell\ell)\PH$ production mode.
Additional selection criteria are imposed to select two leptons consistent with the decay of a $\PZ$ boson, 
including a requirement that the dilepton mass (\mll) is between 60 and 120\GeV.

The so-called \ZH leptonic BDT is used to discriminate the $\PZ(\ell\ell)\PH$ signal events 
from backgrounds including both other Higgs boson production modes and non-Higgs-boson SM processes.
Its input variables are kinematic properties of the photons, leptons, and jets present in the event, 
including angular variables describing the separation between the photons and leptons.
In addition, jet identification variables such as the {\cPqb} tag score are used as inputs, 
which helps to discriminate against backgrounds containing top quarks.

\begin{figure}
  \centering
  \includegraphics[width=0.49\textwidth]{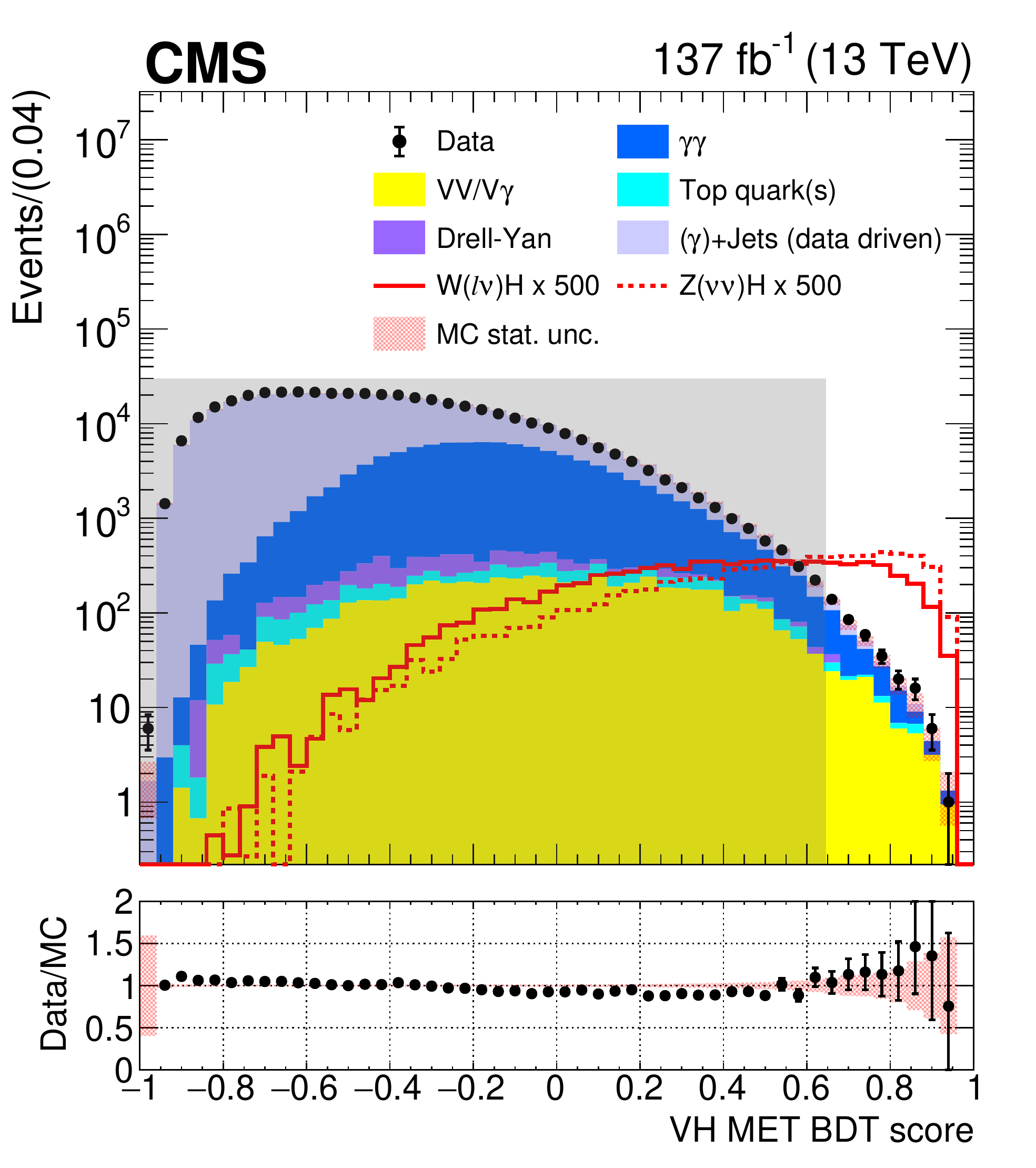}
  \includegraphics[width=0.49\textwidth]{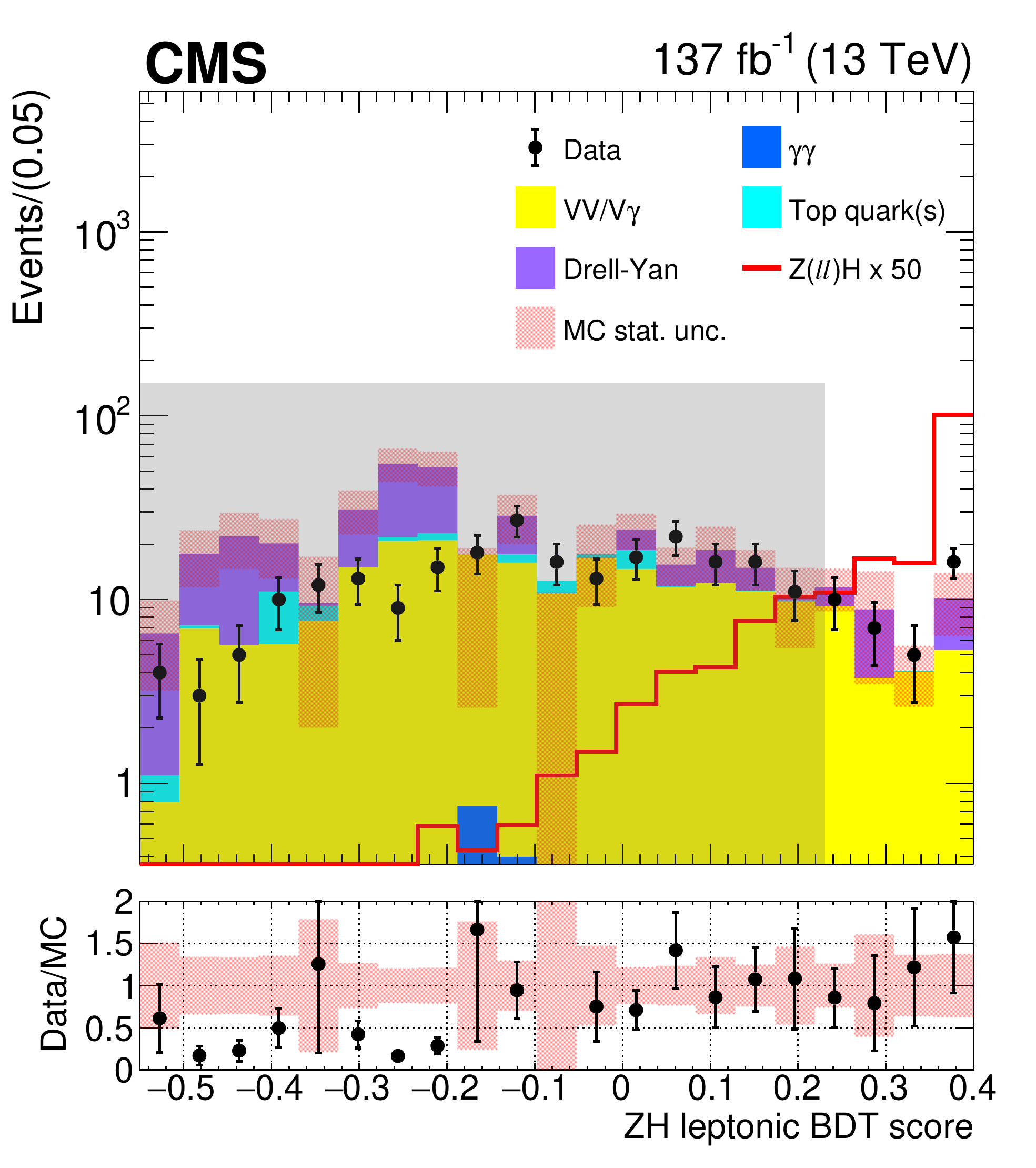}
  \includegraphics[width=0.49\textwidth]{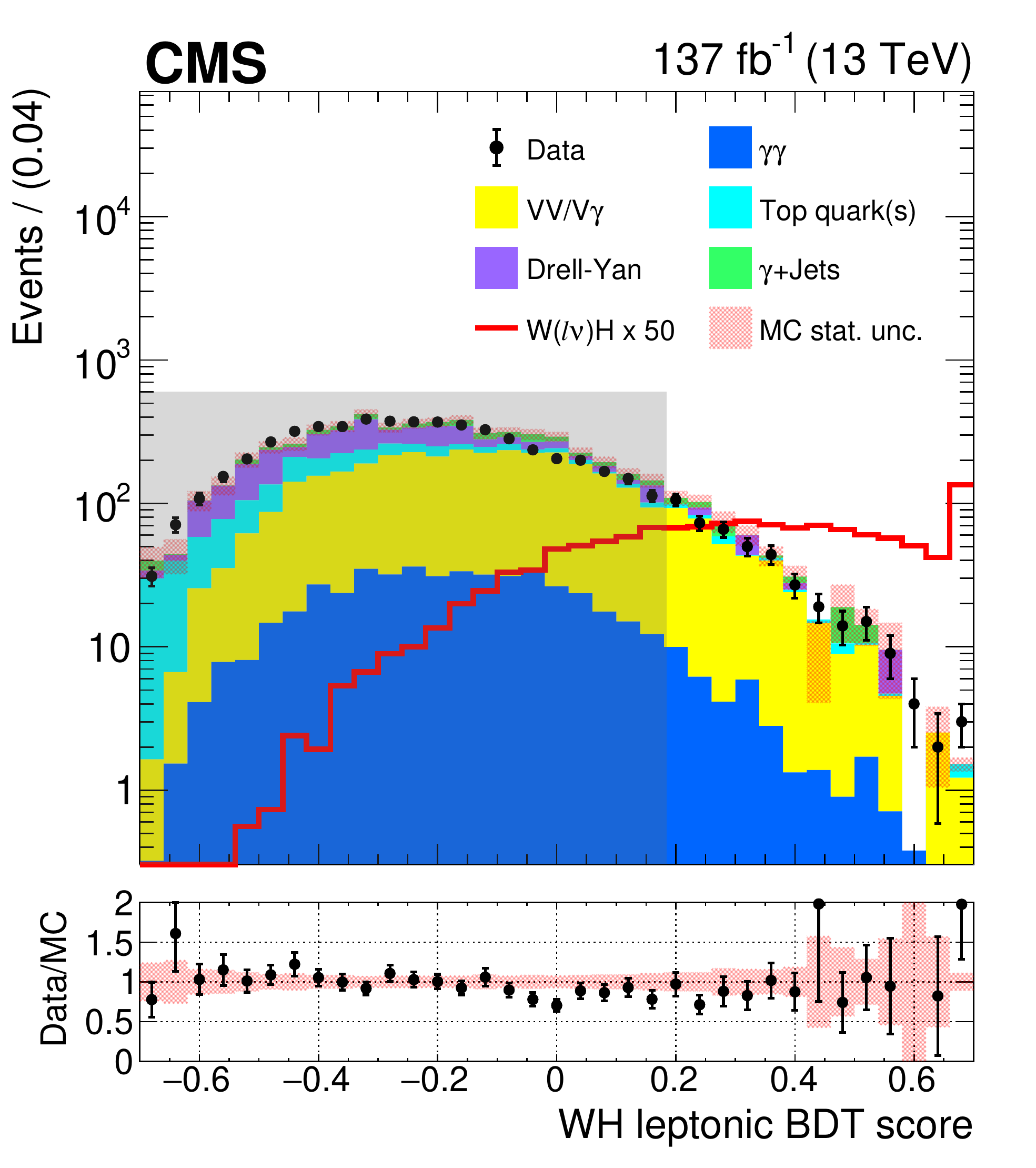}
  \caption{
	   Output scores for the three \VH leptonic BDTs.
     The \VH MET BDT is shown in the upper left, with the \ZH leptonic BDT in the upper right, 
     and the \WH leptonic BDT below.
     In each case, the signal and background simulation are shown as histograms with the data as black points.
     Events are taken from the $\mgg$ sidebands, satisfying either $100 < \mgg < 120\GeV$ or $130 < \mgg < 180\GeV$.
	   The statistical uncertainty in the data points is denoted as vertical bars and that on the background simulation by the pink band.
	   The simulated signal and background distributions are normalised to the luminosity of the data. 
     To increase its visibility, the signal is scaled by a factor of 500 for the \VH MET BDT, 
     with a factor of 50 applied for both \ZH leptonic and \WH leptonic BDTs.
     The regions shaded grey are not considered in the analysis.
     The full data set collected in 2016--2018 and the corresponding simulation are shown.
     }
  \label{fig:vhlep_bdt}
\end{figure}

The distributions of the \ZH leptonic BDT score for simulated signal and background events, 
along with the same for the data sidebands, are shown in Fig.~\ref{fig:vhlep_bdt}. 
With the available data set, this categorisation region is not sensitive to the corresponding individual STXS bins. 
For this reason, further splitting of the analysis categories is not performed. 
The sensitivity to inclusive leptonic \ZH production is maximised 
by defining two analysis categories using the BDT score.

To gain sensitivity to the $\PW(\ell\nu)\PH$ production mode, 
events with one reconstructed lepton are selected.
Additional selection criteria are applied on the photon ID BDT to further reject background events containing nonprompt photons,
and on the invariant mass of the reconstructed lepton with each photon 
to reduce the contamination of Drell--Yan events with an electron misidentified as a photon.

With this selection, the \WH leptonic BDT is trained with simulated $\PW(\ell\nu)\PH$ signal events 
against other Higgs boson modes and non-Higgs-boson SM backgrounds.
The input features of the \WH leptonic BDT are similar to those used in the \ZH leptonic BDT, 
including photon, lepton, and jet kinematic variables.
In addition, the transverse mass of the leading lepton and \ptmiss are used.
The distributions of the \WH leptonic BDT score 
for the signal and background simulation samples and data sidebands is shown Fig.~\ref{fig:vhlep_bdt}.

This single-lepton final state is sensitive to a reduced set of STXS bins. 
Three sets of analysis categories are defined, with \ptgg thresholds at 75 and 150~\GeV. 
The \ptgg variable is used because it provides the most accurate estimate of the particle level 
\ptV used to define the STXS bins;
the presence of a neutrino in the final state 
means that the vector boson itself cannot be fully reconstructed.
The sensitivity to each set of STXS bins is optimised by deriving analysis categories 
based on the \WH leptonic BDT score.
Two analysis categories are constructed with $\ptgg < 75\GeV$ and $75 < \ptgg < 150\GeV$, 
whilst one analysis category is defined with $\ptgg > 150\GeV$.

The analysis categories targeting \VH production where there are no reconstructed leptons in the event
are referred to as the \VH MET tags.
These analysis categories receive contributions 
from both the $\PZ(\nu\nu)\PH$ and $\PW(\ell\nu)\PH$ production modes. 
In addition to vetoing events with leptons, $\ptmiss > 50\GeV$ is required 
and the azimuthal angle between the diphoton system and \ptvecmiss must be greater than two radians.

With this selection the \VH MET BDT is trained to discriminate between signal and background processes.
The input features of the \VH MET BDT rely on the same diphoton variables as in the \ZH and \WH leptonic BDTs, 
together with \ptmiss and jet variables.
One of the dominant backgrounds in this final state consists of \gamplusjets events 
where one of the jets is misidentified as a photon. 
The simulation does not model this process well and the number of such events available is limited.
Hence the \gamplusjets background component is modelled from a sample of data events 
where one of the photon candidates fails to satisfy the photon ID BDT requirement of $-0.9$. 
To enable this, a control sample is constructed by inverting the requirement on the photon ID BDT score.
These events otherwise fulfil the full set of selection requirements for the \VH MET BDT channel.
A new value of the photon ID BDT score is generated for each event. 
This is achieved by assigning a random value drawn from the photon ID BDT distribution of 
simulated \gamplusjets events which pass the full set of selection criteria.
The events are then appropriately weighted and used in the \VH MET BDT training instead of the corresponding simulated samples.
This is the same method first developed for the analysis described in Ref.~\cite{HIG-19-013}, 
but differs to the method used in the training of the VBF and \VH hadronic BDTs.
The resulting increased number of events on which to train, 
as well as the improved modelling of the input variable distributions, 
improves the performance of the \VH MET BDT.

The distributions of the \VH MET BDT output score for the signal and background simulation samples together 
with the same for the data sidebands are shown Fig.~\ref{fig:vhlep_bdt}.

The final expected signal and background yields for each \ZH leptonic, \WH leptonic, and \VH MET 
analysis category are shown in Table~\ref{tab:vh_yields}.

\begin{table}
    \topcaption{The expected number of signal events for $\mH = 125\GeV$
    in analysis categories targeting Higgs boson production in association with a leptonically decaying $\PW$ or $\PZ$ boson,    shown for an integrated luminosity of 137\fbinv.                
    The fraction of the total number of events arising from each production mode in each analysis category is provided, 
    as is the fraction of events originating from the targeted STXS bin or bins. 
    Entries with values less than 0.05\% are not shown. 
    Here \ggH includes contributions from the \ggZHhad and \bbH production modes, 
    \qqH incorporates both VBF and \VH production with hadronic vector boson decays, 
    and ``Top" represents both \ttH and \tH production together.
    The $\seff$, defined as the smallest interval containing 68.3\% of the \mgg distribution, 
    is listed for each analysis category. 
    The final column shows the expected ratio of signal to signal-plus-background, S/(S+B), 
    where S and B are the numbers of expected signal and background events 
    in a $\pm1\seff$ window centred on $\mH$.}
    \label{tab:vh_yields}
    \centering
    \cmsTable{
      \begin{tabular}{lcccccccccc}
          \multirow{3}{*}{Analysis categories} & \multicolumn{9}{c}{SM 125\GeV Higgs boson expected signal} & \multirow{3}{*}{S/(S+B)} \\
           & \multirow{2}{*}{Total} & \multirow{2}{*}{\begin{tabular}[c]{@{}c@{}}Target\\STXS bin(s)\end{tabular}} & \multicolumn{6}{c}{Fraction of total events} & \multirow{2}{*}{\begin{tabular}[c]{@{}c@{}}$\seff$\\(GeV)\end{tabular}} & \\
           & & & \ggH & \qqH & \WH lep & \ZH lep & \ggZH lep & Top & & \\ \hline
           \ZH lep Tag0 & 2.4 & 99.6\% &\NA&\NA&\NA& 82.0\% & 17.7\% & 0.4\% & 1.67 & 0.57 \\
           \ZH lep Tag1 & 0.9 & 97.5\% & 0.1\% &\NA& 0.2\% & 80.7\% & 16.9\% & 2.2\% & 1.85 & 0.32 \\
           [\cmsTabSkip]
           \WH lep $\ptV<75$ Tag0 & 2.0 & 81.1\% &\NA& 0.2\% & 95.0\% & 3.3\% & 0.2\% & 1.3\% & 1.89 & 0.43 \\
           \WH lep $\ptV<75$ Tag1 & 4.5 & 75.7\% & 2.6\% & 0.5\% & 87.2\% & 7.0\% & 0.3\% & 2.4\% & 1.85 & 0.19 \\
           [\cmsTabSkip]
           \WH lep $75<\ptV<150$ Tag0 & 3.0 & 77.7\% & 0.7\% & 0.3\% & 93.2\% & 3.4\% & 0.8\% & 1.6\% & 1.94 & 0.56 \\ 
           \WH lep $75<\ptV<150$ Tag1 & 3.3 & 60.8\% & 1.7\% & 1.4\% & 83.1\% & 7.7\% & 1.6\% & 4.4\% & 2.02 & 0.33 \\ 
           [\cmsTabSkip]
           \WH lep $\ptV>150$ Tag0 & 3.5 & 79.9\% & 0.5\% & 0.4\% & 91.5\% & 3.6\% & 1.1\% & 2.8\% & 1.84 & 0.77 \\
           [\cmsTabSkip]
           \VH MET Tag0 & 2.2 & 97.9\% & 0.4\% & 0.9\% & 23.5\% & 56.9\% & 17.6\% & 0.8\% & 2.22 & 0.48 \\
           \VH MET Tag1 & 3.6 & 90.5\% & 4.6\% & 3.1\% & 28.8\% & 46.0\% & 15.7\% & 1.9\% & 2.30 & 0.34 \\
           \VH MET Tag2 & 6.6 & 72.2\% & 15.5\% & 8.8\% & 27.7\% & 33.5\% & 11.0\% & 3.5\% & 2.15 & 0.18 \\
           [\cmsTabSkip]
      \end{tabular}
    }
\end{table}

\subsection{Event categories for top quark associated production}
\label{sec:TTHtag}

The coupling between the Higgs boson and the top quark affects \Hgg cross sections both
via \ggH production, entering in the gluon loop, and via decay in the diphoton decay loop.
In addition, the coupling can be accessed directly by measuring the rate of \Hgg events 
when the Higgs boson is produced in association with one or more top quarks.
The observation of \ttH production
in the diphoton decay channel was recently reported by CMS and ATLAS~\cite{HIG-19-013,ATLASfullRun2ttH}.
There, multivariate discriminants are trained separately for hadronic and leptonic 
decays of the top quarks to construct analysis categories enriched in \ttH events. 
In this analysis, the same techniques for the event categorisation are used. 
Additional analysis categories are constructed to provide sensitivity to individual STXS bins, 
the definitions of which are given in Table~\ref{tab:top_definitions}.
These correspond to the purple entries in Fig.~\ref{fig:allSTXSbins} for \ttH, 
and the single yellow entry for \tH.

\begin{table}
    \topcaption{
    Definition of the \ttH, \tH, and \bbH STXS bins.
    The product of the cross section and branching fraction ($\mathcal{B}$), 
    evaluated at \sqrts and $\mH = 125\GeV$, is given for each bin in the last column.
    The fraction of the total production mode cross section from each STXS bin is also shown.
    Unless stated otherwise, the STXS bins are defined for $\Hrap < 2.5$.
    Events with $\Hrap > 2.5$ are mostly outside of the experimental acceptance
    and therefore have a negligible contribution to all analysis categories.
    }
    \label{tab:top_definitions}
    \centering
    \cmsTable{
      \begin{tabular}{lccccc}
         \multirow{2}{*}{STXS bin} & \multirow{2}{*}{\begin{tabular}[c]{@{}c@{}}Definition\\ units of $\ptH$ in \GeVns\end{tabular}} & \multicolumn{3}{c}{Fraction of cross section} & \multirow{2}{*}{$\sigma_{\text{SM}}\mathcal{B}$~(fb)} \\ 
          &  & \ttH & \tHq & \tHW & \\ [\cmsTabSkip] \hline
         \ttH forward& \multirow{3}{*}{$\Hrap > 2.5$}& 1.35\%&\NA&\NA& 0.016 \\  
         \tH forward& &\NA& 2.79\%& 1.06\%& 0.005 \\  
         \ttH $\ptH<60$ & No jet requirements, $\ptH<60$ & 22.42\% &\NA&\NA& 0.259 \\ 
         \ttH $60<\ptH<120$ & No jet requirements, $60<\ptH<120$ & 34.61\% &\NA&\NA& 0.400 \\ 
         \ttH $120<\ptH<200$ & No jet requirements, $120<\ptH<200$ & 25.60\% &\NA&\NA& 0.296 \\ 
         \ttH $200<\ptH<300$ & No jet requirements, $200<\ptH<300$ & 10.72\% &\NA&\NA& 0.124 \\ 
         \ttH $\ptH>300$ & No jet requirements, $\ptH>300$ & 5.31\% &\NA&\NA& 0.061 \\ [\cmsTabSkip] 
         \tH & No additional requirements &\NA& 97.21\% & 98.94\% & 0.204 \\ 
      \end{tabular}
    }
\end{table}

Production of the Higgs boson in association with a single top quark is also measured in this analysis. 
A dedicated analysis category enriched in \tHq events where the top decays leptonically is constructed.
The \tHq leptonic and \ttH leptonic final states are very similar; 
an effort is therefore made to distinguish between the two. 

A DNN referred to as the top DNN is trained 
with \ttH as signal and \tHq as background. 
It is used both by the \tHq leptonic tag to reduce \ttH contamination, 
and by the \ttH leptonic analysis categories to reduce the contamination from \tHq. 
The \tHq leptonic tag is considered first in the tag priority sequence because of its lower expected signal yield.
Each of the three categorisation regions (\tHq leptonic, \ttH leptonic, and \ttH hadronic) then uses 
a dedicated discriminant referred to as BDT-bkg.
The purpose of the BDT-bkg is to reduce backgrounds from non-Higgs-boson SM diphoton production
and split events further by expected S/B into the final analysis categories. 

For an event to be considered for the \tHq leptonic analysis category, it must have
at least one lepton, at least one {\cPqb}-tagged jet, and at least one additional jet.
The top DNN and the \tHq leptonic BDT-bkg are trained with these selection criteria applied.
The top DNN takes both kinematic information from individual objects characteristic of top decays 
and global event information as inputs. 
The objects considered are the six leading jets and two leading leptons in \pt. 
The four-momenta, along with the {\cPqb} tagging score and lepton identification scores, are included for each object. 
The global event features include the \ptmiss, number of jets, and photon kinematic variables and identification scores.

The \tHq leptonic BDT-bkg uses similar input variables to distinguish \tHq events 
from non-Higgs boson SM backgrounds, both of which are taken from simulation to perform the training.
Kinematic variables and {\cPqb} tag scores for the three leading jets and {\cPqb}-tagged jets in \pt are considered, 
as well as photon kinematic variables, and angular variables relating the jet and photon directions.

\begin{figure}[htb!]
    \centering
    \includegraphics[width=0.49\textwidth]{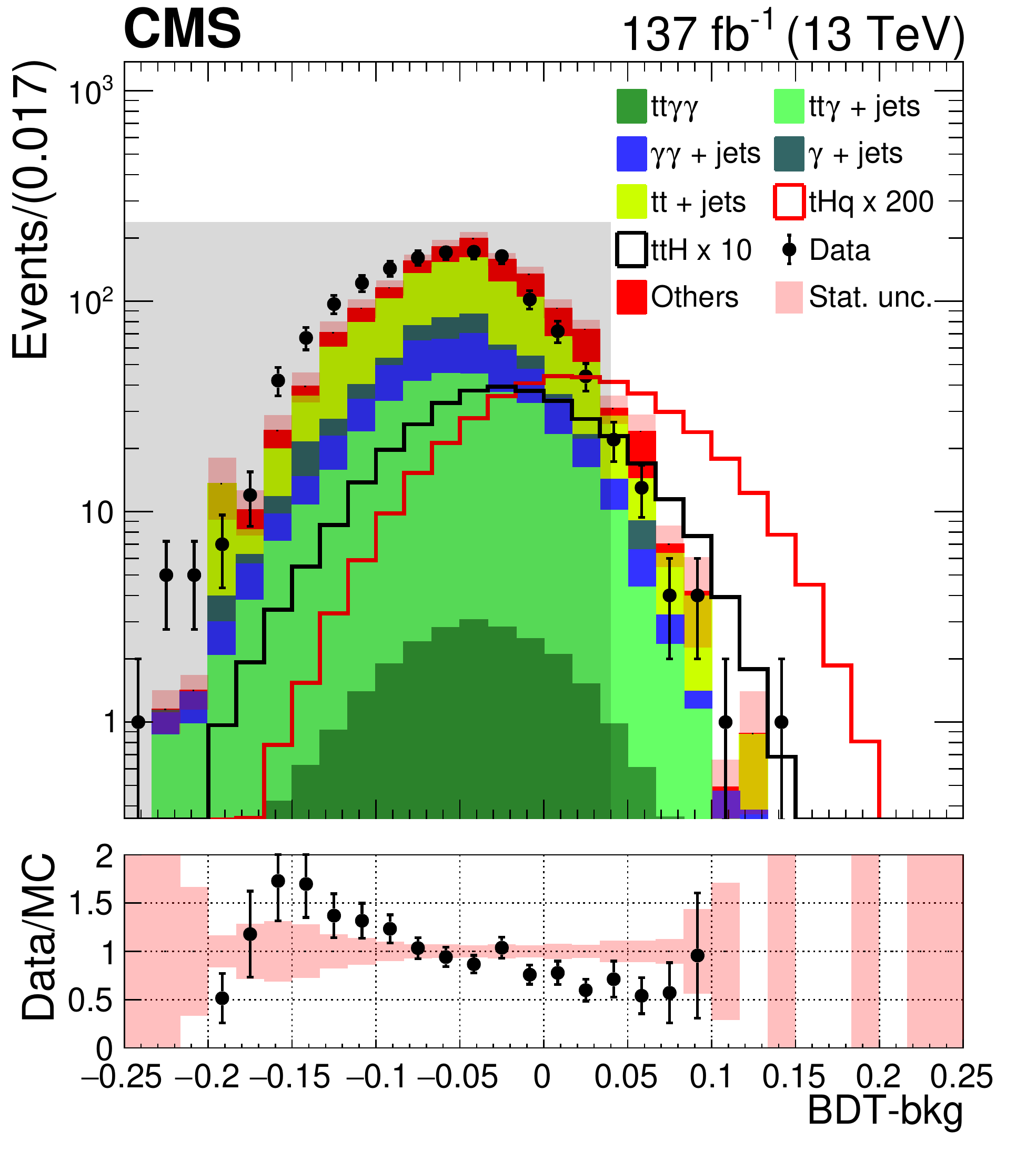}
    \includegraphics[width=0.49\textwidth]{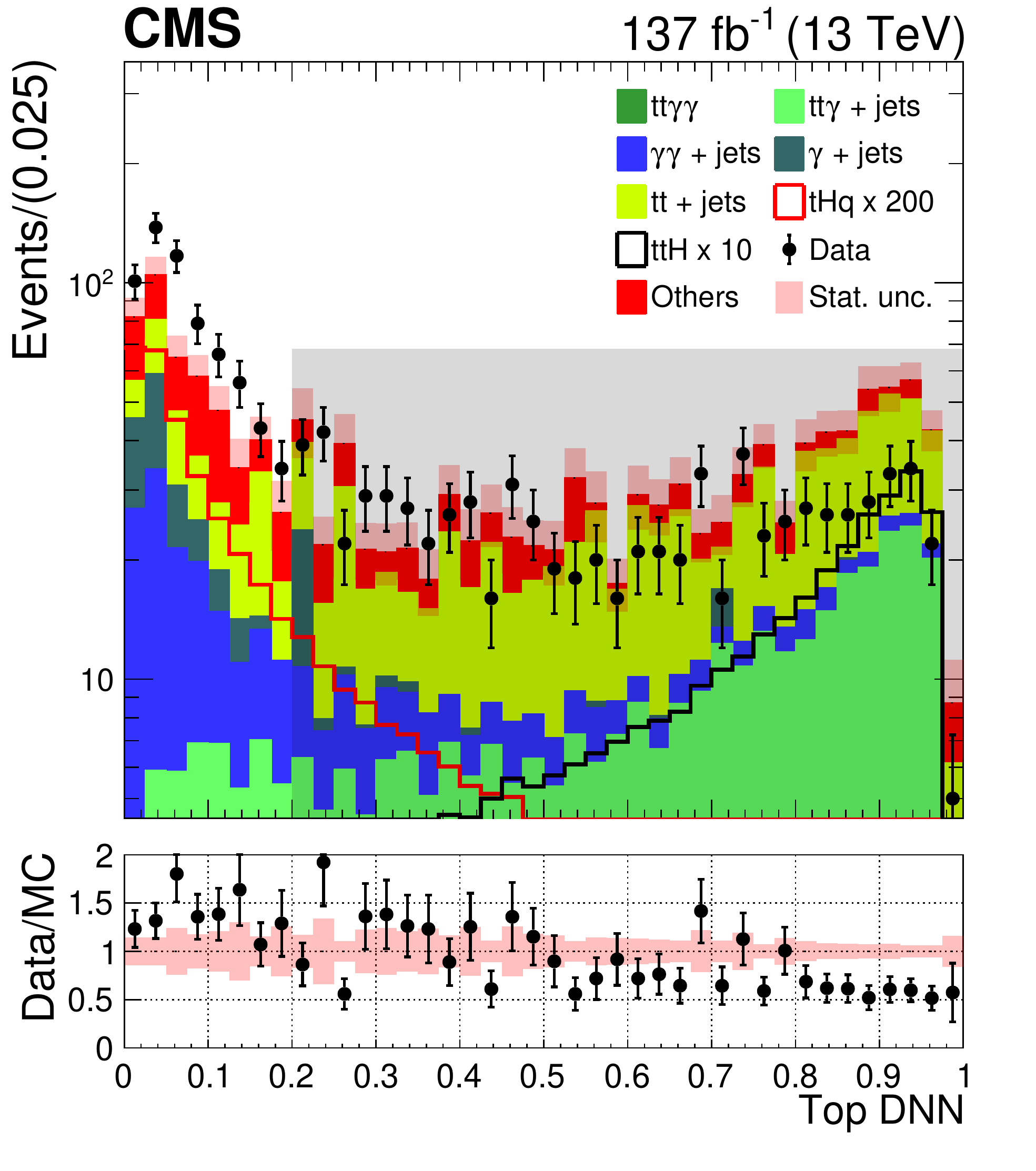}
    \caption{Distributions of \tHq BDT-bkg score (left) and the top DNN (right), 
        which are used together to define the \tHq leptonic analysis category.
        Events are taken from the $\mgg$ sidebands, satisfying either $100 < \mgg < 120\GeV$ or $130 < \mgg < 180\GeV$.
        The statistical uncertainty in the background estimation is represented by the pink band.
        The regions shaded grey contain BDT-bkg and top DNN scores below and above the respective thresholds for the \tHq analysis category.
        The full data set collected in 2016--2018 and the corresponding simulation are shown.
    }
    \label{fig:thq_scores}
\end{figure}

The distributions of the output scores for both the top DNN and the \tHq BDT-bkg are shown
in Fig.~\ref{fig:thq_scores}.
In both cases, the agreement between data and simulation in this background-like region is imperfect. 
However, this does not affect the results of this analysis because the final background model is derived directly from data. 

The final analysis category is defined by placing a requirement on 
both the score of the top DNN and the \tHq leptonic BDT-bkg.
Due to the low expected \tHq signal yield, only one analysis category is constructed.

The analysis categories targeting \ttH production are divided into two channels, 
representing either fully hadronic or leptonic decays of the $\ttbar$ system.
The hadronic channel is defined by having zero isolated leptons, 
whilst the leptonic channel requires one or more isolated leptons, 
meaning it includes events where one or both top quarks decay leptonically. 
In the hadronic channel, three or more jets must be present, 
of which at least one is tagged as originating from a bottom quark.
The leptonic channel requires the presence of one or more jets, 
and also includes a loose requirement on the top DNN to reject \tHq events.
This loose preselection for both channels maximises the available number of events 
for the training of the BDT-bkg in each channel and the top DNN in the leptonic channel.

For each channel, the BDT-bkg is trained on simulated signal and background events. 
The exception is that in the hadronic channel, \gamplusjets events are modelled from data.
This provides both an improved description of the input features 
and a greater number of events on which to train the BDT-bkg.
The procedure used to derive these events is identical to that described in Section~\ref{sec:categorisation_vh}.

The inputs to the \ttH BDT-bkg discriminants in each channel are kinematic properties of the jets, leptons, photons, and diphoton pair. 
It is not possible to infer the diphoton mass from the inputs.
In addition to these features, the outputs of dedicated DNNs designed to reject specific backgrounds 
and the output of a dedicated ``top quark tagger BDT" are used~\cite{TopTagger}.

The additional DNNs are trained with \ttH signal events against one source of background only. 
There are three such DNNs in total:
one for each of the $\PGg\PGg\!+\text{jets}$ and $\ttbar\!+\!\PGg\PGg$ backgrounds in the hadronic channel, 
and one for the $\ttbar\!+\!\PGg\PGg$ background in the leptonic channel.
These backgrounds are chosen because both they are well-modelled in simulation and because
it is possible to generate a high number of simulated events on which to train. 
Furthermore, $\ttbar\!+\!\PGg\PGg$ events in particular are the principal 
background in the analysis categories most sensitive to \ttH production.
With these sufficiently large training samples, the background-specific DNNs 
are able to exploit features such as the full four-momentum vectors of physics objects.
Adding these features directly to the inputs of the BDT-bkg do not improve its performance; 
the DNNs are required as an intermediate step to utilise this information effectively.

The top quark tagger BDT is designed to distinguish events with top quarks decaying into three jets from events that do not contain top quarks.
It is trained on jet triplets from simulation of $\ttbar$ events, 
with inputs related to the kinematics, {\cPqb} tag scores, and jet shape information.
The signal is jet triplets matched at generator-level to a top quark, 
and background is taken as random jet triplets~\cite{TopTagger}.

\begin{figure}
    \centering
		\includegraphics[width=0.49\textwidth]{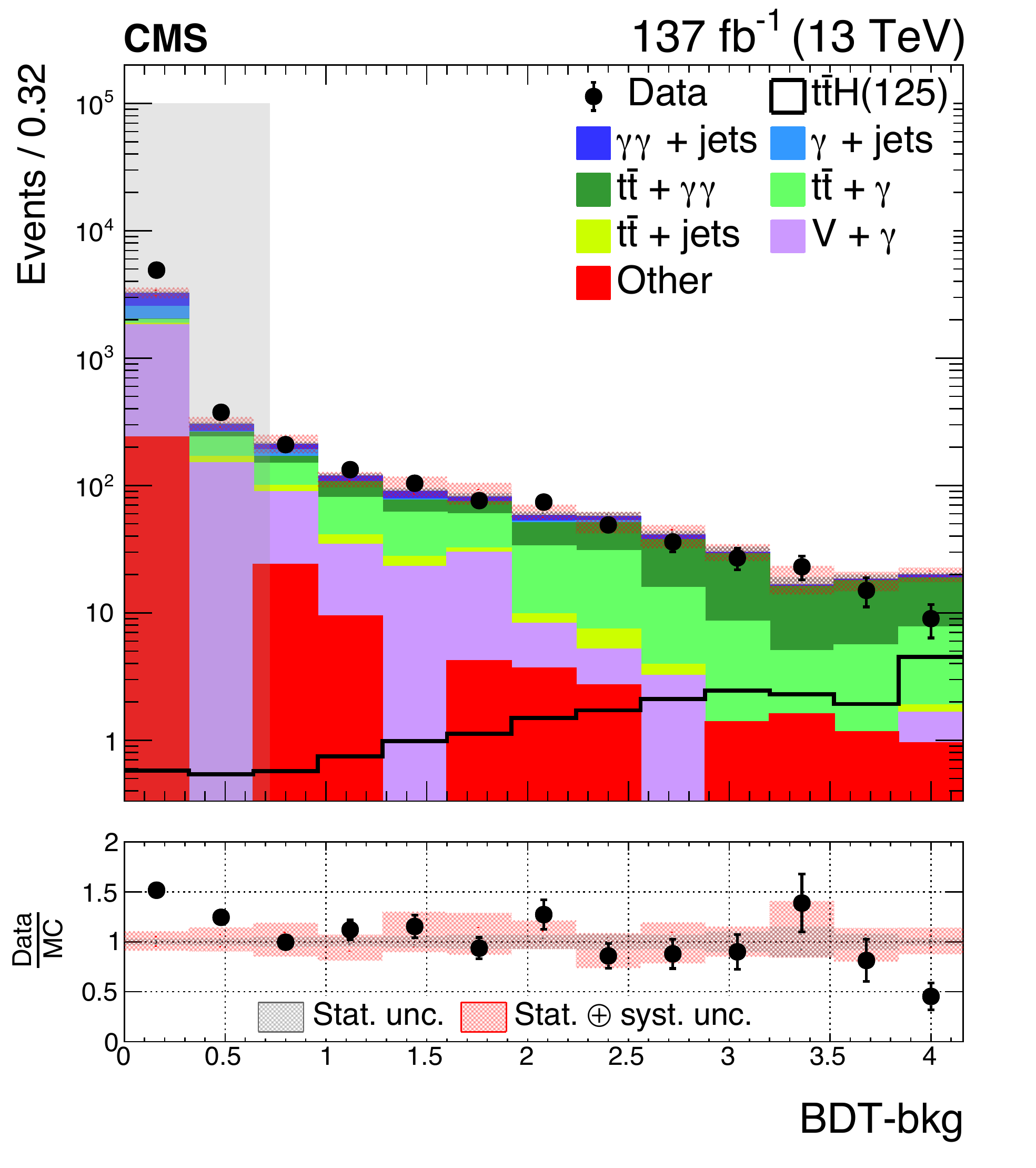} 
    \includegraphics[width=0.49\textwidth]{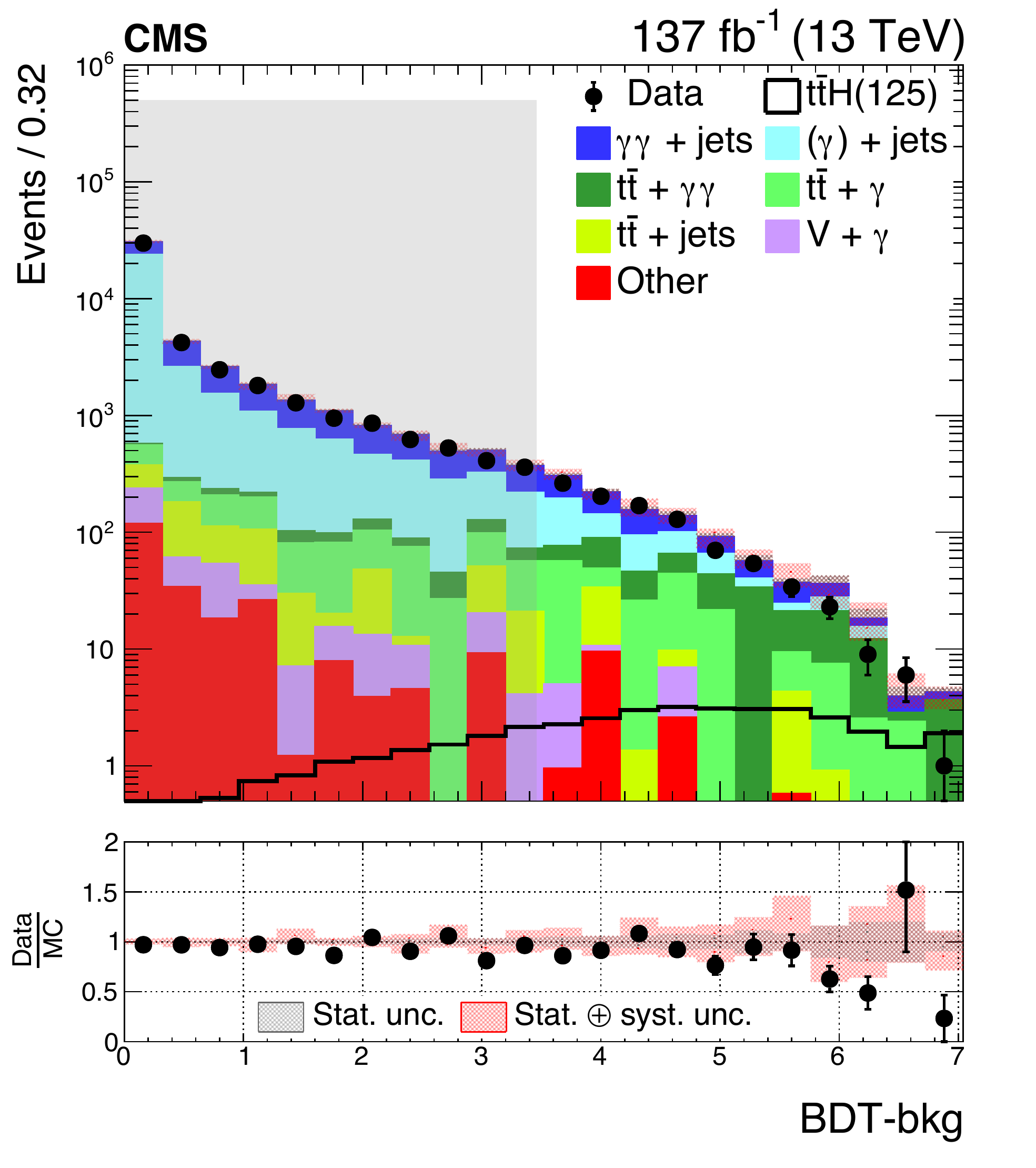} \\
    \includegraphics[width=0.49\textwidth]{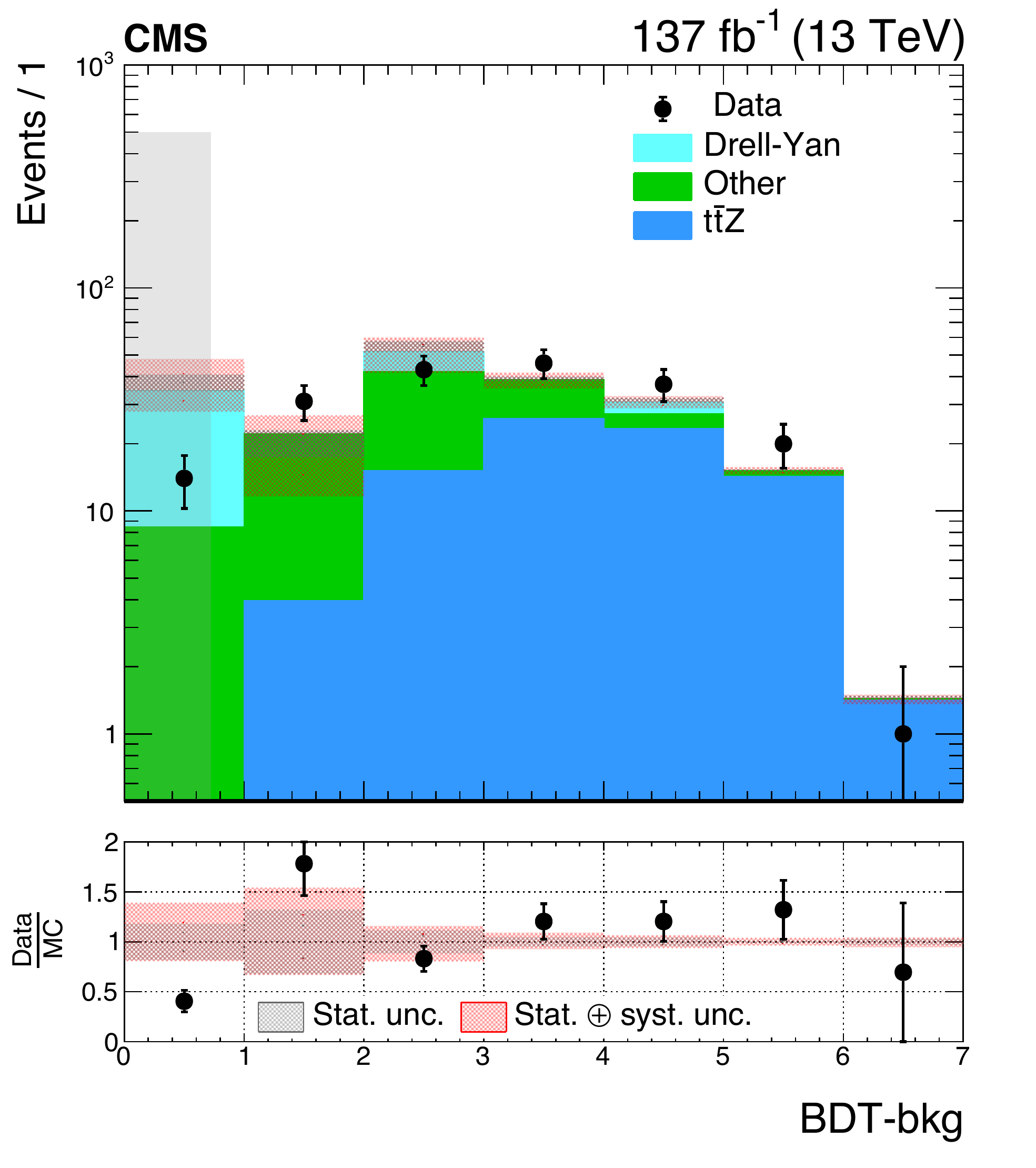}
    \includegraphics[width=0.49\textwidth]{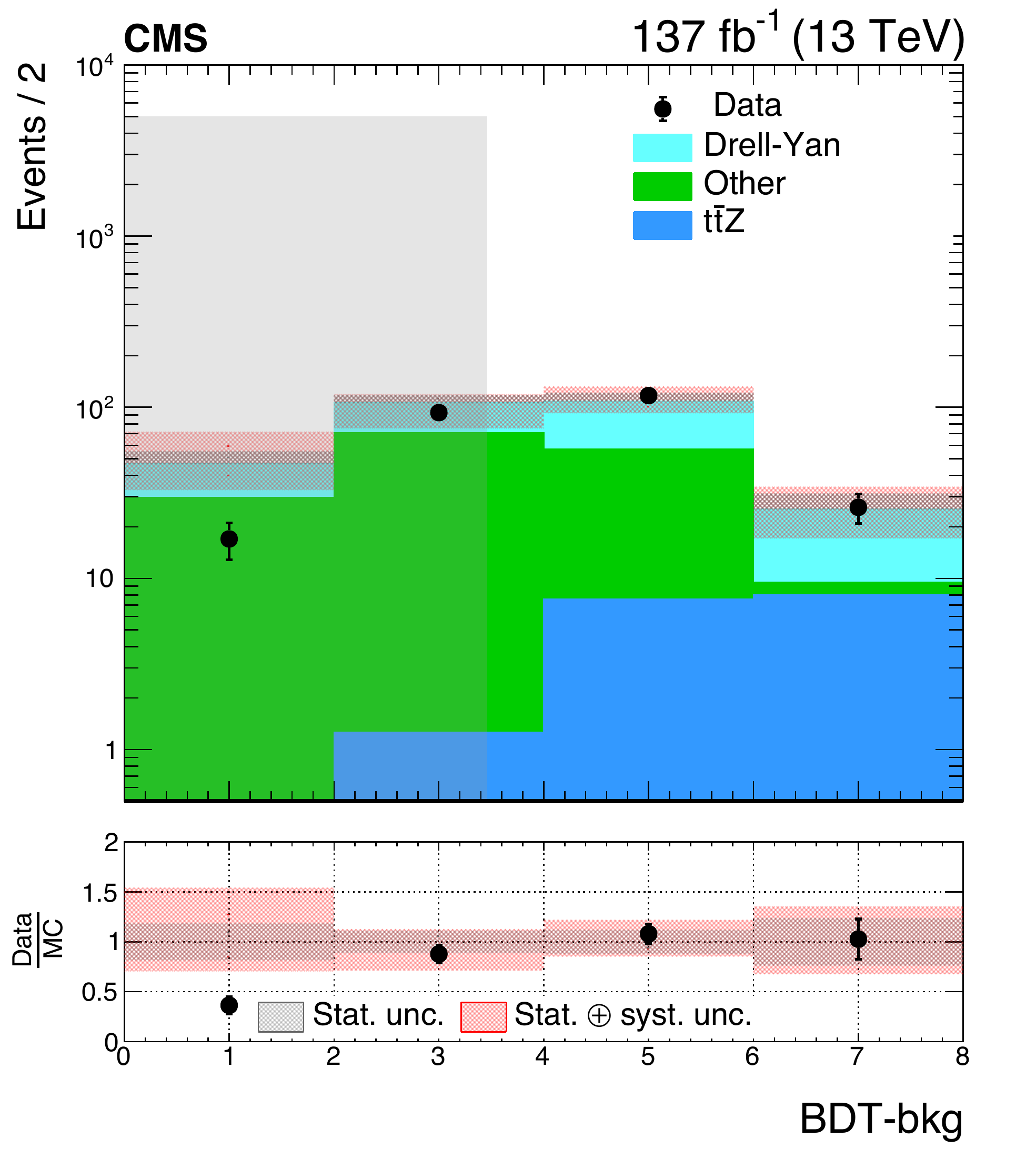}
    \caption{
        Distributions of BDT-bkg output used in the analysis categories targeting \ttH production, for the leptonic (left) and the hadronic (right) channels. 
        The upper two plots show events taken from the $\mgg$ sidebands, satisfying either $100 < \mgg < 120\GeV$ or $130 < \mgg < 180\GeV$.
        The lower two contain events from the $\ttZ$ control regions, described in the text.
        The grey region contains BDT-bkg scores below the lowest threshold for the $\ttH$ analysis categories.
        Total background uncertainties (statistical $\oplus$ systematic) are represented by the black (pink) shaded bands. 
    }
    \label{fig:tth_bdt-bkg}
\end{figure}

The output distributions of the BDT-bkg for both the hadronic and leptonic channels 
are shown in Fig.~\ref{fig:tth_bdt-bkg}.
To validate the modelling of the BDT-bkg in each channel, a $\ttZ,\,\Zee$ control region is used.
The \ttZ events have similar kinematical properties to \ttH events, 
and are therefore suitable for testing the agreement 
between data and simulation in the BDT-bkg score distributions.
Additional requirements on the dielectron kinematics, number of jets, and number of {\cPqb}-tagged jets 
are imposed to increase the $\ttZ$ purity.
The resulting comparisons between data and simulation are shown in Fig.~\ref{fig:tth_bdt-bkg} for the hadronic and leptonic channels.

Finally, events are split using the reconstructed \ptgg value to targeting specific STXS bins.
Analysis categories are then defined through requirements placed on the BDT-bkg output, 
with the boundaries chosen to maximise the expected sensitivity to each bin.

The expected signal and background yields in each analysis category targeting top quark associated Higgs boson
production are shown in Table~\ref{tab:top_yields}.

\begin{table}
    \topcaption{The expected number of signal events for $\mH = 125\GeV$
    in analysis categories targeting Higgs boson production in association with top quark,
    shown for an integrated luminosity of 137\fbinv.
    The fraction of the total number of events arising from each production mode in each analysis category is provided, 
    as is the fraction of events originating from the targeted STXS bin or bins. 
    Entries with values less than 0.05\% are not shown. 
    Here \ggH includes contributions from the \ggZHhad and \bbH production modes, 
    whilst \qqH incorporates both VBF and hadronic \VH production.
    The $\seff$, defined as the smallest interval containing 68.3\% of the \mgg distribution, 
    is listed for each analysis category. 
    The final column shows the expected ratio of signal to signal-plus-background, S/(S+B), 
    where S and B are the numbers of expected signal and background events 
    in a $\pm1\seff$ window centred on $\mH$.}
    \label{tab:top_yields}
    \centering
    \cmsTable{
      \begin{tabular}{lcccccccccc}
          \multirow{3}{*}{Analysis categories} & \multicolumn{9}{c}{SM 125\GeV Higgs boson expected signal} & \multirow{3}{*}{S/(S+B)} \\
           & \multirow{2}{*}{Total} & \multirow{2}{*}{\begin{tabular}[c]{@{}c@{}}Target\\STXS bin(s)\end{tabular}} & \multicolumn{6}{c}{Fraction of total events} & \multirow{2}{*}{\begin{tabular}[c]{@{}c@{}}$\seff$\\(GeV)\end{tabular}} & \\
           & & & \ggH & \qqH & \VH lep & \ttH & \tHq & \tHW & & \\ \hline
           \tHq lep & 1.8 & 23.9\% & 3.5\% & 3.7\% & 34.0\% & 28.8\% & 23.9\% & 6.0\% & 1.62 & 0.42 \\
           [\cmsTabSkip]
           \ttH lep $\ptgg<60$ Tag0 & 0.8 & 93.8\% &\NA&\NA& 0.7\% & 98.2\% & 0.7\% & 0.5\% & 1.71 & 0.72 \\
           \ttH lep $\ptgg<60$ Tag1 & 1.0 & 94.4\% &\NA&\NA& 0.5\% & 97.9\% & 1.5\% & 0.7\% & 1.69 & 0.53 \\
           \ttH lep $\ptgg<60$ Tag2 & 1.8 & 87.7\% &\NA& 0.5\% & 5.1\% & 90.7\% & 3.2\% & 1.1\% & 1.94 & 0.19 \\
           [\cmsTabSkip]
           \ttH lep $60<\ptgg<120$ Tag0 & 1.4 & 95.0\% &\NA&\NA& 1.0\% & 97.3\% & 1.0\% & 0.8\% & 1.60 & 0.64 \\
           \ttH lep $60<\ptgg<120$ Tag1 & 0.6 & 90.8\% &\NA& 0.7\% & 1.0\% & 95.6\% & 1.6\% & 1.1\% & 1.61 & 0.55 \\
           \ttH lep $60<\ptgg<120$ Tag2 & 2.1 & 90.9\% &\NA& 0.1\% & 2.8\% & 93.7\% & 2.5\% & 1.3\% & 1.92 & 0.38 \\
           [\cmsTabSkip]
           \ttH lep $120<\ptgg<200$ Tag0 & 3.6 & 90.1\% & 0.3\% & 0.2\% & 2.7\% & 92.8\% & 2.0\% & 2.0\% & 1.63 & 0.71 \\
           \ttH lep $120<\ptgg<200$ Tag1 & 0.8 & 77.9\% & 2.0\% & 0.5\% & 11.3\% & 80.6\% & 3.2\% & 2.5\% & 1.72 & 0.43 \\
           [\cmsTabSkip]
           \ttH lep $200<\ptgg<300$ Tag0 & 2.5 & 85.9\% & 0.1\% &\NA& 4.1\% & 88.1\% & 3.0\% & 4.8\% & 1.54 & 0.68 \\
           [\cmsTabSkip]
           \ttH lep $\ptgg>300$ Tag0 & 2.1 & 61.7\% & 1.0\% &\NA& 18.0\% & 69.3\% & 3.0\% & 8.7\% & 1.57 & 0.69 \\
           [\cmsTabSkip]
           \ttH had $\ptgg<60$ Tag0 & 1.2 & 94.2\% & 1.7\% & 0.2\% &\NA& 96.6\% & 0.9\% & 0.4\% & 1.68 & 0.49 \\
           \ttH had $\ptgg<60$ Tag1 & 0.4 & 93.5\% & 0.1\% & 0.9\% &\NA& 96.7\% & 1.7\% & 0.6\% & 1.66 & 0.38 \\
           \ttH had $\ptgg<60$ Tag2 & 3.1 & 89.8\% & 1.6\% & 1.5\% & 0.3\% & 92.9\% & 3.0\% & 0.7\% & 1.88 & 0.15 \\
           [\cmsTabSkip]
           \ttH had $60<\ptgg<120$ Tag0 & 1.8 & 92.6\% & 0.6\% &\NA& 0.1\% & 97.6\% & 1.1\% & 0.6\% & 1.55 & 0.77 \\
           \ttH had $60<\ptgg<120$ Tag1 & 0.4 & 90.8\% & 4.6\% & 0.8\% &\NA& 91.9\% & 1.9\% & 0.8\% & 1.35 & 0.39 \\
           \ttH had $60<\ptgg<120$ Tag2 & 5.2 & 88.7\% & 1.0\% & 2.2\% & 0.5\% & 91.8\% & 3.5\% & 1.0\% & 1.90 & 0.23 \\
           [\cmsTabSkip]
           \ttH had $120<\ptgg<200$ Tag0 & 3.6 & 91.4\% & 1.5\% & 0.4\% & 0.1\% & 94.7\% & 2.2\% & 1.3\% & 1.53 & 0.66 \\
           \ttH had $120<\ptgg<200$ Tag1 & 2.1 & 83.3\% & 4.6\% & 2.9\% & 0.5\% & 86.2\% & 4.2\% & 1.7\% & 1.76 & 0.40 \\
           \ttH had $120<\ptgg<200$ Tag2 & 1.7 & 74.3\% & 10.0\% & 4.6\% & 0.6\% & 76.5\% & 6.3\% & 2.0\% & 1.65 & 0.29 \\
           \ttH had $120<\ptgg<200$ Tag3 & 2.6 & 62.2\% & 15.4\% & 8.4\% & 1.2\% & 64.7\% & 8.5\% & 1.9\% & 1.73 & 0.14 \\
           [\cmsTabSkip]
           \ttH had $200<\ptgg<300$ Tag0 & 2.0 & 90.1\% & 0.5\% & 0.4\% & 0.1\% & 92.3\% & 3.8\% & 2.9\% & 1.44 & 0.72 \\
           \ttH had $200<\ptgg<300$ Tag1 & 1.5 & 74.6\% & 8.8\% & 3.1\% & 0.7\% & 77.0\% & 6.8\% & 3.5\% & 1.47 & 0.54 \\
           \ttH had $200<\ptgg<300$ Tag2 & 1.7 & 56.5\% & 18.8\% & 8.4\% & 0.4\% & 58.0\% & 10.5\% & 3.8\% & 1.59 & 0.30 \\ 
           [\cmsTabSkip]
           \ttH had $\ptgg>300$ Tag0 & 2.5 & 73.8\% & 8.3\% & 1.6\% & 0.8\% & 74.9\% & 7.7\% & 6.8\% & 1.44 & 0.77 \\
           \ttH had $\ptgg>300$ Tag1 & 1.9 & 45.6\% & 27.1\% & 7.3\% & 1.4\% & 46.0\% & 11.4\% & 6.7\% & 1.56 & 0.57 \\
           [\cmsTabSkip]
      \end{tabular}
    }
\end{table}

\subsection{Summary of the event categorisation}
\label{sec:categorisation_summary}

The full set of analysis categories targeting the \ggH, VBF, hadronic and leptonic \VH, 
\ttH, and \tHq production mechanisms are summarised in Table~\ref{tab:categorisation_summary}.
The different categorisation regions are shown in descending order of tag priority, 
starting with the \tHq leptonic tag. 
If an event passes the selection criteria for more than one analysis category, 
it is assigned to the tag with the highest priority.
Each STXS bin, or merged group of bins, and the number of analysis categories targeting it are shown.

\begin{table}
    \topcaption{Description of the different categorisation regions, 
           listed in descending order of priority in the first column. 
           The second column shows each targeted STXS bin, or merged group of bins, 
           together with the number of associated analysis categories.
           The last row contains the bins for which no analysis categories are constructed.
          }
    \label{tab:categorisation_summary}
    \centering
    \cmsTableAlt{
      \begin{tabular}{lll}
      Categorisation                 & Particle level STXS bin,                    & Number of  \\
      region                         & (units in \GeVns)                             & categories \\
      \hline
      \multirow{2}{*}{\tHq leptonic}  & \multirow{2}{*}{\tHq}                        & \multirow{2}{*}{1} \\
                                     &                                             &   \\
      [\cmsTabSkip]
      \multirow{5}{*}{\ttH leptonic} & \ttH $\ptH < 60$                            & 3 \\
                                     & \ttH $60 < \ptH < 120$                      & 3 \\
                                     & \ttH $120 < \ptH < 200$                     & 2 \\
                                     & \ttH $200 < \ptH < 400$                     & 1 \\
                                     & \ttH $\ptH > 300$                           & 1 \\
      [\cmsTabSkip]
      \multirow{2}{*}{\ZH leptonic}   & all \ZH lep and                              & \multirow{2}{*}{2} \\
                                     & \ggZH lep bins (10 bins total)               &   \\
      [\cmsTabSkip]
      \multirow{3}{*}{\WH leptonic}   & \WH lep $\ptV < 75$                          & 2 \\
                                     & all \WH lep $75 < \ptV < 150$ (3 bins total) & 2 \\
                                     & \WH lep $\ptV > 150$                         & 1 \\
      [\cmsTabSkip]
      \multirow{2}{*}{\VH MET}        & \multirow{2}{*}{all \VH leptonic bins (15 bins total)}        & \multirow{2}{*}{3} \\
                                     &                                             &   \\
      [\cmsTabSkip]
      \multirow{5}{*}{\ttH hadronic} & \ttH $\ptH < 60$                            & 3 \\
                                     & \ttH $60 < \ptH < 120$                      & 3 \\
                                     & \ttH $120 < \ptH < 200$                     & 4 \\
                                     & \ttH $200 < \ptH < 400$                     & 3 \\
                                     & \ttH $\ptH > 300$                           & 2 \\
      [\cmsTabSkip]
      \multirow{6}{*}{VBF}           & \qqH VBF-like low $\mjj$ low $\ptHjj$        & 2 \\
                                     & \qqH VBF-like low $\mjj$ high $\ptHjj$       & 2 \\
                                     & \qqH VBF-like high $\mjj$ low $\ptHjj$       & 2 \\
                                     & \qqH VBF-like high $\mjj$ high $\ptHjj$      & 2 \\
                                     & \qqH BSM                                     & 2 \\
                                     & all \ggH VBF-like (4 bins total)             & 2 \\
      [\cmsTabSkip]
      \multirow{2}{*}{\VH hadronic}   & \multirow{2}{*}{\qqH \VH-like}                & \multirow{2}{*}{2} \\
                                     &                                             &   \\
      [\cmsTabSkip]
      \multirow{12}{*}{\ggH}          & \ggH 0J low $\ptH$                           & 3 \\
                                     & \ggH 0J high $\ptH$                          & 3 \\
                                     & \ggH 1J low $\ptH$                           & 3 \\
                                     & \ggH 1J med $\ptH$                           & 3 \\
                                     & \ggH 1J high $\ptH$                          & 3 \\
                                     & \ggH $\geq$2J low $\ptH$                     & 3 \\
                                     & \ggH $\geq$2J med $\ptH$                     & 3 \\
                                     & \ggH $\geq$2J high $\ptH$                    & 3 \\
                                     & \ggH $200 < \ptH < 300$                      & 2 \\
                                     & \ggH $300 < \ptH < 450$                      & 2 \\
                                     & \ggH $450 < \ptH < 650$                      & 1 \\
                                     & \ggH $\ptH > 650$                            & 1 \\ 
      [\cmsTabSkip]
      \multirow{2}{*}{No categories} & \qqH 0J, 1J, \mjj \textless 60, 120 \textless \mjj \textless 350, & \multirow{2}{*}{0} \\
                                     & \bbH, \tHW, (6 bins total)                   &                    \\
      \end{tabular}
    }
\end{table}

\section{Statistical procedure}
\label{sec:sig_bkg}

The statistical procedure used in this analysis is identical to that described in Ref.~\cite{Khachatryan:2014jba}, as developed by the ATLAS and CMS Collaborations. 
Simultaneous binned maximum likelihood fits are performed to the $\mgg$ distributions of all analysis categories, in the range $100 < \mgg < 180\GeV$. 
A likelihood function is defined for each analysis category using analytic models to describe the $\mgg$ distributions of signal and background events, 
with nuisance parameters to account for the experimental and theoretical systematic uncertainties.

The analytic signal model is derived from simulation, with a model constructed for each particle level STXS bin in each reconstructed analysis category.
Both the shape and normalisation of the model are parametrised as functions of \mH.

The background model is determined directly from the observed \mgg distribution in data.
The analytic model for each analysis category can take one of a range of different functional forms, 
all of which represent a smoothly falling spectrum.

The best fit values and confidence intervals for the parameters of interest are estimated using a profile likelihood test statistic
\begin{equation}\label{eq:test_statistic}
  q(\vec{\alpha}) = -2 \ln \Bigg( \frac{L(\vec{\alpha},\hat{\vec{\theta}}_{\vec{\alpha}})}{L(\hat{\vec{\alpha}},\hat{\vec{\theta}})} \Bigg).
\end{equation}
The likelihood functions in the numerator and denominator of Eq.~(\ref{eq:test_statistic}) are constructed using the product over the likelihood functions defined for each analysis category. 
The quantities $\hat{\vec{\alpha}}$ and $\hat{\vec{\theta}}$ describe the unconditional maximum likelihood estimates for the parameters of interest and the nuisance parameters, respectively, 
whereas $\hat{\vec{\theta}}_{\vec{\alpha}}$ corresponds to the conditional maximum likelihood estimate for fixed values of the parameters of interest, $\vec{\alpha}$.
In this analysis, the parameters of interest can be signal strengths, cross sections or coupling modifiers, depending on the fit being performed. 
In all fits, $\mH$ is fixed to its most precisely measured value of 125.38\GeV~\cite{HggMass}.
This choice is made to ensure that all measurements are reported 
with respect to the theoretical predictions consistent with the best available knowledge of \mH.
Further discussion of the implications of this choice and the difference with respect to profiling \mH
is given in Section~\ref{sec:results_sigstrengths}.

The best fit parameter values, $\hat{\vec{\alpha}}$, are identified as those that maximise the likelihood. 
For one-dimensional measurements, such as the signal strength and STXS fits, 
the 68 and 95\% confidence intervals are defined by the union of intervals 
for which $q(\vec{\alpha})<0.99$ and $<3.84$, respectively. 
In the case where there are multiple parameters of interest in the fit, 
the intervals are determined treating the other parameters as nuisance parameters. 
For two-dimensional measurements, such as those performed to coupling modifiers in the $\kappa$-framework, 
the 68 and 95\% confidence regions are defined by the set of parameter values 
that satisfy $q(\vec{\alpha})<2.30$ and $<5.99$, respectively. 
To compute the SM expected results, the observed data is replaced by an Asimov data set generated with all parameter values set to the SM expectation~\cite{Cowan}.

The methods used to construct the signal and background models are described in detail in the remainder of this section.

\subsection{Signal model}
\label{sec:signal}

The signal shape for the \mgg distribution in each
analysis category and for a nominal \mH is constructed from
the simulation of each production process.

Since the distribution of \mgg 
depends on whether the vertex associated with the candidate diphoton
was correctly identified within 1\cm, the correct
vertex and wrong vertex scenarios are considered separately when
constructing the signal model.
In a given analysis category, a separate function is constructed for events
originating from each STXS bin in each vertex scenario, by fitting
the \mgg distribution using a sum
of at most five Gaussian functions.
This choice provides sufficient flexibility in the fit whilst maintaining computational efficiency.
The number of Gaussian functions is determined using an $\mathcal{F}$-test~\cite{fTest},
avoiding overfitting statistical fluctuations due to the limited size of the simulated samples.

\begin{figure}[htb!]
  \centering
	\includegraphics[width=0.49\textwidth]{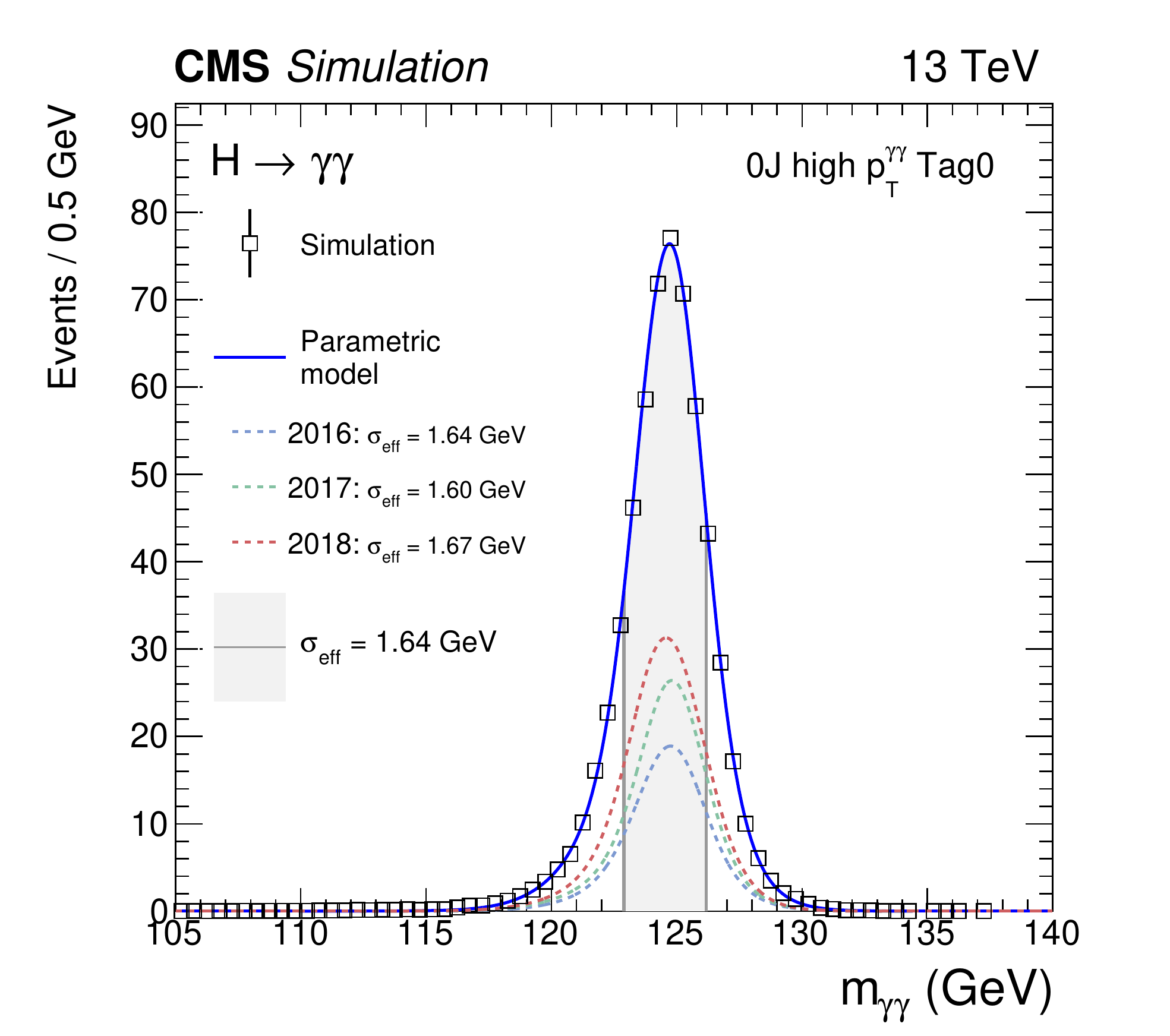}
	\includegraphics[width=0.49\textwidth]{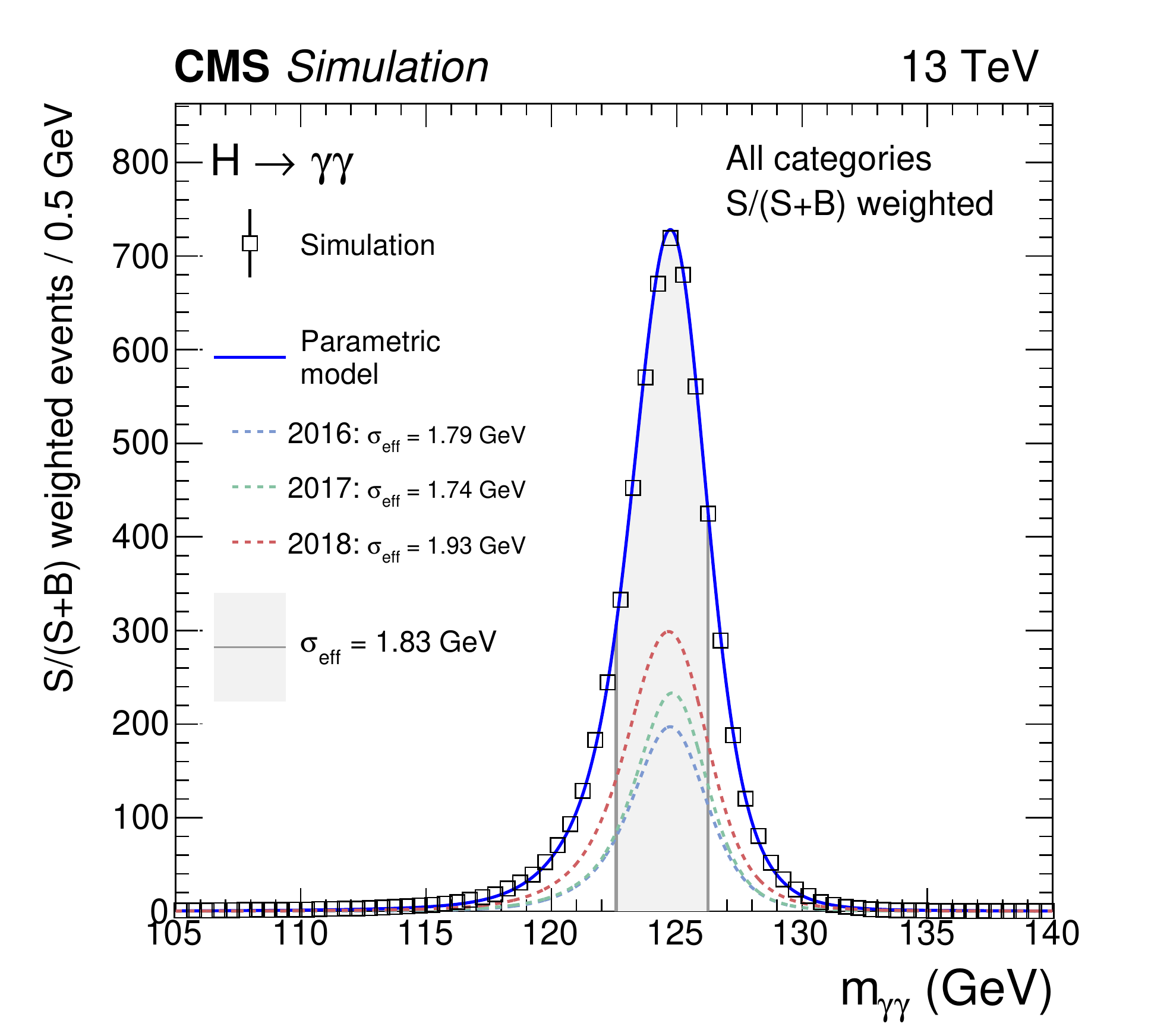}
  \caption{
  The shape of the parametric signal model for each year of simulated data, 
  and for the sum of all years together, is shown.
  The open squares represent weighted
  simulation events and the blue line the corresponding model.
  Also shown is the \seff value (half the width of
  the narrowest interval containing 68.3\% of the \mgg
  distribution) in the grey shaded area.
  The contribution of the signal model from each year of data taking is 
  illustrated with the dotted lines.
  The models are shown for an analysis category targeting \ggH 0J high \ptH production (left), 
  and for the weighted sum of all analysis categories (right).
  Here each analysis category is weighted by S/(S+B),
  where S and B are the numbers of expected signal and background events, respectively,
  in a $\pm 1 \seff$ \mgg window centred on \mH.
  }
  \label{fig:SigBkg_SigPlots}
\end{figure}

The final fit function for each analysis category is obtained by summing the
individual functions for all STXS bins in both vertex scenarios.
Figure~\ref{fig:SigBkg_SigPlots} shows signal models for
each year individually, and for the sum of the three years together. 
The \seff is defined as half of the smallest interval containing 68.3\%
of the \mgg distribution.

\subsection{Background model}
\label{sec:background}

The model used to describe the background is extracted from data using
the discrete profiling method~\cite{Envelope,HggRun1}.
This technique estimates the systematic uncertainty
associated with choosing a particular analytic function to fit the
background \mgg distribution.
The choice of the background function is treated as a discrete 
nuisance parameter in the likelihood fit to the data.

A large set of candidate function families is considered, including
exponential functions, Bernstein polynomials, Laurent series, and power law functions~\cite{Envelope}.
For each family of functions, an $\mathcal{F}$-test~\cite{fTest} is 
performed to determine the maximum order of parameters to be used,
while the minimum order is determined
by placing a requirement on the goodness-of-fit to the data.

When fitting these functions to the \mgg distribution,
the value of twice the negative logarithm of the likelihood (\dNLL) is minimised.
A penalty is added to \dNLL to take into account the
number of floating parameters in each candidate function.
When making a measurement of a given parameter of interest, the discrete profiling
method minimises the overall \dNLL considering all allowed functions for each analysis category.
Checks are performed to ensure that describing the background \mgg distribution in this way 
introduces negligible bias to the final results.

\section{Systematic uncertainties}
\label{sec:systematics}

In this analysis, the systematic uncertainty associated with the background estimation from data 
is handled using the discrete profiling method, as described above.
There are many systematic uncertainties that affect the signal model; 
these are handled in one of two ways.
Uncertainties that modify the shape of the \mgg distribution are incorporated into the signal model
as nuisance parameters, where the mean and width of each Gaussian function can be affected.
These uncertainties are typically experimental uncertainties 
relating to the energy of the individual photons. 
Conversely if the shape of the \mgg distribution is unaffected, 
the uncertainty is treated as a log-normal variation in the event yield.
These uncertainties include theoretical sources
and experimental uncertainties such as those affecting the BDTs used to categorise events.
The magnitude of each uncertainty's impact is determined individually 
for each STXS bin in each analysis category.

\subsection{Theoretical uncertainties}

Theoretical uncertainties affect both the overall cross section prediction for a given STXS bin, and the distributions of kinematic variables used in the event selection and categorisation. 
When measurements of cross sections are performed, the uncertainties in the overall cross sections are omitted, and instead are considered as uncertainties in the SM predictions. 
The uncertainties related to the event kinematic properties, which affect the efficiency and acceptance of the analysis, are still taken into account. 
Uncertainties affecting the overall cross section normalisations and those affecting the event kinematic properties are included when measuring signal strengths and coupling modifiers.
When deriving the effect on the kinematic distributions, the impact on the STXS bin cross section normalisation is factored out to avoid double counting.

The sources of theoretical uncertainty considered in this analysis are as follows:
\begin{itemize}
\item \textit{Renormalisation and factorisation scale uncertainties}: 
  the uncertainty arising from variations of the renormalisation and factorisation scales
  used when computing the expected SM cross section and event kinematic properties.
  These account for the missing higher-order terms in perturbative calculations.
  The recommendations provided in Ref.~\cite{YR4} are followed.
  The uncertainty in the overall normalisation is estimated by: 
  varying the renormalisation by a factor of two; varying the factorisation scale by a factor of two; 
  varying both in the same direction simultaneously.
  Depending on the production process, the size of the uncertainty in the overall normalisation varies from around 0.5\% for VBF production to 15\% for \tHq production.

  To estimate the uncertainty in the event kinematic properties,
  the distribution of events falling into each analysis category is recalculated when
  varying both the renormalisation and factorisation scales by a factor of two in the same direction simultaneously.  The overall cross section for a given STXS bin is kept constant.
  These uncertainties, representing migrations between analysis categories, are decorrelated 
  for different production modes and different regions of the Higgs boson phase space, 
  resulting in 22 independent nuisance parameters.
  The migration uncertainties are in general around 1\%.
   
\item \textit{Uncertainties in the \ggH STXS fractions}: 
  for \ggH production, 
  additional sources are included that account for the uncertainty in the modelling 
  of the \ptH distributions, the number of jets in the event, 
    and the \ggH contamination in the VBF categories.
  A number of sources are introduced to reflect the migration of events around the \ptH bin boundaries,
  at 10\GeV for zero-jet events and 60 and 120\GeV for events with at-least one jet, such that 
  their magnitude depends on the number of jets and the \ptH.
  An additional source covers the uncertainty in \ptH in the Lorentz-boosted region
  arising from the assumption of infinite top quark mass in the \ggH loop. 
  This is determined by comparing the \ptH distribution to the prediction from finite-mass calculations.
  Two further sources account for the migration between the zero, one, and two or more jet bins.
  The uncertainty in the \ggH production of events with a VBF-like dijet system 
  is covered by two sources 
  corresponding to the prediction in the two- and three-jet-like bins.
  In addition, two nuisance parameters are introduced to account for migrations across the \mjj bin boundaries, at 350 and 700\GeV.
  The total magnitudes of these uncertainties vary from around 5 to 30\%, 
  with events that have one or more jets and high values of \ptH 
  typically having the greatest associated uncertainty.
  These uncertainties affect the overall cross section normalisations, 
  and so are attributed to the SM prediction in the cross section measurements.

\item \textit{Uncertainties in the \qqH STXS fractions}:           
  similarly
  for \qqH production, 
  additional sources are introduced to account for the uncertainty in the modelling 
  of the \ptH, \mjj, and \ptHjj distributions, and the number of jets in the event.
  A total of six sources are defined to reflect migrations of events across \mjj boundaries
  at 60, 120, 350, 700, 1000, and 1500\GeV.
  Two additional nuisance parameters account for migrations across the $\ptH=200\GeV$
  and $\ptHjj=25\GeV$ bin boundaries.
  Finally, a single source is defined to account for a migration between the zero and one,
  and the two or more jet bins.
  In each case, the uncertainty is computed by varying the renormalisation and factorisation scales and recalculating
  the fractional breakdown of \qqH STXS stage-1.2 cross sections.
  The total magnitude varies between bins but is at most 8\%.
  Again, these are considered as uncertainties in the SM predictions when performing cross section measurements.

\item \textit{Uncertainties in the \ttH STXS fractions}:           
  for \ttH production, four nuisance parameters are used to account for the 
  uncertainty in the \ptH distributions. 
  Each nuisance parameter represents migration across one of the boundaries at the \ptH values of 
  60, 120, 200, and 300\GeV that define the ttH STXS bins.
  The magnitudes of these uncertainties are derived by varying the renormalisation and factorisation scales, 
  and have values of up to 9\%. 
  When performing cross section measurements, the sources are treated as uncertainties on the SM prediction.

\item \textit{Uncertainties in the \VH leptonic STXS fractions}:           
  for \VH leptonic production, 
  additional sources are introduced to account for the uncertainty in the modelling 
  of the \ptV distributions, and the number of jets in the event.
  Four independent sources are defined to reflect the migrations of events across the \ptV boundaries at 75, 150, and 250\GeV, in addition to the migration between the zero and greater than one-jet bins for events with \ptV of 150-250\GeV. 
  These sources are defined separately for the \WH leptonic, \ZH leptonic, and \ggZH leptonic production modes, leading to 12 independent nuisance parameters.
  In each case, the uncertainty is computed by varying the renormalisation and factorisation scales and recalculating
  the fractional breakdown of \VH leptonic STXS stage-1.2 cross sections.
  The total magnitude varies between bins but is at most 5\% for the dominant \WH and \ZH leptonic production modes.
  Again, these are considered as uncertainties in the SM predictions when performing cross section measurements.

\item \textit{Uncertainty in the \ggH contamination of the top quark associated categories}:
  the theoretical predictions for \ggH are less reliable in a regime 
  where the Higgs boson is produced in association with a large number of jets. 
  {\tolerance=800 Three different contributions are considered: the uncertainty from the parton shower modelling, 
  estimated by taking  the observed difference in the jet multiplicity 
  between \MGvATNLO predictions and data in $\ttbar+\text{jets}$ events \cite{CMS-PAS-TOP-16-011}, 
  the uncertainty in the gluon splitting modelling, 
  estimated by scaling the fraction of events from \ggH with real {\cPqb} quark jets 
  in simulation by the measured difference between data and 
  simulation of $\sigma(\ttbar\bbbar)$/$\sigma(\ttbar\text{jj})$~\cite{Sirunyan:2017snr} 
  and the uncertainty due to the limited size of the simulated samples. \par} 
  The combined impact of these uncertainties in the top quark associated signal strength is about 2\%.

\item \textit{Parton distribution function uncertainties}:
  these account for the uncertainty due to imperfect knowledge of the composition of the proton, 
  which affects the partons that are most likely to initiate high energy events.
  The overall normalisation uncertainties for each Higgs boson production process 
  also include the uncertainty in the value of the strong coupling constant $\alpS$, 
  and are taken from Ref.~\cite{YR4}.
  Uncertainties in the event kinematic properties are calculated 
  following the PDF4LHC\textunderscore100~prescription~\cite{PDF4LHC,Dulat:2015mca,Harland-Lang:2014zoa,NNPDF3} 
  using the {\sc MC2hessian}~procedure~\cite{MC2Hessian,Gao:2013bia}.
  As with the renormalisation and factorisation scale uncertainties, 
  the normalisation for a given STXS bin is kept constant 
  when calculating the migrations between analysis categories.
  The overall normalisation uncertainties are 1--5\%, 
  with the migrations significantly smaller, usually less than 1\%.

\item \textit{Uncertainty in the strong coupling constant}: 
  the uncertainty in the value of $\alpS$
  is included in the treatment of the PDF uncertainties, following the PDF4LHC prescription.
  The impact on the overall normalisation is largest for \ggH production, with a value of 2.6\%.
  An additional source is included to account for changes in the event kinematic properties
  due to the uncertainty in $\alpS$.
  This is calculated using a similar procedure to the renormalisation and factorisation scale migration uncertainties,
  but instead varying the value of $\alpS$,
  and corresponds to uncertainties that are in general less than 1\%.

\item \textit{Uncertainty in the \Hgg branching fraction}: 
  the probability of the Higgs boson decaying to two photons is required to calculate
  the SM expected cross section, but this branching fraction is not known exactly.
  The uncertainty is currently estimated to be around 3\%~\cite{YR4}.
  This uncertainty is included in the signal strength and coupling modifier measurements,
  and is considered an uncertainty in the SM predictions for cross section measurements.

\item \textit{Underlying-event and parton shower uncertainties}: 
  these uncertainties are obtained using dedicated simulated samples.
  The parton shower uncertainties originating from the modelling of the hadronization
  are evaluated by varying the renormalisation scale for QCD emissions in
  initial-state and final-state radiation by a factor of 2 and 0.5.
  The uncertainties in the modelling of the underlying-event are evaluated
  by varying the {\PYTHIA}8 tune from that used in the nominal simulation samples,
  introduced in Section~\ref{sec:samples}.
  Both these uncertainties are treated as variations in the relative contributions from each
  STXS bin for a given production mode, and therefore affect the STXS bin cross section normalisation. 
  The impact is in general around 5\%, but can be as large as 30\% for bins 
  corresponding to high \ptH and high jet multiplicity.
\end{itemize}

As described in Section \ref{sec:results_STXS},
it is necessary to merge certain STXS bins when measuring cross sections
to avoid large uncertainties or very high correlations between parameters.
If two bins are measured individually, the theoretical uncertainty representing event migrations
between the two bins are not included since both cross sections are being fitted.
The act of merging bins across a boundary means the measurement is sensitive to
the relative fraction of the two bins, and an uncertainty must be included to model this.
As a result, the uncertainty sources accounting for migrations across the merged boundaries are included in the relevant cross section measurements.

\subsection{Experimental uncertainties}

The uncertainties that affect the shape of the signal \mgg distribution are listed below.
These include uncertainties that account for the difference between photon showers 
and the electron showers used to derive the energy scale corrections.
The combined effect of all signal model shape uncertainties in the measurement of the inclusive Higgs boson signal strength modifier is found to be about 2\%.
\begin{itemize}
\item \textit{Photon energy scale and resolution}: 
  the uncertainties associated with the corrections applied to the photon energy scale in data
  and the resolution in simulation are evaluated using \Zee events.
  The estimate is computed by varying the regression training scheme, the distribution of \RNINE, 
  and the electron selection criteria. 
  For the majority of photons the resulting uncertainty in the energy scale is 0.05--0.15\%, 
  although for those with very high \pt the effect can be 0.5--3.0\%.
\item \textit{Nonlinearity of the photon energy scale}: 
  a further source of uncertainty covers possible remaining differences in the linearity 
  of the photon energy scale between data and simulation.
  The uncertainty is estimated using Lorentz-boosted \Zee events.
  In this analysis, an uncertainty of 0.2\% on the photon energy scale is assigned, 
  which accounts for the nonlinearity across the full range of photon \pt values.
\item \textit{Shower shape corrections}: 
  an uncertainty in the shower shape corrections accounts for the imperfect modelling 
  of shower shapes in simulation.
  The impact is estimated by comparing the energy scale before and after 
  the corrections to shower shape variables, 
  as described in Section~\ref{sec:event_reco}, are applied.
  The magnitude of the uncertainty in the energy scale ranges from 0.01--0.15\%, 
  depending on the photon $\abs{\eta}$ and {\RNINE} values.
\item \textit{Longitudinal nonuniformity of light collection}: 
  an uncertainty is associated with the modelling of the light collection 
  as a function of emission depth within a given ECAL crystal.
  The calculation of this uncertainty is described in detail in Ref.~\cite{HggMass}.
  The uncertainty is 0.16--0.25\% for photons with $\RNINE > 0.96$,
  whilst the magnitude for low {\RNINE} photons is below 0.07\%.
\item \textit{Modelling of material in front of the ECAL}: 
  the amount of material through which objects pass before reaching the ECAL 
  affects the behaviour of the electromagnetic showers, 
  and may not be perfectly modelled in simulation.
  Dedicated samples with variations in the amount of upstream material are used to 
  estimate the impact on the photon energy scale.
  The magnitude of the resulting uncertainty ranges from 0.02--0.05\% for the most central photons, 
  increasing to as much as 0.24\% for those in the endcap.
\item \textit{Vertex assignment}: 
  the largest contribution to the uncertainty in the fraction of events where the chosen vertex
  is smaller than $1\unit{cm}$ from the true vertex comes from the modelling of the underlying-event. 
  In addition, the uncertainty in the ratio of data and simulation obtained using \Zmumu events is incorporated.
  A nuisance parameter is included in the signal model that allows the fraction
  of events in each vertex scenario to vary by $\pm 2\%$.
\end{itemize}

The uncertainties that only modify the event yield have an effect of around 4\% on the inclusive Higgs boson signal strength modifier measurement.
They include the set of sources described below.
\begin{itemize}
\item \textit{Integrated luminosity}: 
  uncertainties of 2.5, 2.3, and 2.5\% are determined by the CMS luminosity monitoring 
  for the 2016, 2017, and 2018 data sets~\cite{CMSlumi2016,CMSlumi2017,CMSlumi2018}, respectively, 
  whilst the uncertainty on the total integrated luminosity of the three years together is 1.8\%.
  The uncertainties for each data set are partially correlated 
  to account for common sources in the luminosity measurement schemes.
\item \textit{Photon identification BDT score}: 
  the uncertainty arising from the photon identification BDT score 
  is estimated by varying the set of events used to train the quantile regression corrections.
  It is seen to cover the residual discrepancies between data and simulation. 
  The uncertainty in the signal yields is
  estimated by propagating this uncertainty through the full category selection procedure.
  The impact in the most sensitive analysis categories is around 3\%.
\item \textit{Jet energy scale and smearing corrections}: 
  The energy scale of jets is measured using the \pt balance of jets with $\PZ$ bosons and photons in
  \Zee, \Zmumu, and \gamplusjets events, as well as the \pt balance between jets 
  in dijet and multijet events \cite{JetsInRun2}. The uncertainty in the jet energy scale
  is a few percent and depends on \pt and $\eta$. The impact of jet energy scale uncertainties in 
  event yields is evaluated by varying the jet energy corrections within their uncertainties and 
  propagating the effect to the final result.
  Correlations between years are introduced for the different jet energy scale uncertainty sources,
  ranging between 0 and 100\%.
  The impact on the category yields is largest for those targeting VBF, hadronic \VH and top quark associated production 
  and can be as high as 22\% 
  for the scale uncertainties, but is less than around 8\% for the resolution.
\item \textit{Per-photon energy resolution estimate}: 
  the uncertainty in the per-photon resolution is
  parametrised as a rescaling of the resolution by
  $\pm 5\%$ about its nominal value. 
  This is designed to cover all differences between data and simulation 
  in the distribution, which is an output of the energy regression.
  The maximum yield variation in an analysis category is around 5\%, 
  however for most categories the impact is below the percent level.
\item \textit{Trigger efficiency}: 
  the efficiency of the trigger selection is measured with 
  $\Zee$ events using the tag-and-probe technique.
  The size of its uncertainty is less than 1\%.
  An additional uncertainty is introduced to account for a 
  gradual shift in the timing of the inputs of the ECAL first level trigger in the region at $\abs{\eta} > 2.0$, 
  which caused a specific trigger inefficiency during 2016 and 2017 data taking~\cite{CMS_L1T}. 
  Both photons and to a greater extent jets can be affected by this inefficiency. 
  The resulting uncertainty is largest for the categories targeting VBF production,
  with a maximum impact on the yield of 1.4\%.
\item \textit{Photon preselection}: 
  the uncertainty in the preselection efficiency
  is computed as the ratio between the efficiency measured in data and in simulation.
  Its magnitude is less than 1\%.
\item \textit{Missing transverse momentum}: 
  this uncertainty is computed by shifting the
  reconstructed $\pt$ of the particle candidates entering the
  \ptmiss computation, 
  within the momentum scale and resolution uncertainties appropriate 
  to each type of reconstructed object,
  as described in Ref.~\cite{JetsInRun2}.
  In this analysis, the impact on the category yields is never larger than 5\%,
  even for analysis categories that explicitly use the \ptmiss in their definition.
\item \textit{Pileup jet identification}: 
  the uncertainty in the pileup jet classification output score is estimated by
  comparing the score of jets in events with a $\PZ$ boson and one balanced jet
  in data and simulation. 
  The magnitude is of the order 1\%.
\item \textit{Lepton isolation and identification}: 
  this uncertainty affecting electrons and muons
  is computed by varying the ratio of the efficiency in simulation to the efficiency in data and
  using the tag-and-probe technique in \Zee events. 
  The resulting impact on the categories selecting leptons is up to around 1\%.
\item \textit{{\cPqb} jet tagging}: 
  uncertainties in the {\cPqb} tagging efficiency are evaluated 
  by comparing data and simulated distributions for the {\cPqb} tagging discriminator.
  The uncertainties include the statistical component in the
  estimate of the fraction of heavy- and
  light-flavour jets in data and simulation.
  Its magnitude is around 3\% for the analysis categories targeting top quark associated production,
  which make use of the {\cPqb} tagging discriminant.
\end{itemize}

Most of the experimental uncertainties are left uncorrelated among the different years.
The exceptions are the partial correlations introduced for the integrated luminosity and jet energy correction uncertainties.

\section{Results}
\label{sec:results}

\begin{figure}
  \centering
  \includegraphics[width=1.\textwidth]{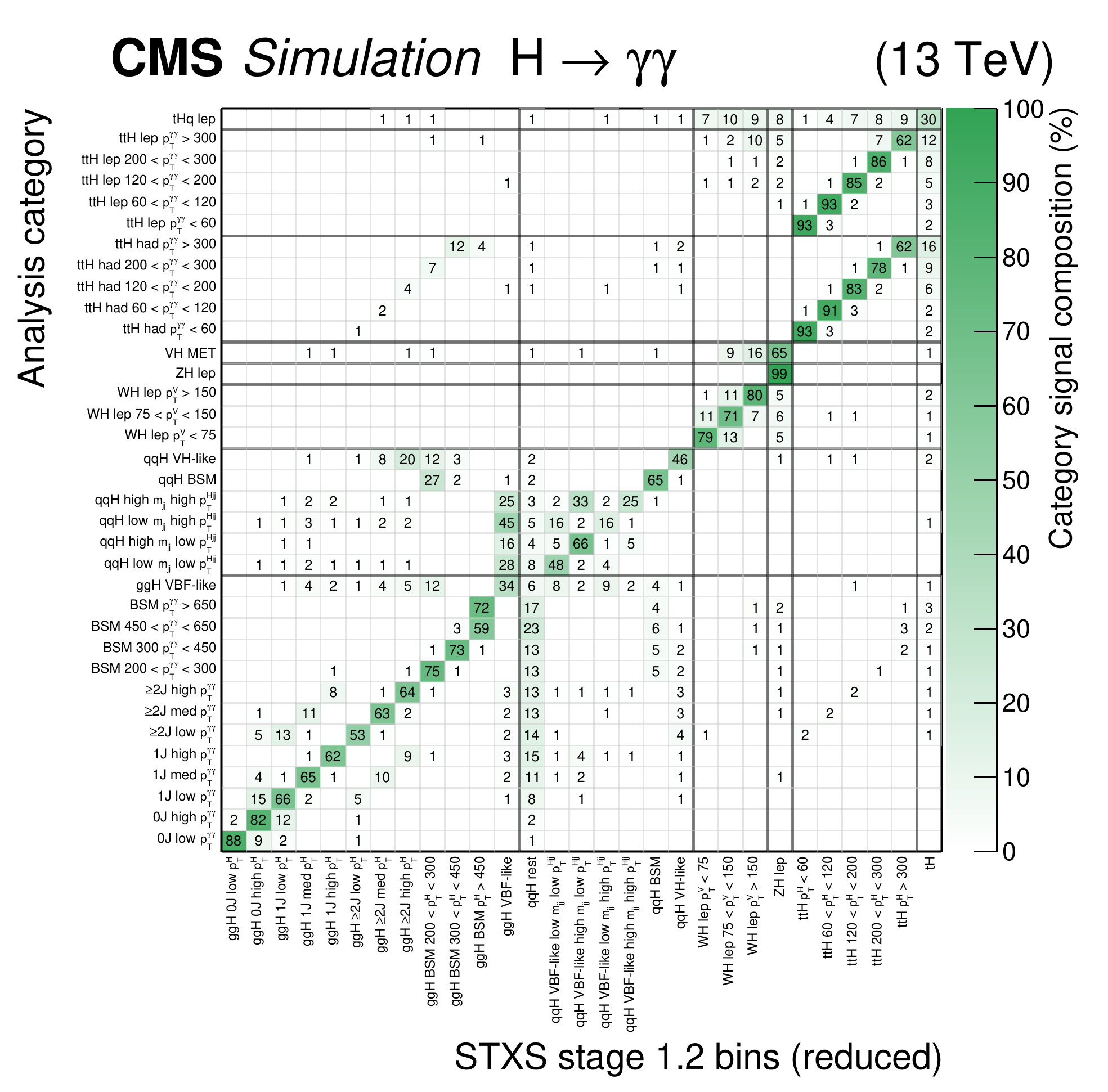}
  \caption{The composition of the analysis categories in terms of a merged set of STXS bins
    is shown. The granularity of the STXS bin merging corresponds to the 
    finest granularity used for the cross section measurements in this analysis.
    Analysis categories targeting a common STXS region are summed, where the
    signal compositions of the individual categories are weighted in the sum by the 
    expected ratio of signal to signal-plus-background events.
    The colour scale corresponds to the fractional yield in each analysis category group
    (rows) accounted for by each STXS process (columns). Each row therefore
    sums to 100\%. Entries with values less than 0.5\% are not shown.
    Simulated events for each year in the period 2016--2018 are combined with appropriate weights
    corresponding to their relative integrated luminosity in data.
    The column labelled as ``qqH rest" includes contributions from the \qqH 0J, \qqH 1J, \qqH $\mjj<60\GeV$ and \qqH $120<\mjj<350\GeV$ STXS bins.
    }
  \label{fig:results_purityMatrix}
\end{figure}

The expected signal composition of the analysis categories in terms of a set of merged STXS bins is shown in Fig.~\ref{fig:results_purityMatrix}. In the plot, the analysis categories targeting a common STXS region are summed, such that the signal compositions of the individual analysis categories are weighted according to the ratio of the numbers of signal to signal-plus-background events (S/S+B). The fractional contribution of the total signal yield in a given analysis category group arising from each process is shown.

\begin{figure}[htb!]
  \centering
  \includegraphics[width=0.7\textwidth]{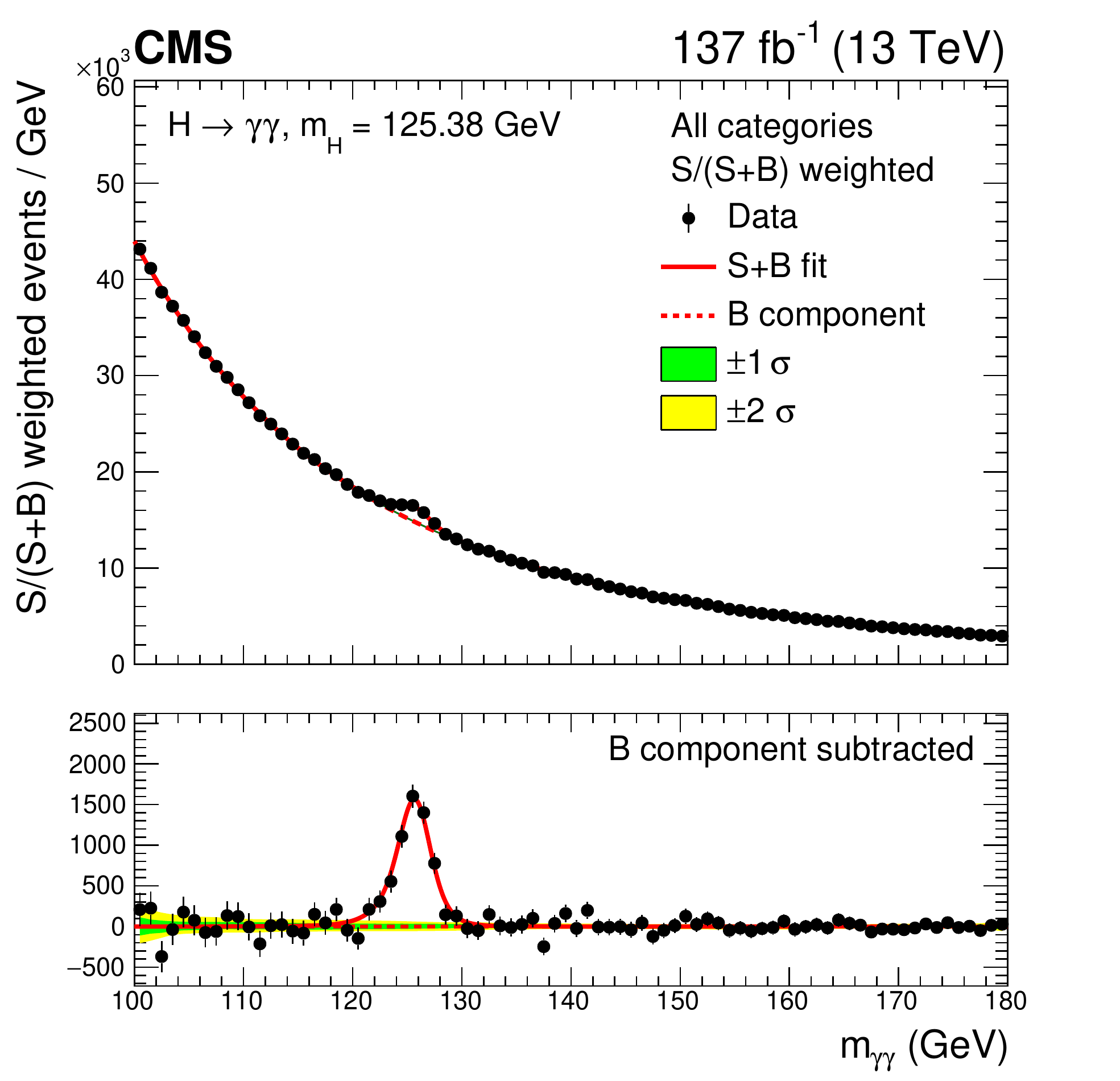}
  \caption{
    Data points (black) and signal-plus-background model fit for the sum of all analysis categories is shown.
    Each analysis category is weighted by S/(S+B),
    where S and B are the numbers of expected signal and background events, respectively,
    in a $\pm 1 \seff$ \mgg window centred on \mH.
    The one (green) standard deviation and two (yellow) standard deviation bands show
    the uncertainties in the background component of the fit.
    The solid red line shows the total signal-plus-background contribution, whereas the dashed red line shows the background component only. The lower 
    panel shows the residuals after subtraction of this background component.}
  \label{fig:results_sPlusBplot}
\end{figure}

The best fit signal-plus-background model is shown with data for the sum of all analysis categories in Fig.~\ref{fig:results_sPlusBplot}.
Again each analysis category is weighted by (S/S+B), such that the absolute signal yield is kept constant.

\subsection{Signal strength modifiers}
\label{sec:results_sigstrengths}

A common signal strength modifier, $\mu$, is defined as the ratio of the observed product of the Higgs boson cross section and diphoton branching fraction to the SM expectation.
It is measured to be
\begin{equation*}
  \mu = 1.12 ^{+0.09}_{-0.09} = 1.12 ^{+0.06}_{-0.06}\thy ^{+0.03}_{-0.03}\syst ^{+0.07}_{-0.06}\stat.
\end{equation*}
The uncertainty is decomposed into theoretical systematic, experimental systematic, and statistical components.
The statistical component includes the uncertainty in the background modelling.
The compatibility of this fit with respect to the SM prediction, expressed as a $p$-value, is approximately 17\%.

In this fit, and in all subsequent fits, \mH is fixed to its most precisely measured value of $125.38\GeV$~\cite{HggMass}.
The precise determination of \mH and the systematic uncertainties that enter its measurement are beyond the scope of this analysis.
Nonetheless, the dependence of the measured signal strengths on \mH is checked.                              
Profiling \mH without constraint, rather than fixing it to 125.38\GeV, has a small impact on the measured results;        
the best fit signal strength values change by 0.7--1.8\%.                                             
In each case, the change is less than 10\% of the measured uncertainty. 

Signal strength modifiers for each Higgs boson production mode are also measured.
Unlike the subsequent STXS fits described in Section \ref{sec:results_STXS}, the \VH hadronic and \VH leptonic processes are grouped to scale according to $\mu_{\VH}$,
whereas the VBF production mode scales with $\mu_{\text{VBF}}$. 
The parameter $\mu_{\text{top}}$ scales the \ttH, \tHq and \tHW production modes equally and $\mu_{\ggH}$ scales both \ggH and \bbH production.

\begin{figure}
  \centering
  \includegraphics[width=0.49\textwidth]{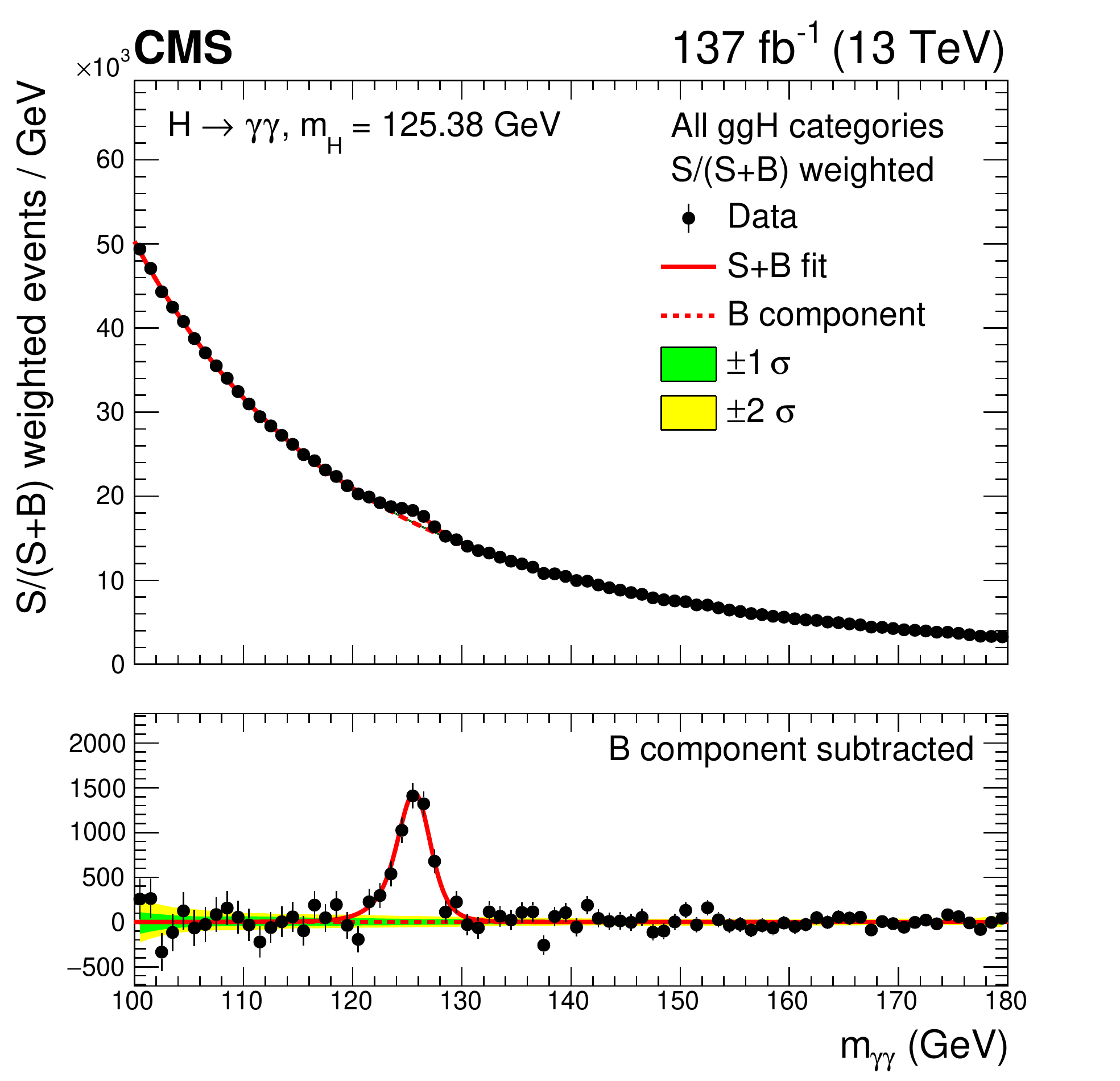}
  \includegraphics[width=0.49\textwidth]{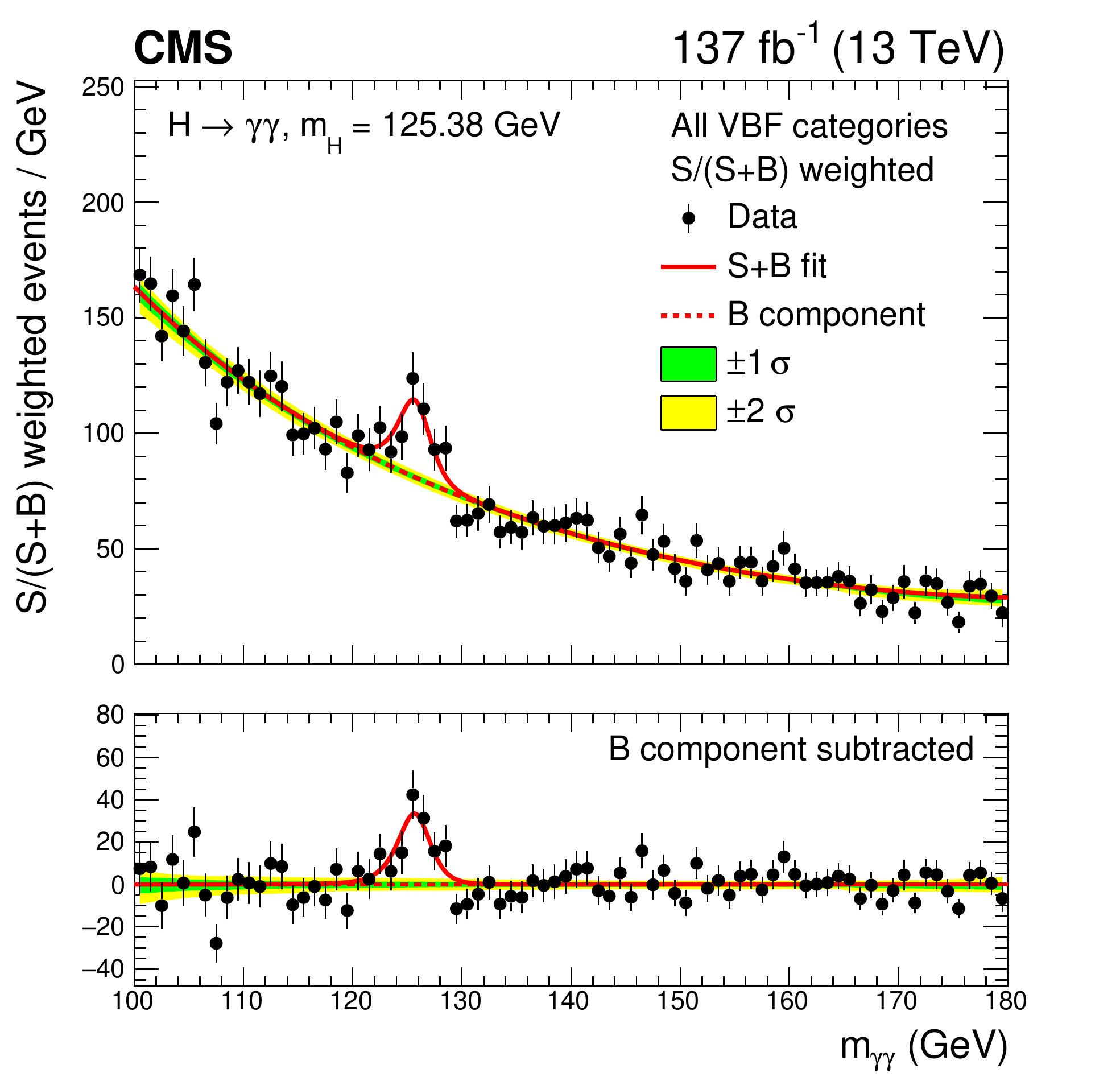}
  \includegraphics[width=0.49\textwidth]{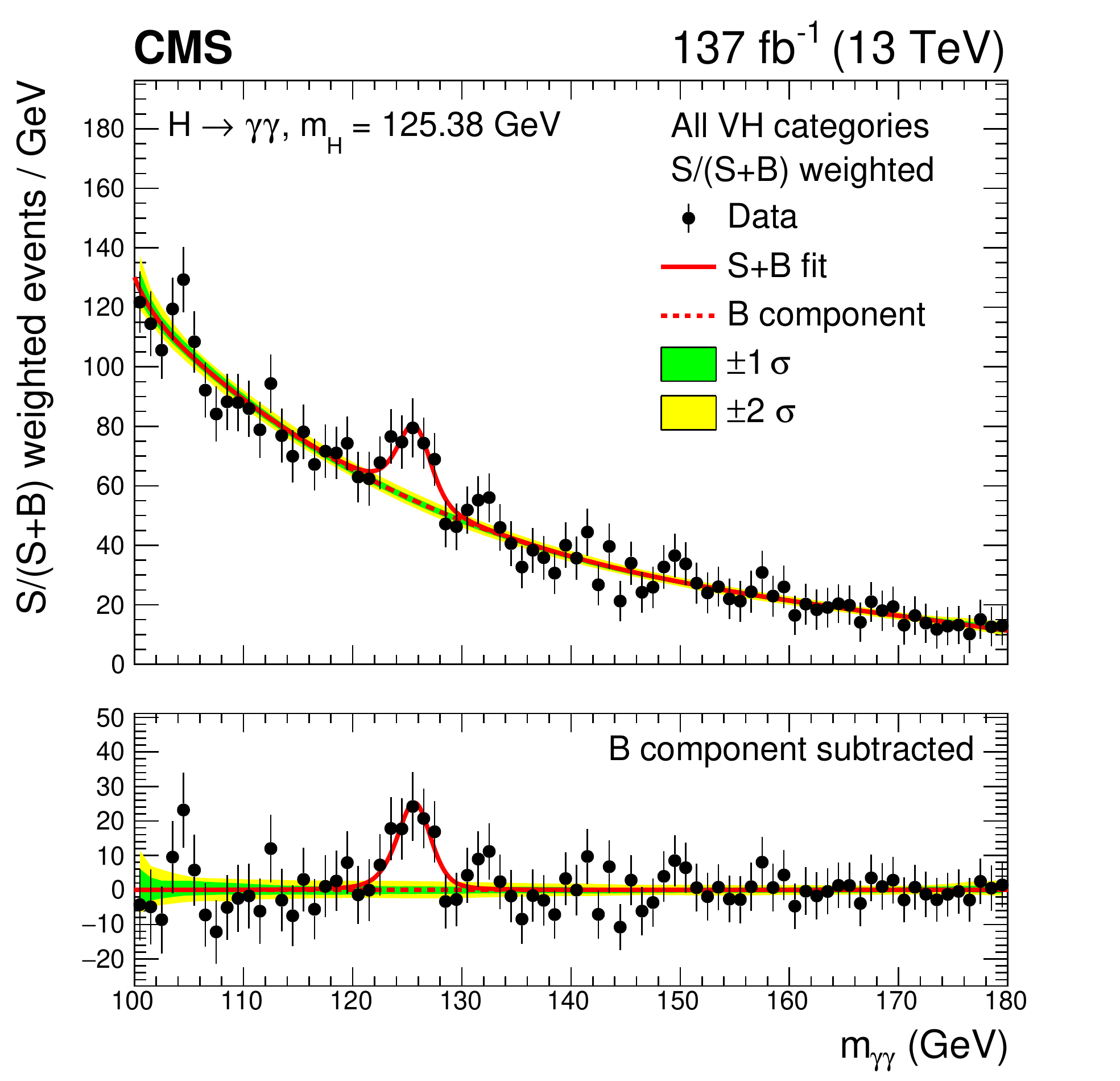}
  \includegraphics[width=0.49\textwidth]{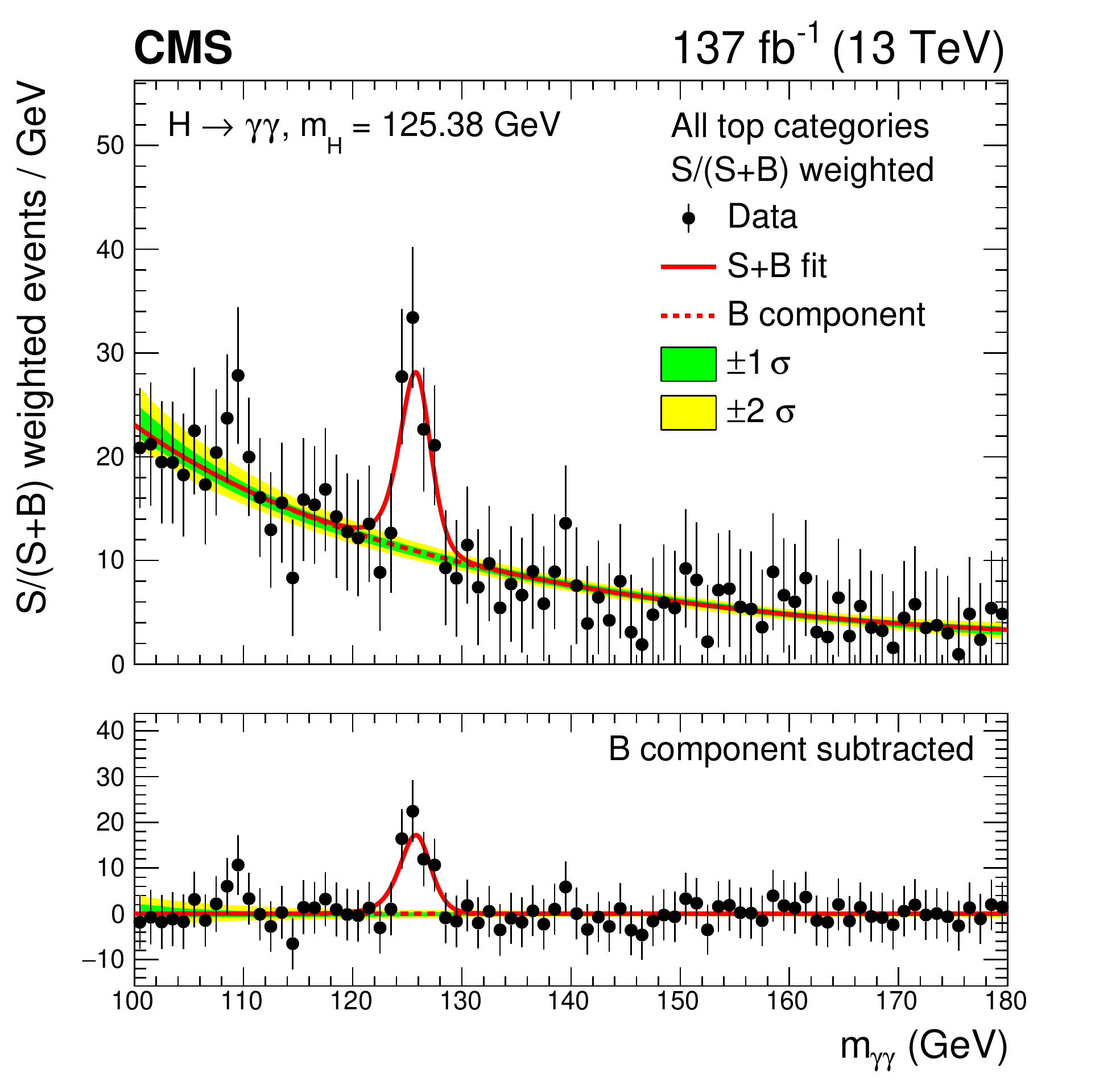}
  \caption{
    The best fit signal-plus-background model with data points (black) in the fit to signal strength modifiers of the four principal production modes.
    The model is shown separately for groups of analysis categories targeting the \ggH (upper left), VBF (upper right), \VH (lower left) and top quark associated (lower right) production modes.
    Here, the analysis categories in each group are summed after weighting by S/(S+B), 
    where S and B are the numbers of expected signal and background events in a $\pm 1 \seff$ \mgg window centred on \mH.
    The one standard deviation (green) and two standard deviation (yellow) bands show
    the uncertainties in the background component of the fit.
    The solid red line shows the total signal-plus-background contribution, whereas the dashed red line represents the background component only. 
    The lower panel in each plot shows the residuals after subtraction of this background component.}
  \label{fig:results_sPlusBplot_mu}
\end{figure}

The resulting signal and background \mgg distributions after the fit using this parameter scheme are shown in Fig.~\ref{fig:results_sPlusBplot_mu}. 
Analysis categories are divided into four groups, corresponding to those targeting the \ggH, VBF, \VH, and top quark production modes. 
In each group, the individual analysis categories are summed after weighting by S/(S+B).

\begin{figure}
  \centering
  \includegraphics[width=0.7\textwidth]{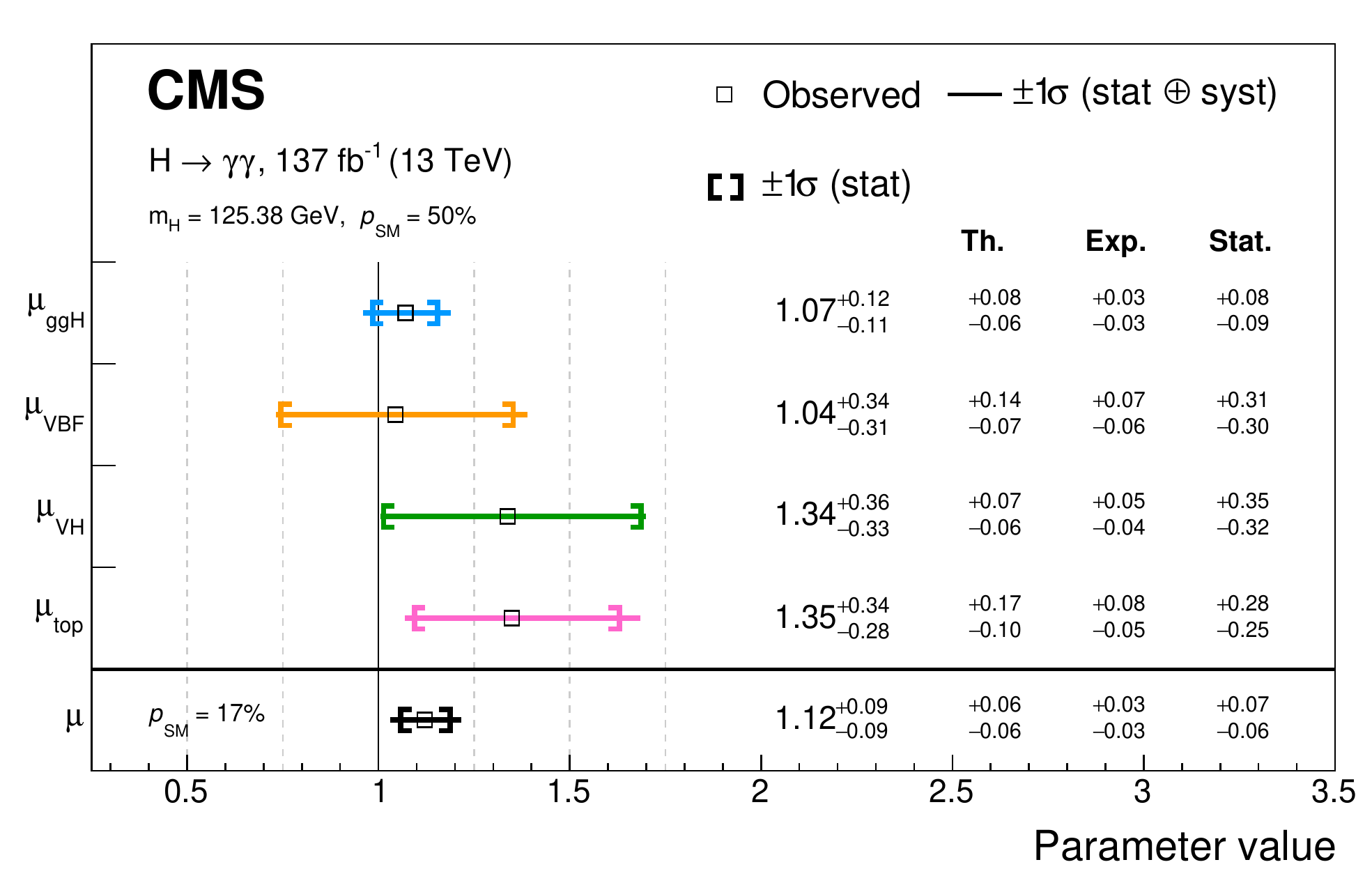}
  \caption{
    Observed results of the fit to signal strength modifiers of the four principal production modes. 
    The contributions to the total uncertainty in each parameter from the theoretical systematic, 
    experimental systematic, and statistical components are shown. 
    The colour scheme is chosen to match the diagram presented in Fig.~\ref{fig:allSTXSbins}.
    The compatibility of this fit with respect to the SM prediction, expressed as a $p$-value, is approximately 50\%.
    Also shown in black is the result of the fit to the inclusive signal strength modifier, 
    which has a $p$-value of 17\%.
   }
  \label{fig:results_perproc_mu}
\end{figure}

The values of the production mode signal strength modifiers and their uncertainties are displayed in Fig.~\ref{fig:results_perproc_mu}. 
The precision of these measurements is significantly improved from previous analyses
performed by the CMS Collaboration in the \Hgg decay channel.
In particular, the measurement of the $\mu_{\VH}$ signal strength modifier
has improved substantially from that shown in Ref.~\cite{HIG-16-040},
beyond what would be expected from the increase in the size of the data set alone.
The $p$-value of the production mode signal strength modifier fit with respect to the SM prediction is approximately 50\%.

\begin{figure}
  \centering
  \includegraphics[width=.99\textwidth]{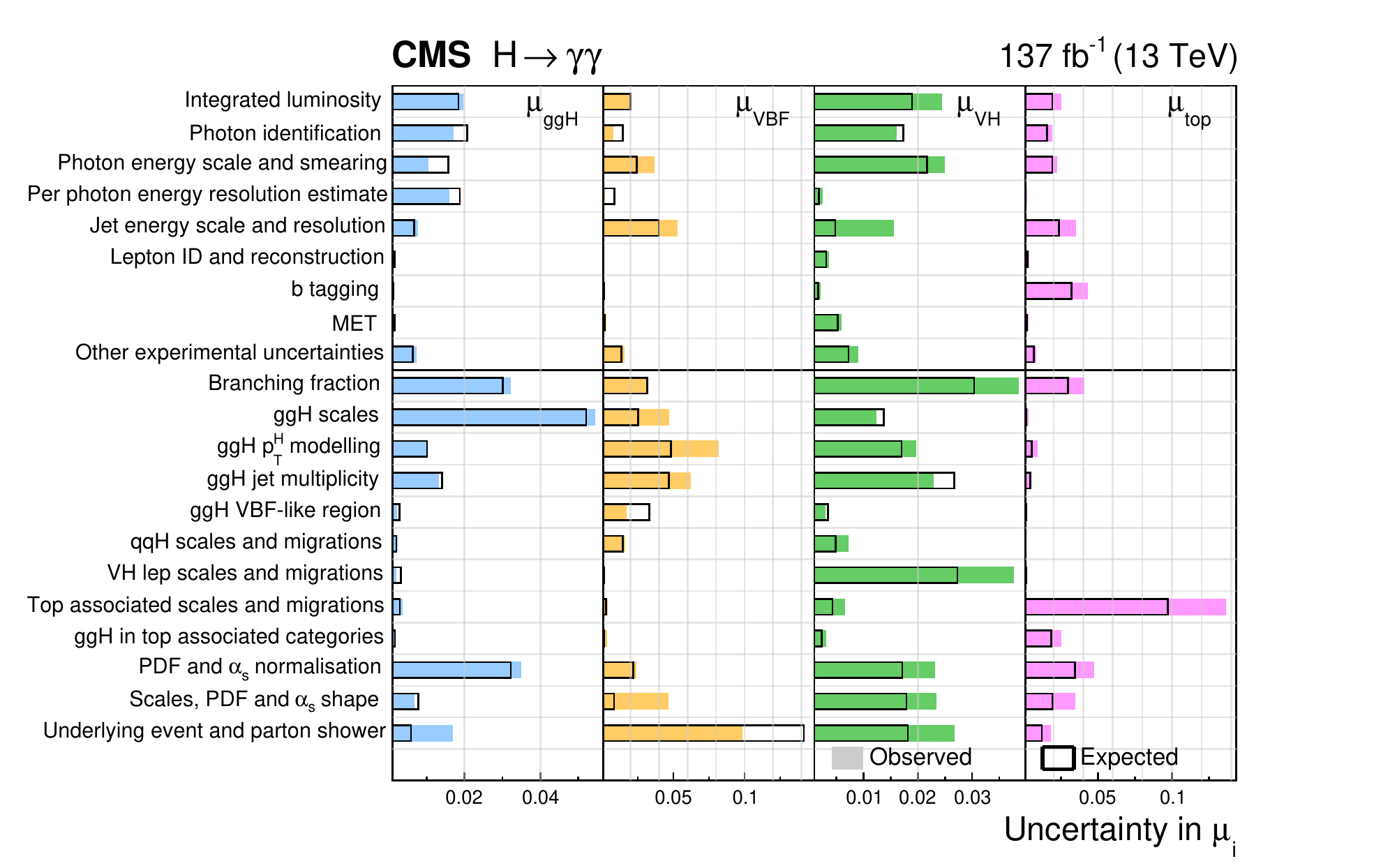}
  \caption{
    A summary of the impact of the main sources of systematic uncertainty 
    in the fit to signal strength modifiers of the four principal production modes. 
    The observed (expected) impacts are shown by the solid (empty) bars.
    The colour scheme is chosen to match the diagram presented in Fig.~\ref{fig:allSTXSbins}.
  }
  \label{fig:mu_impacts}
\end{figure}

The main sources of systematic uncertainty affecting the signal strength modifier 
in each production mode are summarised in Fig.~\ref{fig:mu_impacts}. 
The dominant contributions to the measurement uncertainty in the 
$\mu_{\ggH}$, $\mu_{\VH}$ and $\mu_{\text{top}}$ signal strength modifiers originate 
from the corresponding renormalisation and factorisation scale uncertainties, 
whereas the underlying event and parton shower uncertainties 
are the dominant sources of uncertainty in the $\mu_{\text{VBF}}$ measurement. 
The largest experimental uncertainties originate from the integrated luminosity, 
the photon identification, and the photon energy measurement 
for the $\mu_{\ggH}$ and $\mu_{\VH}$ signal strength modifiers. 
The uncertainties in the jet energy scale and resolution have a larger impact on 
$\mu_{\text{VBF}}$ and $\mu_{\text{top}}$, where $\mu_{\text{top}}$ 
has an additional large contribution from the uncertainty in the {\cPqb} tagging.

\subsection{Simplified template cross sections}
\label{sec:results_STXS}

This section details the fits performed to extract cross sections within the STXS framework and their respective 68\% confidence level (\CL) intervals. 
The theoretical uncertainties in the normalisation of the signal parameters are not included in the cross section measurements. 
In each fit, \ggZH events in which the $\PZ$ boson decays hadronically are grouped with the corresponding \ggH STXS bin. 
All \bbH events are treated as part of the \ggH 0J high \ptH bin.
The hadronic \VH processes are grouped with VBF production to form the \qqH parameters. 
Parameters which are not measured are constrained to their SM prediction, within theoretical uncertainties. 
These are the zero jet, one jet, $\mjj<60\GeV$, and $120<\mjj<350\GeV$ bins in the \qqH binning scheme.

Two different parameterisations are considered, with varying levels of granularity defined by the merging of certain STXS bins. 
It is necessary to merge bins to avoid either very large uncertainties in some parameters or very high correlations between parameters.
Merging fewer bins keeps the model-dependence of the results as low as possible, as no additional 
assumptions are made about the relative contributions of different STXS bins. 
The results with reduced model-dependence however have larger uncertainties in the measured cross section parameters.

In this paper, the results of two different fits to the cross sections of partially merged STXS bins are reported.
The first is referred to as the ``maximal" merging scenario, 
where in general STXS bins are merged until their expected uncertainty is less than 150\% of the SM prediction.
The second ``minimal" merging fit instead merges as few bins as possible whilst ensuring that 
parameters do not become too anti-correlated, meaning values of less than around 90\%.

The maximal merging scheme defines 17 parameters of interest. 
The VBF-like regions ($\geq$2-jets, $\mjj > 350\GeV$) in the \ggH and \qqH schemes are merged to define the \ggH VBF-like and \qqH VBF-like parameters, respectively. 
The four bins with $\ptH > 200\GeV$ in the \ggH scheme are merged into a single bin, labelled as \ggH BSM. 
Additionally, the \WH leptonic, \ZH leptonic and \ttH bins are all fully merged into single parameters. 
The \ZH leptonic parameter groups both \ZH and \ggZH production.

The minimal merging scheme defines a more granular fit with 27 parameters of interest. 
The \qqH VBF-like region is fully split into the four STXS bins defined by the boundaries at $\mjj=700\GeV$ and $\ptHjj=25\GeV$. 
To avoid large correlations between the fitted parameters, the four \ggH VBF-like bins are merged with the corresponding bins in the \qqH scheme. 
Additional splittings are included in the \ggH scheme at $\ptH=300$ and $450\GeV$, and the \WH leptonic scheme at $\ptV=75$ and $150\GeV$.  
Furthermore, the \ttH region is split into five parameters according to the boundaries at $\ptH=60$, 120, 200, and 300\GeV.

Table \ref{tab:param_merging_scenarios} summarises the maximal and minimal merging schemes by listing the STXS bins that contribute to each parameter of interest. 
The STXS bins that are constrained to their respective SM predictions in both fits are also listed.

\begin{table}
  \centering
  \caption{
    A summary of the maximal and minimal parameter merging scenarios. 
    The STXS bins that contribute to each parameter are listed. 
    Furthermore, the bins that are constrained to their respective SM predictions 
    in the fits are listed in the final row.
  }
  \label{tab:param_merging_scenarios}
  \cmsTable{
    \begin{tabular}{cll}
      Scheme & Parameters & STXS stage 1.2 bins (total number of bins) \\ \hline
      \multirow{17}{*}{\begin{tabular}{c}Maximal \\ (17 parameters)\end{tabular}} & \ggH 0J low $\ptH$ & \ggH 0J low $\ptH$ (1) \\ 
       & \ggH 0J high $\ptH$ & \ggH 0J high $\ptH$, \bbH (2) \\ [\cmsTabSkip]
       & \ggH 1J low $\ptH$ & \ggH 1J low $\ptH$ (1) \\ 
       & \ggH 1J med $\ptH$ & \ggH 1J med $\ptH$ (1) \\ 
       & \ggH 1J high $\ptH$ & \ggH 1J high $\ptH$ (1) \\ [\cmsTabSkip]
       & \ggH $\geq2$J low $\ptH$ & \ggH $\geq$2J low $\ptH$ (1)\\ 
       & \ggH $\geq2$J med $\ptH$ & \ggH $\geq$2J med $\ptH$ (1)\\ 
       & \ggH $\geq2$J high $\ptH$ & \ggH $\geq$2J high $\ptH$ (1)\\ [\cmsTabSkip]
       & \ggH BSM & \Bigg\{\! \begin{tabular}{@{}l}\ggH BSM $200<\ptH<300$, \ggH BSM $300<\ptH<450$\\ \ggH BSM $450<\ptH<650$, \ggH BSM $\ptH>650$\end{tabular} \!\!\Bigg\} (4) \\  [\cmsTabSkip]
      & \ggH VBF-like & $\Bigg\{\!$ \begin{tabular}{@{}l}\ggH VBF-like low $\mjj$ low $\ptHjj$, \ggH VBF-like low $\mjj$ high $\ptHjj$\\\ggH VBF-like high $\mjj$ low $\ptHjj$, \ggH VBF-like high $\mjj$ high $\ptHjj$\end{tabular} $\!\!\Bigg\}$ (4) \\ 
      & \qqH VBF-like & $\Bigg\{\!$ \begin{tabular}{@{}l}\qqH VBF-like low $\mjj$ low $\ptHjj$, \qqH VBF-like low $\mjj$ high $\ptHjj$\\qqH VBF-like high $\mjj$ low $\ptHjj$, \qqH VBF-like high $\mjj$ high $\ptHjj$\end{tabular} $\!\!\Bigg\}$  (4) \\  [\cmsTabSkip]
       & \qqH \VH-like & \qqH \VH-like (1)\\ 
       & \qqH BSM & \qqH BSM (1)\\  [\cmsTabSkip]
       & \WH lep & All \WH lep (5) \\ 
       & \ZH lep & All \ZH lep and \ggZH lep (10) \\  [\cmsTabSkip]
       & \ttH & All \ttH (5) \\ 
       & \tH & $\tH=\tHq+\tHW$ (1) \\
      \hline
      \multirow{27}{*}{\begin{tabular}{c}Minimal \\ (27 parameters)\end{tabular}} & \ggH 0J low $\ptH$ & \ggH 0J low $\ptH$ (1) \\ 
       & \ggH 0J high $\ptH$ & \ggH 0J high $\ptH$, \bbH (2) \\  [\cmsTabSkip]
       & \ggH 1J low $\ptH$ & \ggH 1J low $\ptH$ (1) \\ 
       & \ggH 1J med $\ptH$ & \ggH 1J med $\ptH$ (1) \\ 
       & \ggH 1J high $\ptH$ & \ggH 1J high $\ptH$ (1) \\  [\cmsTabSkip]
       & \ggH $\geq2$J low $\ptH$ & \ggH $\geq$2J low $\ptH$ (1)\\ 
       & \ggH $\geq2$J med $\ptH$ & \ggH $\geq$2J med $\ptH$ (1)\\ 
       & \ggH $\geq2$J high $\ptH$ & \ggH $\geq$2J high $\ptH$ (1)\\  [\cmsTabSkip]
       & \ggH BSM $200<\ptH<300$ & \ggH BSM $200<\ptH<~300$ (1)\\ 
       & \ggH BSM $300<\ptH<450$ & \ggH BSM $300<\ptH<~450$ (1)\\ 
       & \ggH BSM $\ptH>450$ & \ggH BSM $450<\ptH<650$, \ggH BSM $\ptH>650$ (2) \\  [\cmsTabSkip]
       & VBF-like low $\mjj$ low $\ptHjj$ & \ggH + \qqH VBF-like low $\mjj$ low $\ptHjj$ (2) \\ 
       & VBF-like low $\mjj$ high $\ptHjj$ & \ggH + \qqH VBF-like low $\mjj$ high $\ptHjj$ (2) \\ 
       & VBF-like high $\mjj$ low $\ptHjj$ & \ggH + \qqH VBF-like high $\mjj$ low $\ptHjj$ (2) \\ 
       & VBF-like high $\mjj$ high $\ptHjj$ & \ggH + \qqH VBF-like high $\mjj$ high $\ptHjj$ (2) \\  [\cmsTabSkip]
       & \qqH \VH-like & \qqH \VH-like (1)\\ 
       & \qqH BSM & \qqH BSM (1)\\  [\cmsTabSkip]
       & \WH lep $\ptV<75$ & \WH lep $\ptV<75$ (1)\\ 
       & \WH lep $75<\ptV<150$ & \WH lep $75<\ptV<150$ (1)\\ 
       & \WH lep $\ptV>150$ & \Bigg\{\! \begin{tabular}{@{}l}\WH lep 0J $150<\ptV<250$, \WH lep $\geq$1J $150<\ptV<250$ \\ \WH lep $\ptV>250$ \end{tabular} \!\!\Bigg\} (3) \\ 
       & \ZH lep & All \ZH lep and \ggZH lep (10) \\  [\cmsTabSkip]
       & \ttH $\ptH<60$ & \ttH $\ptH<60$ (1) \\ 
       & \ttH $60<\ptH<120$ & \ttH $60<\ptH<120$ (1) \\ 
       & \ttH $120<\ptH<200$ & \ttH $120<\ptH<200$ (1) \\ 
       & \ttH $200<\ptH<300$ & \ttH $200<\ptH<300$ (1) \\ 
       & \ttH $\ptH>300$ & \ttH $\ptH>300$ (1) \\  [\cmsTabSkip]
       & \tH & $\tH=\tHq+\tHW$ (1) \\
      \hline
      \multicolumn{2}{c}{\rule{0pt}{4ex} Constrained to SM prediction} & \qqH 0J, \qqH 1J, \qqH $\mjj<60$, \qqH $120<\mjj<350$ (4) \\
    \end{tabular}
  }
\end{table}

\begin{figure}[htb!]
  \centering
  \includegraphics[width=1\textwidth]{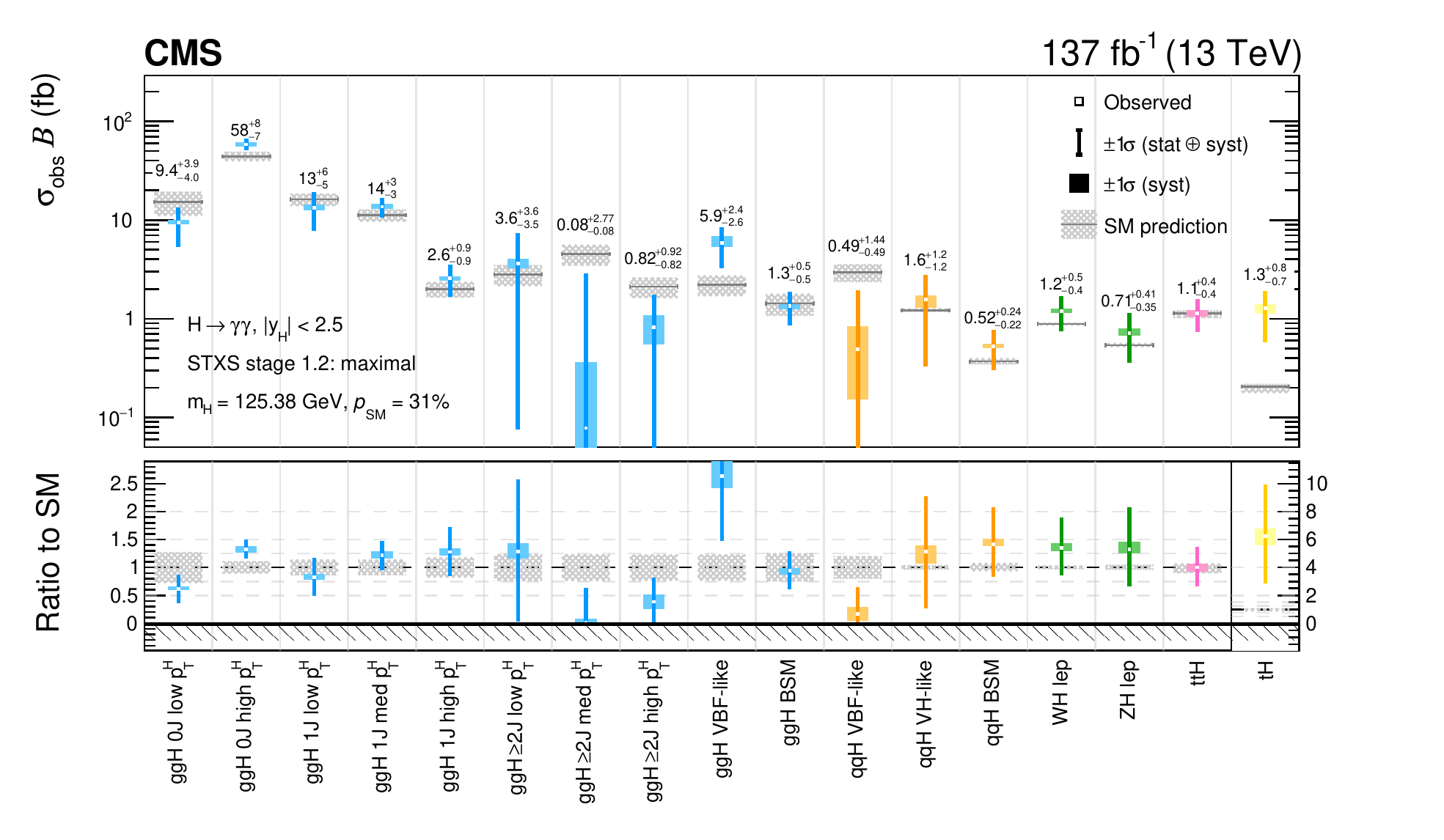}
  \caption{
    Observed results of the maximal merging scheme STXS fit. 
    The best fit cross sections are plotted together with the respective 68\% \CL intervals. 
    The systematic components of the uncertainty in each parameter are shown by the coloured boxes. 
    The hatched grey boxes demonstrate the theoretical uncertainties in the SM predictions. 
    The lower panel shows the ratio of the fitted values to the SM predictions. 
    Here the \tH cross section ratio has a different scale, due to its high best fit value and uncertainty.
    The cross sections are constrained to be non-negative, as indicated by the hashed pattern below zero. 
    The parameters whose best fit values are at zero are known to have 68\% \CL intervals which slightly under-cover; 
    this is checked to be a small effect using pseudo-experiments.
    The colour scheme is chosen to match the diagram presented in Fig.~\ref{fig:allSTXSbins}.
    The compatibility of this fit with respect to the SM prediction, expressed as a $p$-value, is approximately 31\%.
  }
  \label{fig:results_stage1p2_maximal_dist}
\end{figure}

\begin{figure}
  \centering
  \includegraphics[width=1\textwidth]{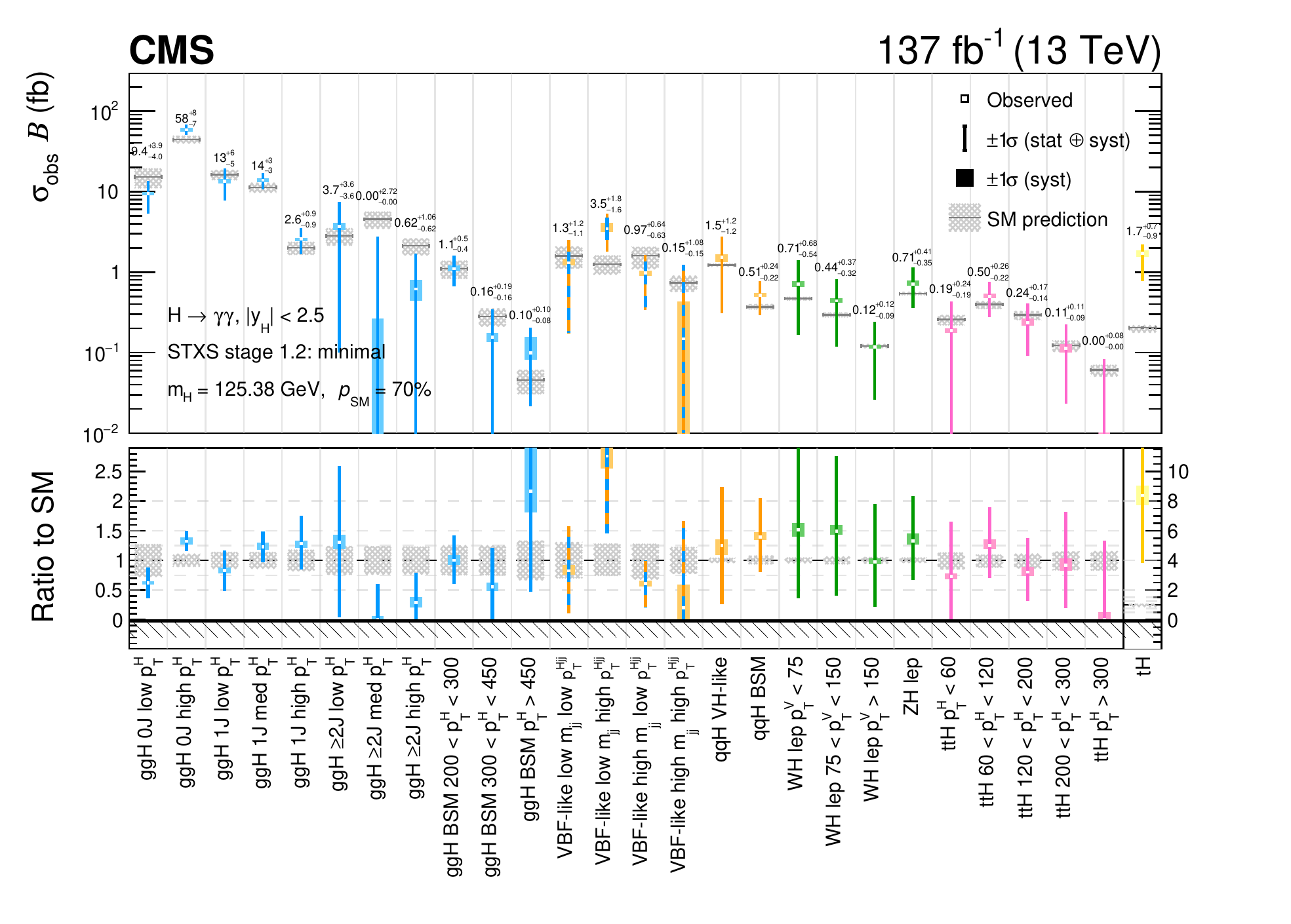}
  \caption{
    Observed results of the minimal merging scheme STXS fit. 
    The best fit cross sections are plotted together with the respective 68\% \CL intervals. 
    The systematic components of the uncertainty in each parameter are shown by the coloured boxes. 
    The hatched grey boxes demonstrate the theoretical uncertainties in the SM predictions. 
    The lower panel shows the ratio of the fitted values to the SM predictions. 
    Here the \tH cross section ratio has a different scale, due to its high best fit value and uncertainty.
    The cross sections are constrained to be non-negative, as indicated by the hashed pattern below zero. 
    The parameters whose best fit values are at zero are known to have 68\% \CL intervals which slightly under-cover; 
    this is checked to be a small effect using pseudo-experiments.
    The colour scheme is chosen to match the diagram presented in Fig.~\ref{fig:allSTXSbins}. 
    The orange lines dashed with blue for the VBF-like parameters represent contributions from both the \ggH and the \qqH STXS bins.
    The compatibility of this fit with respect to the SM prediction, expressed as a $p$-value, is approximately 70\%.
  }
  \label{fig:results_stage1p2_minimal_dist}
\end{figure}

The best fit cross sections and 68\% \CL intervals are shown for the two merging schemes in Figs.~\ref{fig:results_stage1p2_maximal_dist} and ~\ref{fig:results_stage1p2_minimal_dist}.
The corresponding numerical values are given in Tables~\ref{tab:results_stage1p2_maximal_summary}~and~\ref{tab:results_stage1p2_minimal_summary}.
For both the maximal and minimal fits, the statistical component of the uncertainty dominates for all measured cross sections.
Overall, the results from both merging scenario fits are in agreement with SM predictions; 
the $p$-values with respect to the SM predictions are 31 and 70\% for the maximal and minimal merging scenarios, respectively.

In the maximal merging scenario, \ggH production with $\ptH > 200$\GeV,
which is particularly sensitive to BSM physics entering the \ggH loop, 
is measured to a precision of less than 50\%, relative to the SM prediction.
The cross section is found to be consistent with the SM, 
with a measured value of $0.9_{-0.3}^{+0.4}$ relative to the SM prediction.
In addition, the product of the \tH production cross section times \Hgg branching fraction 
is measured to be $1.3_{-0.7}^{+0.8}\unit{fb}$, corresponding to an excess of $6.3_{-3.7}^{+3.4}$ times the SM expectation. 
Using the \CLs procedure~\cite{CLs}, a rate of \tH production of $14$ ($8$) times the SM expectation is observed (expected) to be excluded at the 95\% \CL.

The minimal merging scenario fit represents the current most granular cross section measurement performed in a single Higgs boson decay channel, 
showing reasonable sensitivity to many different regions of Higgs boson production phase space. 
In particular, the results contain the first measurements of \ttH production in bins of \ptH.
The size of the uncertainty in each of the four \ttH bins with $\ptH < 300$\GeV is less than 100\% of the SM prediction.
Additionally, \ggH production with $\ptH > 200$\GeV is measured in three separate regions.
The three corresponding cross sections are all measured to be within one standard deviation of the respective SM expectations.

\begin{figure}
  \centering
  \includegraphics[width=1\textwidth]{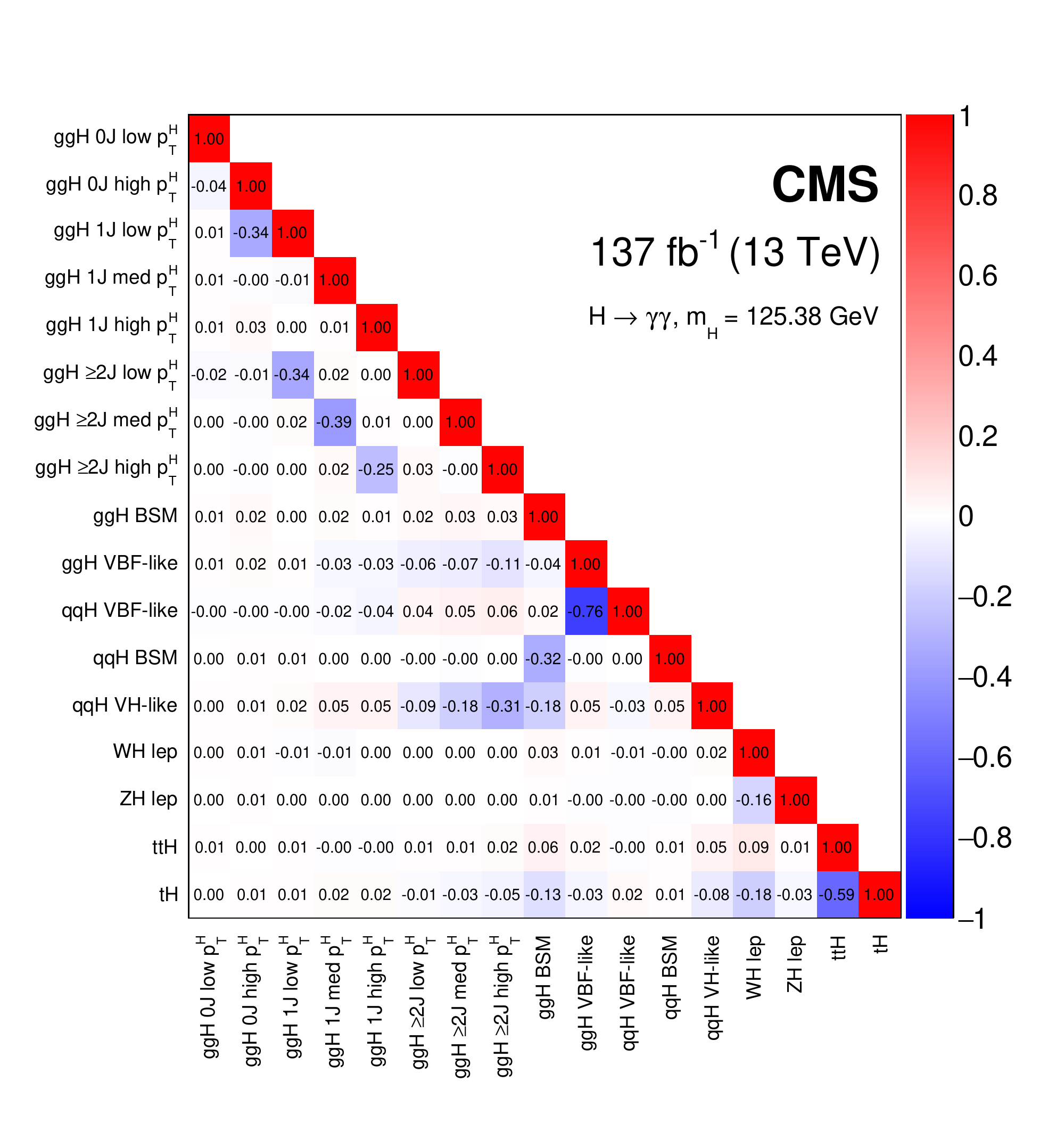}
  \caption{
    Observed correlations between the 17 parameters considered in the maximal merging STXS fit. 
    The size of the correlations is indicated by the colour scale.
  }
  \label{fig:results_stage1p2_maximal_correlations}
\end{figure}

\begin{figure}
  \centering
  \includegraphics[width=1\textwidth]{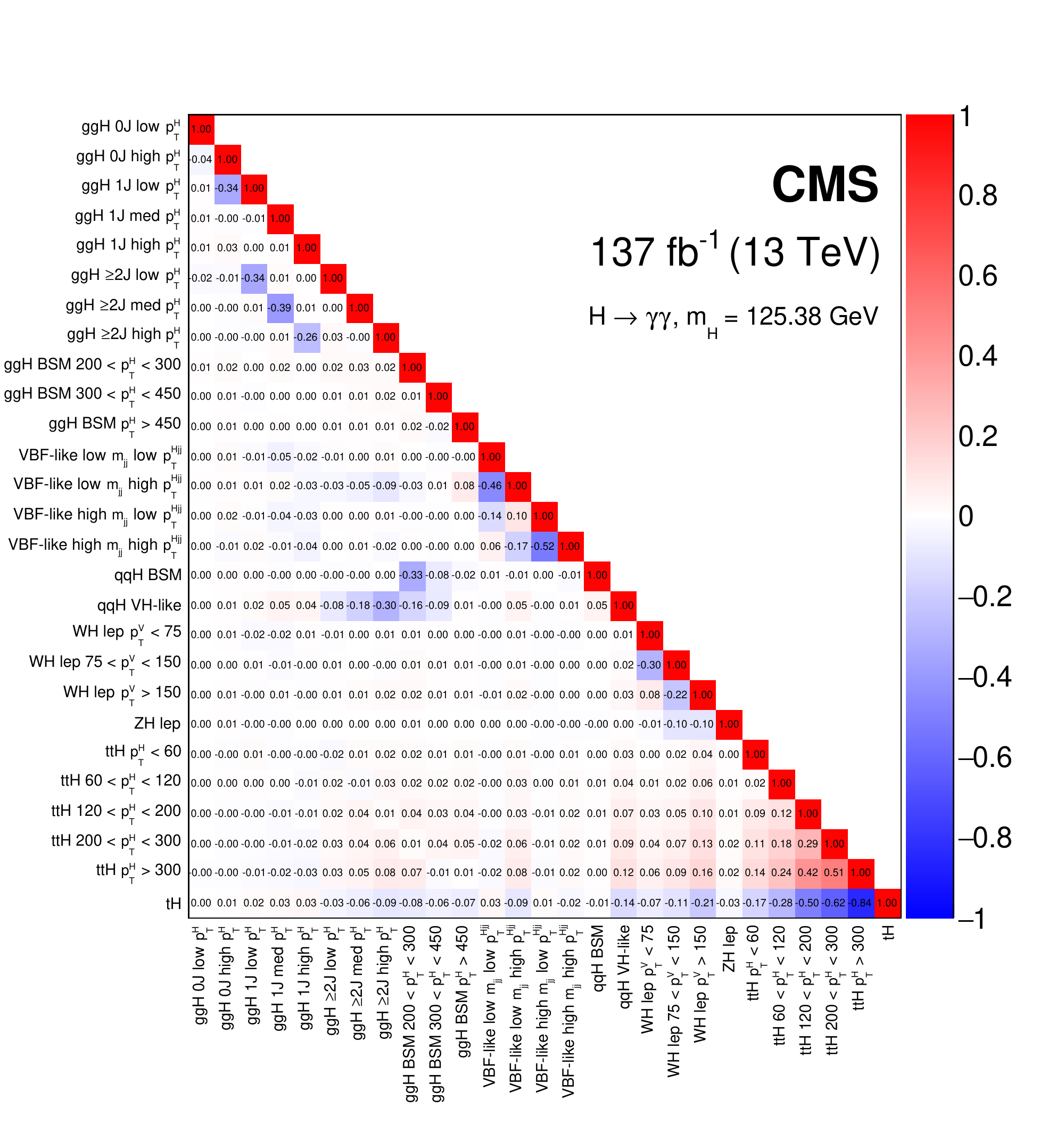}
  \caption{
    Observed correlations between the 27 parameters considered in the minimal merging STXS fit. 
    The size of the correlations is indicated by the colour scale.
  }
  \label{fig:results_stage1p2_minimal_correlations}
\end{figure}

Correlations between the fitted parameters are presented in Figs.~\ref{fig:results_stage1p2_maximal_correlations} and ~\ref{fig:results_stage1p2_minimal_correlations}. 
The correlations for the \ggH parameters are observed to be small between adjacent \ptH bins and larger between adjacent number of jet bins. 
This results from the fact that \ptgg is a well-measured quantity, whereas reconstructing the number of jets 
in an event is more difficult.
Nevertheless, the application of the \ggH BDT in the event categorisation helps to minimise these correlations.
The largest correlations in the maximal merging scheme exist between the \ggH VBF-like and \qqH VBF-like cross sections and the \ttH and \tH cross sections, with values of $-0.76$ and $-0.59$, respectively. 
These result from the sizeable contamination of \ggH VBF-like events in the \qqH analysis categories,
and the contamination of \ttH events in the \tHq leptonic category.
The act of splitting \ttH production into five separate parameters in the minimal merging scenario 
introduces larger correlations into the measurement.

\begin{table}
    \centering
    \topcaption{
    Results of the maximal merging scheme STXS fit. 
    The best fit cross sections are shown together with the respective 68\% \CL intervals. 
    The uncertainty is decomposed into the systematic and statistical components. 
    The expected uncertainties on the fitted parameters, computed assuming the SM predicted cross section values, are given in brackets.
    Also listed are the SM predictions for the cross sections 
    and the theoretical uncertainty in those predictions.}
    \label{tab:results_stage1p2_maximal_summary}
    \cmsTable{
      \begin{tabular}{cccccc}
        \multirow{3}{*}{Parameters} & \multicolumn{4}{c}{$\sigma\mathcal{B}$~(fb)} & $\sigma\mathcal{B}$/$(\sigma\mathcal{B})_{\text{SM}}$ \\ 
        & SM prediction & \multicolumn{3}{c}{Observed (Expected)} & Observed (Expected) \\ 
        & ($\mH=125.38\GeV$) & Best fit & Stat. unc. & Syst. unc. & Best fit \\ \hline 
        \ggH 0J low $\ptH$ & \begin{tabular}{r@{}l}$15.21$ & {}$^{+4.14}_{-4.18}$\end{tabular} & \begin{tabular}{r@{}l@{}l}$9.41$ & {}$^{+3.92}_{-3.99}$ & $\Big($$^{+4.20}_{-4.06}$$\Big)$ \end{tabular} & \begin{tabular}{@{}l@{}l}{}$^{+3.90}_{-3.98}$ & $\Big($$^{+4.16}_{-4.05}$$\Big)$ \end{tabular} & \begin{tabular}{@{}l@{}l}{}$^{+0.44}_{-0.25}$ & $\Big($$^{+0.51}_{-0.33}$$\Big)$ \end{tabular} & \begin{tabular}{r@{}l@{}l}$0.62$ & {}$^{+0.26}_{-0.26}$ & $\Big($$^{+0.28}_{-0.27}$$\Big)$ \end{tabular} \\ 
        \ggH 0J high $\ptH$ & \begin{tabular}{r@{}l}$44.25$ & {}$^{+4.84}_{-4.61}$\end{tabular} & \begin{tabular}{r@{}l@{}l}$58.50$ & {}$^{+8.10}_{-7.17}$ & $\Big($$^{+7.87}_{-7.77}$$\Big)$ \end{tabular} & \begin{tabular}{@{}l@{}l}{}$^{+7.70}_{-6.91}$ & $\Big($$^{+7.67}_{-7.63}$$\Big)$ \end{tabular} & \begin{tabular}{@{}l@{}l}{}$^{+2.50}_{-1.92}$ & $\Big($$^{+1.78}_{-1.42}$$\Big)$ \end{tabular} & \begin{tabular}{r@{}l@{}l}$1.32$ & {}$^{+0.18}_{-0.16}$ & $\Big($$^{+0.18}_{-0.18}$$\Big)$ \end{tabular} \\ 
        \ggH 1J low $\ptH$ & \begin{tabular}{r@{}l}$16.20$ & {}$^{+2.25}_{-2.27}$\end{tabular} & \begin{tabular}{r@{}l@{}l}$13.39$ & {}$^{+5.58}_{-5.49}$ & $\Big($$^{+5.67}_{-5.59}$$\Big)$ \end{tabular} & \begin{tabular}{@{}l@{}l}{}$^{+5.52}_{-5.45}$ & $\Big($$^{+5.61}_{-5.56}$$\Big)$ \end{tabular} & \begin{tabular}{@{}l@{}l}{}$^{+0.80}_{-0.63}$ & $\Big($$^{+0.77}_{-0.48}$$\Big)$ \end{tabular} & \begin{tabular}{r@{}l@{}l}$0.83$ & {}$^{+0.34}_{-0.34}$ & $\Big($$^{+0.35}_{-0.34}$$\Big)$ \end{tabular} \\ 
        \ggH 1J med $\ptH$ & \begin{tabular}{r@{}l}$11.23$ & {}$^{+1.56}_{-1.55}$\end{tabular} & \begin{tabular}{r@{}l@{}l}$13.66$ & {}$^{+2.91}_{-2.96}$ & $\Big($$^{+3.15}_{-3.39}$$\Big)$ \end{tabular} & \begin{tabular}{@{}l@{}l}{}$^{+2.83}_{-2.92}$ & $\Big($$^{+3.09}_{-3.36}$$\Big)$ \end{tabular} & \begin{tabular}{@{}l@{}l}{}$^{+0.70}_{-0.50}$ & $\Big($$^{+0.59}_{-0.45}$$\Big)$ \end{tabular} & \begin{tabular}{r@{}l@{}l}$1.22$ & {}$^{+0.26}_{-0.26}$ & $\Big($$^{+0.28}_{-0.30}$$\Big)$ \end{tabular} \\ 
        \ggH 1J high $\ptH$ & \begin{tabular}{r@{}l}$2.00$ & {}$^{+0.36}_{-0.36}$\end{tabular} & \begin{tabular}{r@{}l@{}l}$2.56$ & {}$^{+0.90}_{-0.87}$ & $\Big($$^{+0.91}_{-0.92}$$\Big)$ \end{tabular} & \begin{tabular}{@{}l@{}l}{}$^{+0.90}_{-0.87}$ & $\Big($$^{+0.90}_{-0.90}$$\Big)$ \end{tabular} & \begin{tabular}{@{}l@{}l}{}$^{+0.11}_{-0.11}$ & $\Big($$^{+0.15}_{-0.19}$$\Big)$ \end{tabular} & \begin{tabular}{r@{}l@{}l}$1.28$ & {}$^{+0.45}_{-0.44}$ & $\Big($$^{+0.46}_{-0.46}$$\Big)$ \end{tabular} \\ 
        \ggH $\geq$2J low $\ptH$ & \begin{tabular}{r@{}l}$2.82$ & {}$^{+0.68}_{-0.68}$\end{tabular} & \begin{tabular}{r@{}l@{}l}$3.62$ & {}$^{+3.65}_{-3.55}$ & $\Big($$^{+3.73}_{-2.82}$$\Big)$ \end{tabular} & \begin{tabular}{@{}l@{}l}{}$^{+3.62}_{-3.53}$ & $\Big($$^{+3.69}_{-2.82}$$\Big)$ \end{tabular} & \begin{tabular}{@{}l@{}l}{}$^{+0.41}_{-0.31}$ & $\Big($$^{+0.55}_{-0.55}$$\Big)$ \end{tabular} & \begin{tabular}{r@{}l@{}l}$1.29$ & {}$^{+1.29}_{-1.26}$ & $\Big($$^{+1.32}_{-1.00}$$\Big)$ \end{tabular} \\ 
        \ggH $\geq$2J med $\ptH$ & \begin{tabular}{r@{}l}$4.53$ & {}$^{+1.07}_{-1.07}$\end{tabular} & \begin{tabular}{r@{}l@{}l}$0.08$ & {}$^{+2.77}_{-0.08}$ & $\Big($$^{+2.87}_{-2.82}$$\Big)$ \end{tabular} & \begin{tabular}{@{}l@{}l}{}$^{+2.76}_{-0.08}$ & $\Big($$^{+2.84}_{-2.82}$$\Big)$ \end{tabular} & \begin{tabular}{@{}l@{}l}{}$^{+0.28}_{-0.08}$ & $\Big($$^{+0.38}_{-0.14}$$\Big)$ \end{tabular} & \begin{tabular}{r@{}l@{}l}$0.02$ & {}$^{+0.61}_{-0.02}$ & $\Big($$^{+0.63}_{-0.62}$$\Big)$ \end{tabular} \\ 
        \ggH $\geq$2J high $\ptH$ & \begin{tabular}{r@{}l}$2.12$ & {}$^{+0.49}_{-0.50}$\end{tabular} & \begin{tabular}{r@{}l@{}l}$0.82$ & {}$^{+0.92}_{-0.82}$ & $\Big($$^{+1.15}_{-1.10}$$\Big)$ \end{tabular} & \begin{tabular}{@{}l@{}l}{}$^{+0.88}_{-0.82}$ & $\Big($$^{+1.11}_{-1.09}$$\Big)$ \end{tabular} & \begin{tabular}{@{}l@{}l}{}$^{+0.26}_{-0.26}$ & $\Big($$^{+0.31}_{-0.14}$$\Big)$ \end{tabular} & \begin{tabular}{r@{}l@{}l}$0.39$ & {}$^{+0.43}_{-0.39}$ & $\Big($$^{+0.54}_{-0.52}$$\Big)$ \end{tabular} \\ 
        \ggH VBF-like & \begin{tabular}{r@{}l}$2.22$ & {}$^{+0.52}_{-0.52}$\end{tabular} & \begin{tabular}{r@{}l@{}l}$5.86$ & {}$^{+2.45}_{-2.59}$ & $\Big($$^{+2.90}_{-2.22}$$\Big)$ \end{tabular} & \begin{tabular}{@{}l@{}l}{}$^{+2.27}_{-2.55}$ & $\Big($$^{+2.81}_{-2.22}$$\Big)$ \end{tabular} & \begin{tabular}{@{}l@{}l}{}$^{+0.92}_{-0.48}$ & $\Big($$^{+0.71}_{-0.71}$$\Big)$ \end{tabular} & \begin{tabular}{r@{}l@{}l}$2.64$ & {}$^{+1.10}_{-1.17}$ & $\Big($$^{+1.31}_{-1.00}$$\Big)$ \end{tabular} \\ 
        \ggH BSM & \begin{tabular}{r@{}l}$1.43$ & {}$^{+0.36}_{-0.35}$\end{tabular} & \begin{tabular}{r@{}l@{}l}$1.34$ & {}$^{+0.50}_{-0.47}$ & $\Big($$^{+0.59}_{-0.49}$$\Big)$ \end{tabular} & \begin{tabular}{@{}l@{}l}{}$^{+0.49}_{-0.46}$ & $\Big($$^{+0.58}_{-0.49}$$\Big)$ \end{tabular} & \begin{tabular}{@{}l@{}l}{}$^{+0.05}_{-0.09}$ & $\Big($$^{+0.09}_{-0.05}$$\Big)$ \end{tabular} & \begin{tabular}{r@{}l@{}l}$0.94$ & {}$^{+0.35}_{-0.33}$ & $\Big($$^{+0.41}_{-0.35}$$\Big)$ \end{tabular} \\ 
        \qqH VBF-like & \begin{tabular}{r@{}l}$2.96$ & {}$^{+0.59}_{-0.59}$\end{tabular} & \begin{tabular}{r@{}l@{}l}$0.49$ & {}$^{+1.44}_{-0.49}$ & $\Big($$^{+1.49}_{-1.53}$$\Big)$ \end{tabular} & \begin{tabular}{@{}l@{}l}{}$^{+1.40}_{-0.49}$ & $\Big($$^{+1.47}_{-1.47}$$\Big)$ \end{tabular} & \begin{tabular}{@{}l@{}l}{}$^{+0.34}_{-0.34}$ & $\Big($$^{+0.25}_{-0.43}$$\Big)$ \end{tabular} & \begin{tabular}{r@{}l@{}l}$0.17$ & {}$^{+0.49}_{-0.17}$ & $\Big($$^{+0.50}_{-0.52}$$\Big)$ \end{tabular} \\ 
        \qqH \VH-like & \begin{tabular}{r@{}l}$1.22$ & {}$^{+0.05}_{-0.04}$\end{tabular} & \begin{tabular}{r@{}l@{}l}$1.57$ & {}$^{+1.20}_{-1.24}$ & $\Big($$^{+1.15}_{-1.23}$$\Big)$ \end{tabular} & \begin{tabular}{@{}l@{}l}{}$^{+1.19}_{-1.21}$ & $\Big($$^{+1.15}_{-1.23}$$\Big)$ \end{tabular} & \begin{tabular}{@{}l@{}l}{}$^{+0.13}_{-0.26}$ & $\Big($$^{+0.07}_{-0.04}$$\Big)$ \end{tabular} & \begin{tabular}{r@{}l@{}l}$1.29$ & {}$^{+0.98}_{-1.01}$ & $\Big($$^{+0.94}_{-1.01}$$\Big)$ \end{tabular} \\ 
        \qqH BSM & \begin{tabular}{r@{}l}$0.37$ & {}$^{+0.03}_{-0.02}$\end{tabular} & \begin{tabular}{r@{}l@{}l}$0.52$ & {}$^{+0.24}_{-0.22}$ & $\Big($$^{+0.26}_{-0.23}$$\Big)$ \end{tabular} & \begin{tabular}{@{}l@{}l}{}$^{+0.24}_{-0.22}$ & $\Big($$^{+0.25}_{-0.23}$$\Big)$ \end{tabular} & \begin{tabular}{@{}l@{}l}{}$^{+0.03}_{-0.01}$ & $\Big($$^{+0.03}_{-0.01}$$\Big)$ \end{tabular} & \begin{tabular}{r@{}l@{}l}$1.42$ & {}$^{+0.65}_{-0.59}$ & $\Big($$^{+0.69}_{-0.62}$$\Big)$ \end{tabular} \\ 
        \WH lep & \begin{tabular}{r@{}l}$0.88$ & {}$^{+0.03}_{-0.03}$\end{tabular} & \begin{tabular}{r@{}l@{}l}$1.19$ & {}$^{+0.49}_{-0.44}$ & $\Big($$^{+0.51}_{-0.42}$$\Big)$ \end{tabular} & \begin{tabular}{@{}l@{}l}{}$^{+0.48}_{-0.43}$ & $\Big($$^{+0.50}_{-0.41}$$\Big)$ \end{tabular} & \begin{tabular}{@{}l@{}l}{}$^{+0.07}_{-0.04}$ & $\Big($$^{+0.05}_{-0.05}$$\Big)$ \end{tabular} & \begin{tabular}{r@{}l@{}l}$1.35$ & {}$^{+0.55}_{-0.49}$ & $\Big($$^{+0.57}_{-0.47}$$\Big)$ \end{tabular} \\ 
        \ZH lep & \begin{tabular}{r@{}l}$0.54$ & {}$^{+0.03}_{-0.02}$\end{tabular} & \begin{tabular}{r@{}l@{}l}$0.71$ & {}$^{+0.41}_{-0.35}$ & $\Big($$^{+0.42}_{-0.35}$$\Big)$ \end{tabular} & \begin{tabular}{@{}l@{}l}{}$^{+0.40}_{-0.35}$ & $\Big($$^{+0.41}_{-0.35}$$\Big)$ \end{tabular} & \begin{tabular}{@{}l@{}l}{}$^{+0.07}_{-0.03}$ & $\Big($$^{+0.06}_{-0.03}$$\Big)$ \end{tabular} & \begin{tabular}{r@{}l@{}l}$1.32$ & {}$^{+0.76}_{-0.65}$ & $\Big($$^{+0.78}_{-0.65}$$\Big)$ \end{tabular} \\ 
        \ttH & \begin{tabular}{r@{}l}$1.13$ & {}$^{+0.08}_{-0.11}$\end{tabular} & \begin{tabular}{r@{}l@{}l}$1.13$ & {}$^{+0.42}_{-0.39}$ & $\Big($$^{+0.42}_{-0.41}$$\Big)$ \end{tabular} & \begin{tabular}{@{}l@{}l}{}$^{+0.42}_{-0.38}$ & $\Big($$^{+0.41}_{-0.40}$$\Big)$ \end{tabular} & \begin{tabular}{@{}l@{}l}{}$^{+0.07}_{-0.07}$ & $\Big($$^{+0.09}_{-0.05}$$\Big)$ \end{tabular} & \begin{tabular}{r@{}l@{}l}$1.00$ & {}$^{+0.37}_{-0.35}$ & $\Big($$^{+0.37}_{-0.36}$$\Big)$ \end{tabular} \\ 
        \tH & \begin{tabular}{r@{}l}$0.20$ & {}$^{+0.01}_{-0.03}$\end{tabular} & \begin{tabular}{r@{}l@{}l}$1.27$ & {}$^{+0.76}_{-0.69}$ & $\Big($$^{+0.76}_{-0.20}$$\Big)$ \end{tabular} & \begin{tabular}{@{}l@{}l}{}$^{+0.75}_{-0.68}$ & $\Big($$^{+0.76}_{-0.20}$$\Big)$ \end{tabular} & \begin{tabular}{@{}l@{}l}{}$^{+0.10}_{-0.13}$ & $\Big($$^{+0.08}_{-0.08}$$\Big)$ \end{tabular} & \begin{tabular}{r@{}l@{}l}$6.24$ & {}$^{+3.72}_{-3.37}$ & $\Big($$^{+3.73}_{-1.00}$$\Big)$ \end{tabular} \\ 
      \end{tabular}
    }
\end{table}

\begin{table}
    \centering
    \topcaption{
    Results of the minimal merging scheme STXS fit. 
    The best fit cross sections are shown together with the respective 68\% \CL intervals. 
    The uncertainty is decomposed into the systematic and statistical components. 
    The expected uncertainties on the fitted parameters, computed assuming the SM predicted cross section values, are given in brackets.
    Also listed are the SM predictions for the cross sections 
    and the theoretical uncertainty in those predictions.}
    \label{tab:results_stage1p2_minimal_summary}
    \cmsTable{
      \begin{tabular}{cccccc}
        \multirow{3}{*}{Parameters} & \multicolumn{4}{c}{$\sigma\mathcal{B}$~(fb)} & $\sigma\mathcal{B}$/$(\sigma\mathcal{B})_{\text{SM}}$ \\ 
        & SM prediction & \multicolumn{3}{c}{Observed (Expected)} & Observed (Expected) \\ 
        & ($\mH=125.38\GeV$) & Best fit & Stat. unc. & Syst. unc. & Best fit \\ \hline 
        \ggH 0J low $\ptH$ & \begin{tabular}{r@{}l}$15.21$ & {}$^{+4.14}_{-4.18}$\end{tabular} & \begin{tabular}{r@{}l@{}l}$9.41$ & {}$^{+3.91}_{-4.00}$ & $\Big($$^{+4.19}_{-4.06}$$\Big)$ \end{tabular} & \begin{tabular}{@{}l@{}l}{}$^{+3.90}_{-3.99}$ & $\Big($$^{+4.16}_{-4.05}$$\Big)$ \end{tabular} & \begin{tabular}{@{}l@{}l}{}$^{+0.37}_{-0.30}$ & $\Big($$^{+0.50}_{-0.36}$$\Big)$ \end{tabular} & \begin{tabular}{r@{}l@{}l}$0.62$ & {}$^{+0.26}_{-0.26}$ & $\Big($$^{+0.28}_{-0.27}$$\Big)$ \end{tabular} \\ 
        \ggH 0J high $\ptH$ & \begin{tabular}{r@{}l}$44.25$ & {}$^{+4.84}_{-4.61}$\end{tabular} & \begin{tabular}{r@{}l@{}l}$58.46$ & {}$^{+8.12}_{-7.17}$ & $\Big($$^{+7.87}_{-7.78}$$\Big)$ \end{tabular} & \begin{tabular}{@{}l@{}l}{}$^{+7.69}_{-6.91}$ & $\Big($$^{+7.66}_{-7.63}$$\Big)$ \end{tabular} & \begin{tabular}{@{}l@{}l}{}$^{+2.60}_{-1.94}$ & $\Big($$^{+1.78}_{-1.50}$$\Big)$ \end{tabular} & \begin{tabular}{r@{}l@{}l}$1.32$ & {}$^{+0.18}_{-0.16}$ & $\Big($$^{+0.18}_{-0.18}$$\Big)$ \end{tabular} \\ 
        \ggH 1J low $\ptH$ & \begin{tabular}{r@{}l}$16.20$ & {}$^{+2.25}_{-2.27}$\end{tabular} & \begin{tabular}{r@{}l@{}l}$13.40$ & {}$^{+5.59}_{-5.50}$ & $\Big($$^{+5.70}_{-5.58}$$\Big)$ \end{tabular} & \begin{tabular}{@{}l@{}l}{}$^{+5.53}_{-5.46}$ & $\Big($$^{+5.64}_{-5.55}$$\Big)$ \end{tabular} & \begin{tabular}{@{}l@{}l}{}$^{+0.79}_{-0.67}$ & $\Big($$^{+0.77}_{-0.56}$$\Big)$ \end{tabular} & \begin{tabular}{r@{}l@{}l}$0.83$ & {}$^{+0.34}_{-0.34}$ & $\Big($$^{+0.35}_{-0.34}$$\Big)$ \end{tabular} \\ 
        \ggH 1J med $\ptH$ & \begin{tabular}{r@{}l}$11.23$ & {}$^{+1.56}_{-1.55}$\end{tabular} & \begin{tabular}{r@{}l@{}l}$13.80$ & {}$^{+2.90}_{-2.94}$ & $\Big($$^{+3.14}_{-3.41}$$\Big)$ \end{tabular} & \begin{tabular}{@{}l@{}l}{}$^{+2.82}_{-2.90}$ & $\Big($$^{+3.08}_{-3.37}$$\Big)$ \end{tabular} & \begin{tabular}{@{}l@{}l}{}$^{+0.68}_{-0.51}$ & $\Big($$^{+0.59}_{-0.50}$$\Big)$ \end{tabular} & \begin{tabular}{r@{}l@{}l}$1.23$ & {}$^{+0.26}_{-0.26}$ & $\Big($$^{+0.28}_{-0.30}$$\Big)$ \end{tabular} \\ 
        \ggH 1J high $\ptH$ & \begin{tabular}{r@{}l}$2.00$ & {}$^{+0.36}_{-0.36}$\end{tabular} & \begin{tabular}{r@{}l@{}l}$2.57$ & {}$^{+0.94}_{-0.88}$ & $\Big($$^{+0.92}_{-0.90}$$\Big)$ \end{tabular} & \begin{tabular}{@{}l@{}l}{}$^{+0.94}_{-0.87}$ & $\Big($$^{+0.91}_{-0.88}$$\Big)$ \end{tabular} & \begin{tabular}{@{}l@{}l}{}$^{+0.08}_{-0.12}$ & $\Big($$^{+0.13}_{-0.16}$$\Big)$ \end{tabular} & \begin{tabular}{r@{}l@{}l}$1.28$ & {}$^{+0.47}_{-0.44}$ & $\Big($$^{+0.46}_{-0.45}$$\Big)$ \end{tabular} \\ 
        \ggH $\geq2$J low $\ptH$ & \begin{tabular}{r@{}l}$2.82$ & {}$^{+0.68}_{-0.68}$\end{tabular} & \begin{tabular}{r@{}l@{}l}$3.67$ & {}$^{+3.63}_{-3.57}$ & $\Big($$^{+3.74}_{-2.82}$$\Big)$ \end{tabular} & \begin{tabular}{@{}l@{}l}{}$^{+3.62}_{-3.56}$ & $\Big($$^{+3.71}_{-2.82}$$\Big)$ \end{tabular} & \begin{tabular}{@{}l@{}l}{}$^{+0.34}_{-0.30}$ & $\Big($$^{+0.49}_{-0.49}$$\Big)$ \end{tabular} & \begin{tabular}{r@{}l@{}l}$1.30$ & {}$^{+1.29}_{-1.27}$ & $\Big($$^{+1.33}_{-1.00}$$\Big)$ \end{tabular} \\ 
        \ggH $\geq2$J med $\ptH$ & \begin{tabular}{r@{}l}$4.53$ & {}$^{+1.07}_{-1.07}$\end{tabular} & \begin{tabular}{r@{}l@{}l}$0.00$ & {}$^{+2.72}_{-0.00}$ & $\Big($$^{+2.90}_{-2.80}$$\Big)$ \end{tabular} & \begin{tabular}{@{}l@{}l}{}$^{+2.71}_{-0.00}$ & $\Big($$^{+2.86}_{-2.78}$$\Big)$ \end{tabular} & \begin{tabular}{@{}l@{}l}{}$^{+0.26}_{-0.00}$ & $\Big($$^{+0.45}_{-0.27}$$\Big)$ \end{tabular} & \begin{tabular}{r@{}l@{}l}$0.00$ & {}$^{+0.60}_{-0.00}$ & $\Big($$^{+0.64}_{-0.62}$$\Big)$ \end{tabular} \\ 
        \ggH $\geq2$J high $\ptH$ & \begin{tabular}{r@{}l}$2.12$ & {}$^{+0.49}_{-0.50}$\end{tabular} & \begin{tabular}{r@{}l@{}l}$0.62$ & {}$^{+1.06}_{-0.62}$ & $\Big($$^{+1.15}_{-1.10}$$\Big)$ \end{tabular} & \begin{tabular}{@{}l@{}l}{}$^{+1.04}_{-0.62}$ & $\Big($$^{+1.11}_{-1.10}$$\Big)$ \end{tabular} & \begin{tabular}{@{}l@{}l}{}$^{+0.17}_{-0.17}$ & $\Big($$^{+0.30}_{-0.13}$$\Big)$ \end{tabular} & \begin{tabular}{r@{}l@{}l}$0.29$ & {}$^{+0.50}_{-0.29}$ & $\Big($$^{+0.54}_{-0.52}$$\Big)$ \end{tabular} \\ 
        \ggH BSM $200<\ptH<300$ & \begin{tabular}{r@{}l}$1.10$ & {}$^{+0.28}_{-0.27}$\end{tabular} & \begin{tabular}{r@{}l@{}l}$1.11$ & {}$^{+0.47}_{-0.44}$ & $\Big($$^{+0.56}_{-0.45}$$\Big)$ \end{tabular} & \begin{tabular}{@{}l@{}l}{}$^{+0.46}_{-0.43}$ & $\Big($$^{+0.56}_{-0.45}$$\Big)$ \end{tabular} & \begin{tabular}{@{}l@{}l}{}$^{+0.08}_{-0.07}$ & $\Big($$^{+0.05}_{-0.03}$$\Big)$ \end{tabular} & \begin{tabular}{r@{}l@{}l}$1.00$ & {}$^{+0.42}_{-0.40}$ & $\Big($$^{+0.51}_{-0.41}$$\Big)$ \end{tabular} \\ 
        \ggH BSM $300<\ptH<450$ & \begin{tabular}{r@{}l}$0.28$ & {}$^{+0.07}_{-0.07}$\end{tabular} & \begin{tabular}{r@{}l@{}l}$0.16$ & {}$^{+0.19}_{-0.16}$ & $\Big($$^{+0.20}_{-0.18}$$\Big)$ \end{tabular} & \begin{tabular}{@{}l@{}l}{}$^{+0.18}_{-0.16}$ & $\Big($$^{+0.19}_{-0.18}$$\Big)$ \end{tabular} & \begin{tabular}{@{}l@{}l}{}$^{+0.02}_{-0.02}$ & $\Big($$^{+0.03}_{-0.01}$$\Big)$ \end{tabular} & \begin{tabular}{r@{}l@{}l}$0.55$ & {}$^{+0.66}_{-0.55}$ & $\Big($$^{+0.69}_{-0.65}$$\Big)$ \end{tabular} \\ 
        \ggH BSM $\ptH>450$ & \begin{tabular}{r@{}l}$0.05$ & {}$^{+0.02}_{-0.02}$\end{tabular} & \begin{tabular}{r@{}l@{}l}$0.10$ & {}$^{+0.10}_{-0.08}$ & $\Big($$^{+0.10}_{-0.05}$$\Big)$ \end{tabular} & \begin{tabular}{@{}l@{}l}{}$^{+0.09}_{-0.08}$ & $\Big($$^{+0.09}_{-0.05}$$\Big)$ \end{tabular} & \begin{tabular}{@{}l@{}l}{}$^{+0.06}_{-0.02}$ & $\Big($$^{+0.04}_{-0.04}$$\Big)$ \end{tabular} & \begin{tabular}{r@{}l@{}l}$2.16$ & {}$^{+2.25}_{-1.69}$ & $\Big($$^{+2.19}_{-1.00}$$\Big)$ \end{tabular} \\ 
        VBF-like low $\mjj$ low $\ptHjj$ & \begin{tabular}{r@{}l}$1.59$ & {}$^{+0.49}_{-0.48}$\end{tabular} & \begin{tabular}{r@{}l@{}l}$1.31$ & {}$^{+1.19}_{-1.13}$ & $\Big($$^{+1.22}_{-1.16}$$\Big)$ \end{tabular} & \begin{tabular}{@{}l@{}l}{}$^{+1.18}_{-1.13}$ & $\Big($$^{+1.21}_{-1.16}$$\Big)$ \end{tabular} & \begin{tabular}{@{}l@{}l}{}$^{+0.14}_{-0.09}$ & $\Big($$^{+0.13}_{-0.05}$$\Big)$ \end{tabular} & \begin{tabular}{r@{}l@{}l}$0.82$ & {}$^{+0.75}_{-0.71}$ & $\Big($$^{+0.77}_{-0.73}$$\Big)$ \end{tabular} \\ 
        VBF-like low $\mjj$ high $\ptHjj$ & \begin{tabular}{r@{}l}$1.25$ & {}$^{+0.35}_{-0.32}$\end{tabular} & \begin{tabular}{r@{}l@{}l}$3.46$ & {}$^{+1.76}_{-1.64}$ & $\Big($$^{+1.79}_{-1.25}$$\Big)$ \end{tabular} & \begin{tabular}{@{}l@{}l}{}$^{+1.65}_{-1.62}$ & $\Big($$^{+1.76}_{-1.25}$$\Big)$ \end{tabular} & \begin{tabular}{@{}l@{}l}{}$^{+0.61}_{-0.25}$ & $\Big($$^{+0.32}_{-0.32}$$\Big)$ \end{tabular} & \begin{tabular}{r@{}l@{}l}$2.76$ & {}$^{+1.40}_{-1.31}$ & $\Big($$^{+1.43}_{-1.00}$$\Big)$ \end{tabular} \\ 
        VBF-like high $\mjj$ low $\ptHjj$ & \begin{tabular}{r@{}l}$1.60$ & {}$^{+0.45}_{-0.51}$\end{tabular} & \begin{tabular}{r@{}l@{}l}$0.97$ & {}$^{+0.64}_{-0.63}$ & $\Big($$^{+0.72}_{-0.63}$$\Big)$ \end{tabular} & \begin{tabular}{@{}l@{}l}{}$^{+0.63}_{-0.62}$ & $\Big($$^{+0.71}_{-0.63}$$\Big)$ \end{tabular} & \begin{tabular}{@{}l@{}l}{}$^{+0.07}_{-0.07}$ & $\Big($$^{+0.11}_{-0.06}$$\Big)$ \end{tabular} & \begin{tabular}{r@{}l@{}l}$0.61$ & {}$^{+0.40}_{-0.39}$ & $\Big($$^{+0.45}_{-0.39}$$\Big)$ \end{tabular} \\ 
        VBF-like high $\mjj$ high $\ptHjj$ & \begin{tabular}{r@{}l}$0.73$ & {}$^{+0.16}_{-0.16}$\end{tabular} & \begin{tabular}{r@{}l@{}l}$0.15$ & {}$^{+1.08}_{-0.15}$ & $\Big($$^{+0.93}_{-0.73}$$\Big)$ \end{tabular} & \begin{tabular}{@{}l@{}l}{}$^{+1.04}_{-0.15}$ & $\Big($$^{+0.92}_{-0.73}$$\Big)$ \end{tabular} & \begin{tabular}{@{}l@{}l}{}$^{+0.28}_{-0.15}$ & $\Big($$^{+0.14}_{-0.14}$$\Big)$ \end{tabular} & \begin{tabular}{r@{}l@{}l}$0.20$ & {}$^{+1.47}_{-0.20}$ & $\Big($$^{+1.26}_{-1.00}$$\Big)$ \end{tabular} \\ 
        \qqH \VH-like & \begin{tabular}{r@{}l}$1.22$ & {}$^{+0.05}_{-0.05}$\end{tabular} & \begin{tabular}{r@{}l@{}l}$1.53$ & {}$^{+1.20}_{-1.21}$ & $\Big($$^{+1.13}_{-1.27}$$\Big)$ \end{tabular} & \begin{tabular}{@{}l@{}l}{}$^{+1.20}_{-1.20}$ & $\Big($$^{+1.12}_{-1.27}$$\Big)$ \end{tabular} & \begin{tabular}{@{}l@{}l}{}$^{+0.11}_{-0.19}$ & $\Big($$^{+0.05}_{-0.08}$$\Big)$ \end{tabular} & \begin{tabular}{r@{}l@{}l}$1.25$ & {}$^{+0.98}_{-0.99}$ & $\Big($$^{+0.92}_{-1.04}$$\Big)$ \end{tabular} \\ 
        \qqH BSM & \begin{tabular}{r@{}l}$0.37$ & {}$^{+0.03}_{-0.02}$\end{tabular} & \begin{tabular}{r@{}l@{}l}$0.51$ & {}$^{+0.24}_{-0.22}$ & $\Big($$^{+0.24}_{-0.24}$$\Big)$ \end{tabular} & \begin{tabular}{@{}l@{}l}{}$^{+0.24}_{-0.22}$ & $\Big($$^{+0.24}_{-0.24}$$\Big)$ \end{tabular} & \begin{tabular}{@{}l@{}l}{}$^{+0.03}_{-0.01}$ & $\Big($$^{+0.03}_{-0.02}$$\Big)$ \end{tabular} & \begin{tabular}{r@{}l@{}l}$1.40$ & {}$^{+0.66}_{-0.60}$ & $\Big($$^{+0.66}_{-0.65}$$\Big)$ \end{tabular} \\ 
        \WH lep $\ptV<75$ & \begin{tabular}{r@{}l}$0.47$ & {}$^{+0.02}_{-0.02}$\end{tabular} & \begin{tabular}{r@{}l@{}l}$0.71$ & {}$^{+0.68}_{-0.54}$ & $\Big($$^{+0.75}_{-0.47}$$\Big)$ \end{tabular} & \begin{tabular}{@{}l@{}l}{}$^{+0.68}_{-0.54}$ & $\Big($$^{+0.75}_{-0.47}$$\Big)$ \end{tabular} & \begin{tabular}{@{}l@{}l}{}$^{+0.05}_{-0.05}$ & $\Big($$^{+0.05}_{-0.05}$$\Big)$ \end{tabular} & \begin{tabular}{r@{}l@{}l}$1.51$ & {}$^{+1.45}_{-1.15}$ & $\Big($$^{+1.60}_{-1.00}$$\Big)$ \end{tabular} \\ 
        \WH lep $75<\ptV<150$ & \begin{tabular}{r@{}l}$0.29$ & {}$^{+0.02}_{-0.02}$\end{tabular} & \begin{tabular}{r@{}l@{}l}$0.44$ & {}$^{+0.37}_{-0.32}$ & $\Big($$^{+0.37}_{-0.29}$$\Big)$ \end{tabular} & \begin{tabular}{@{}l@{}l}{}$^{+0.37}_{-0.32}$ & $\Big($$^{+0.37}_{-0.29}$$\Big)$ \end{tabular} & \begin{tabular}{@{}l@{}l}{}$^{+0.03}_{-0.02}$ & $\Big($$^{+0.02}_{-0.02}$$\Big)$ \end{tabular} & \begin{tabular}{r@{}l@{}l}$1.49$ & {}$^{+1.26}_{-1.08}$ & $\Big($$^{+1.25}_{-1.00}$$\Big)$ \end{tabular} \\ 
        \WH lep $\ptV>150$ & \begin{tabular}{r@{}l}$0.12$ & {}$^{+0.01}_{-0.01}$\end{tabular} & \begin{tabular}{r@{}l@{}l}$0.12$ & {}$^{+0.12}_{-0.09}$ & $\Big($$^{+0.13}_{-0.10}$$\Big)$ \end{tabular} & \begin{tabular}{@{}l@{}l}{}$^{+0.12}_{-0.09}$ & $\Big($$^{+0.13}_{-0.10}$$\Big)$ \end{tabular} & \begin{tabular}{@{}l@{}l}{}$^{+0.01}_{-0.00}$ & $\Big($$^{+0.01}_{-0.01}$$\Big)$ \end{tabular} & \begin{tabular}{r@{}l@{}l}$0.98$ & {}$^{+0.98}_{-0.76}$ & $\Big($$^{+1.05}_{-0.79}$$\Big)$ \end{tabular} \\ 
        \ZH lep & \begin{tabular}{r@{}l}$0.54$ & {}$^{+0.03}_{-0.02}$\end{tabular} & \begin{tabular}{r@{}l@{}l}$0.71$ & {}$^{+0.41}_{-0.35}$ & $\Big($$^{+0.42}_{-0.35}$$\Big)$ \end{tabular} & \begin{tabular}{@{}l@{}l}{}$^{+0.40}_{-0.35}$ & $\Big($$^{+0.41}_{-0.35}$$\Big)$ \end{tabular} & \begin{tabular}{@{}l@{}l}{}$^{+0.07}_{-0.03}$ & $\Big($$^{+0.04}_{-0.04}$$\Big)$ \end{tabular} & \begin{tabular}{r@{}l@{}l}$1.32$ & {}$^{+0.76}_{-0.65}$ & $\Big($$^{+0.77}_{-0.65}$$\Big)$ \end{tabular} \\ 
        \ttH $\ptH<60$ & \begin{tabular}{r@{}l}$0.26$ & {}$^{+0.03}_{-0.04}$\end{tabular} & \begin{tabular}{r@{}l@{}l}$0.19$ & {}$^{+0.24}_{-0.19}$ & $\Big($$^{+0.23}_{-0.19}$$\Big)$ \end{tabular} & \begin{tabular}{@{}l@{}l}{}$^{+0.24}_{-0.19}$ & $\Big($$^{+0.23}_{-0.19}$$\Big)$ \end{tabular} & \begin{tabular}{@{}l@{}l}{}$^{+0.01}_{-0.01}$ & $\Big($$^{+0.03}_{-0.03}$$\Big)$ \end{tabular} & \begin{tabular}{r@{}l@{}l}$0.73$ & {}$^{+0.92}_{-0.73}$ & $\Big($$^{+0.89}_{-0.75}$$\Big)$ \end{tabular} \\ 
        \ttH $60<\ptH<120$ & \begin{tabular}{r@{}l}$0.40$ & {}$^{+0.04}_{-0.05}$\end{tabular} & \begin{tabular}{r@{}l@{}l}$0.50$ & {}$^{+0.26}_{-0.22}$ & $\Big($$^{+0.28}_{-0.22}$$\Big)$ \end{tabular} & \begin{tabular}{@{}l@{}l}{}$^{+0.25}_{-0.22}$ & $\Big($$^{+0.28}_{-0.22}$$\Big)$ \end{tabular} & \begin{tabular}{@{}l@{}l}{}$^{+0.04}_{-0.02}$ & $\Big($$^{+0.03}_{-0.01}$$\Big)$ \end{tabular} & \begin{tabular}{r@{}l@{}l}$1.25$ & {}$^{+0.65}_{-0.55}$ & $\Big($$^{+0.72}_{-0.55}$$\Big)$ \end{tabular} \\ 
        \ttH $120<\ptH<200$ & \begin{tabular}{r@{}l}$0.29$ & {}$^{+0.03}_{-0.04}$\end{tabular} & \begin{tabular}{r@{}l@{}l}$0.24$ & {}$^{+0.17}_{-0.14}$ & $\Big($$^{+0.17}_{-0.16}$$\Big)$ \end{tabular} & \begin{tabular}{@{}l@{}l}{}$^{+0.17}_{-0.14}$ & $\Big($$^{+0.17}_{-0.16}$$\Big)$ \end{tabular} & \begin{tabular}{@{}l@{}l}{}$^{+0.02}_{-0.02}$ & $\Big($$^{+0.02}_{-0.01}$$\Big)$ \end{tabular} & \begin{tabular}{r@{}l@{}l}$0.80$ & {}$^{+0.58}_{-0.49}$ & $\Big($$^{+0.58}_{-0.53}$$\Big)$ \end{tabular} \\ 
        \ttH $200<\ptH<300$ & \begin{tabular}{r@{}l}$0.12$ & {}$^{+0.02}_{-0.02}$\end{tabular} & \begin{tabular}{r@{}l@{}l}$0.11$ & {}$^{+0.11}_{-0.09}$ & $\Big($$^{+0.10}_{-0.09}$$\Big)$ \end{tabular} & \begin{tabular}{@{}l@{}l}{}$^{+0.11}_{-0.09}$ & $\Big($$^{+0.10}_{-0.09}$$\Big)$ \end{tabular} & \begin{tabular}{@{}l@{}l}{}$^{+0.01}_{-0.01}$ & $\Big($$^{+0.01}_{-0.01}$$\Big)$ \end{tabular} & \begin{tabular}{r@{}l@{}l}$0.92$ & {}$^{+0.89}_{-0.73}$ & $\Big($$^{+0.81}_{-0.75}$$\Big)$ \end{tabular} \\ 
        \ttH $\ptH>300$ & \begin{tabular}{r@{}l}$0.06$ & {}$^{+0.01}_{-0.01}$\end{tabular} & \begin{tabular}{r@{}l@{}l}$0.00$ & {}$^{+0.08}_{-0.00}$ & $\Big($$^{+0.07}_{-0.06}$$\Big)$ \end{tabular} & \begin{tabular}{@{}l@{}l}{}$^{+0.08}_{-0.00}$ & $\Big($$^{+0.07}_{-0.06}$$\Big)$ \end{tabular} & \begin{tabular}{@{}l@{}l}{}$^{+0.01}_{-0.00}$ & $\Big($$^{+0.02}_{-0.02}$$\Big)$ \end{tabular} & \begin{tabular}{r@{}l@{}l}$0.00$ & {}$^{+1.34}_{-0.00}$ & $\Big($$^{+1.21}_{-1.00}$$\Big)$ \end{tabular} \\ 
        \tH & \begin{tabular}{r@{}l}$0.20$ & {}$^{+0.01}_{-0.03}$\end{tabular} & \begin{tabular}{r@{}l@{}l}$1.71$ & {}$^{+0.71}_{-0.93}$ & $\Big($$^{+1.01}_{-0.20}$$\Big)$ \end{tabular} & \begin{tabular}{@{}l@{}l}{}$^{+0.70}_{-0.92}$ & $\Big($$^{+1.00}_{-0.20}$$\Big)$ \end{tabular} & \begin{tabular}{@{}l@{}l}{}$^{+0.13}_{-0.13}$ & $\Big($$^{+0.11}_{-0.11}$$\Big)$ \end{tabular} & \begin{tabular}{r@{}l@{}l}$8.38$ & {}$^{+3.48}_{-4.55}$ & $\Big($$^{+4.93}_{-1.00}$$\Big)$ \end{tabular} \\ 
      \end{tabular}
    }
\end{table}

\subsection{Coupling modifiers}
\label{sec:kappas}
The $\kappa$-framework defines coupling modifiers to directly parametrise deviations from the SM expectation in the couplings of the Higgs boson to other particles \cite{YR3}. 
Two different likelihood scans, each with two dimensions, are performed. 
Full details of each parameterisation are given in Ref.~\cite{Sirunyan:2018koj}.

\begin{figure}[htbp!]
  \centering
  \includegraphics[width=.7\textwidth]{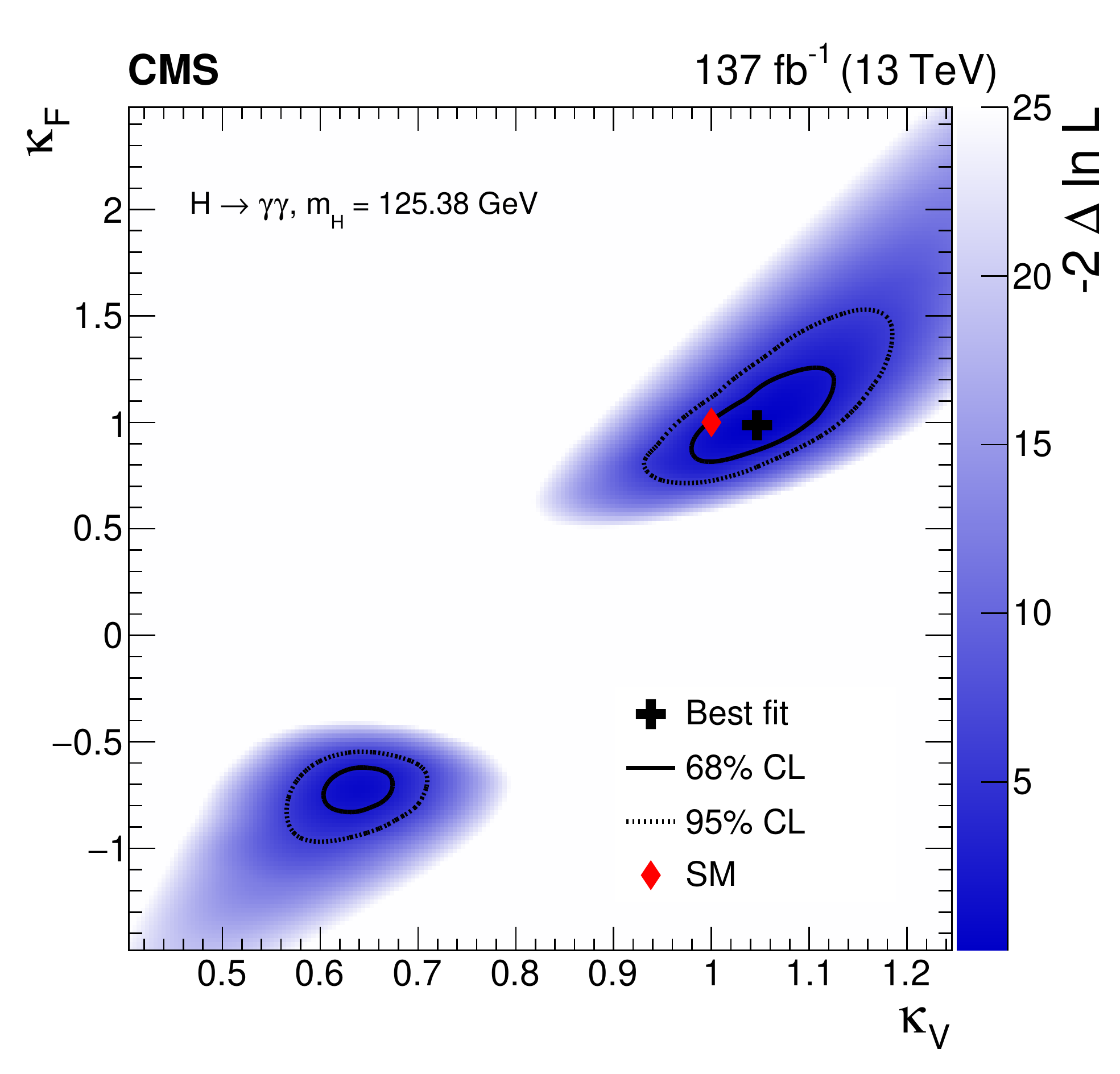}
  \includegraphics[width=.7\textwidth]{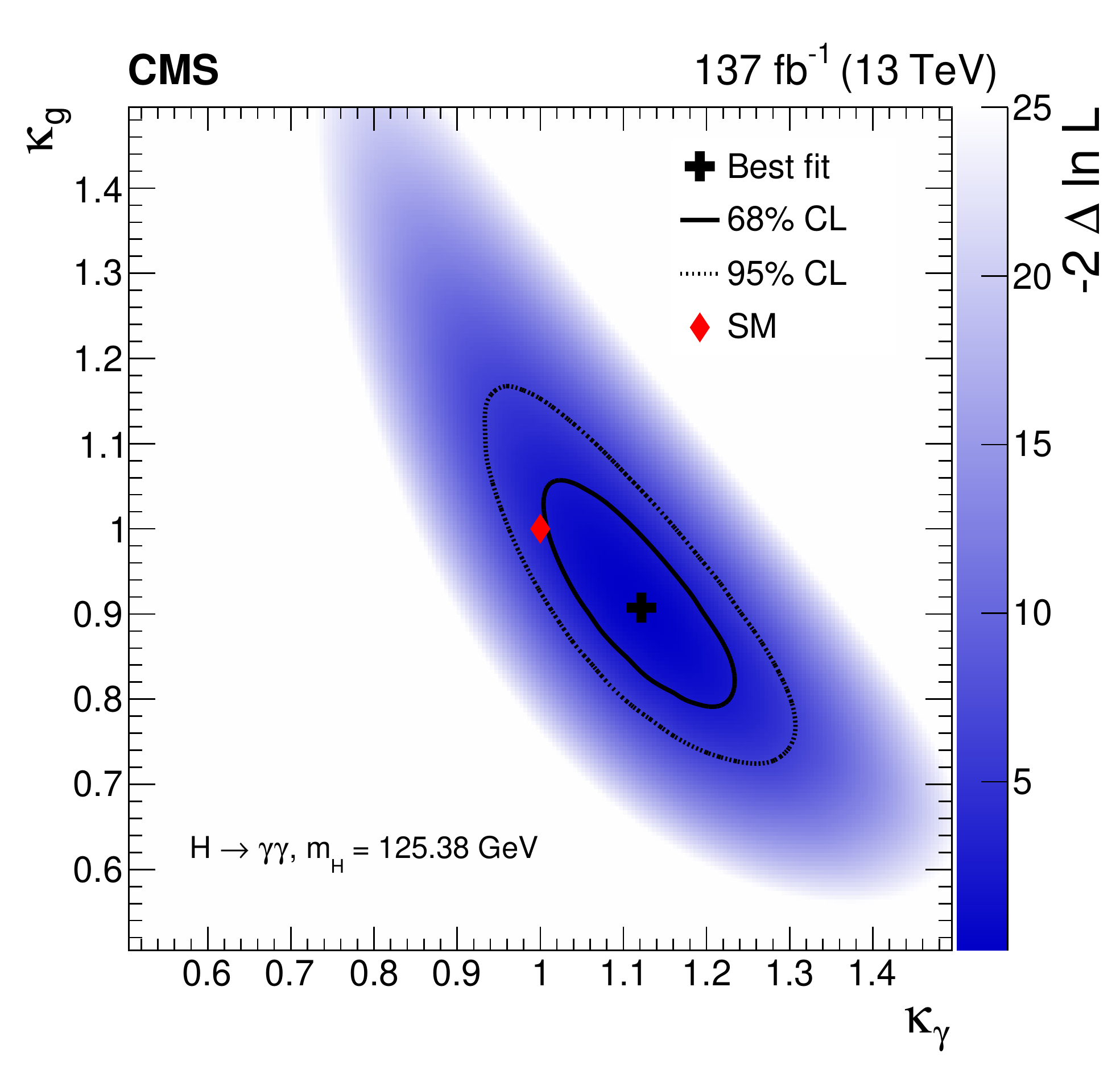}
  \caption{
    Observed two-dimensional likelihood scans performed in the $\kappa$-framework: 
    $\kv$-vs-$\kf$ in the resolved $\kappa$ model (upper) 
    and $\kgam$-vs-$\kglu$ in the unresolved $\kappa$ model (lower) .
    The 68 and 95\% \CL regions are given by the solid and dashed contours, respectively.
    The best fit and SM points are shown by the black cross and red diamond, respectively.
    The colour scale indicates the value of the test statistic.
  }
  \label{fig:scan2D_kappas}
\end{figure}

In the first fit, the resolved $\kappa$ model is used.  
Here the scaling factors of loops present in Higgs boson production and decay are resolved into their SM components, 
in terms of the other $\kappa$ parameters. 
The most important of these are in \ggH production and \Hgg decay, 
but others, such as the loop in \ggZH production, are also resolved.
The results of a two-dimensional scan in $\kv$ and $\kf$, scaling the Higgs boson coupling to vector bosons 
and to fermions, respectively, are shown in the upper plot of Fig.~\ref{fig:scan2D_kappas}.
The \Hgg decay rate contains an interference term proportional to $\kv\kf$.
This means that the rate of \ggH and \ttH production ($\propto\kf^2$), 
relative to the rate of VBF and \VH production ($\propto\kv^2$),
can be used to gain sensitivity to the relative sign of the tt-H and VV-H couplings.
In addition, the \tHq and \tHW production modes also include a term proportional to $\kv\kf$.
This analysis explicitly targets \tHq production via the \tHq leptonic analysis category,
the inclusion of which helps to further reduce the degeneracy between positive and negative $\kf$ values.
The region with negative values of $\kf$ is observed (expected) to be excluded with a significance of $0.5$ ($2.4$) standard deviations.
The reduction in the observed significance with respect to the expected is due to the observed excess in the \tH production mode cross section.

A second fit is performed using the unresolved $\kappa$ model, where the \ggH and \Hgg loops
are given their own effective scaling factors denoted $\kglu$ and $\kgam$, respectively.
The $\kglu$ and $\kgam$ parameters are particularly sensitive to additional BSM states,
that contribute towards the rate of Higgs boson production and decay via loop processes.
The observed result of a two-dimensional scan in these two parameters is shown in the lower plot of Fig.~\ref{fig:scan2D_kappas}.
In the scan, the other $\kappa$ parameters in the unresolved model are fixed to unity.
The best fit point is consistent with the SM expectation at approximately the 68\% \CL.

\section{Summary}
\label{sec:summary}

Measurements of Higgs boson properties with the Higgs boson decaying into a pair of photons are reported. 
Events with two photons are selected from a sample of proton-proton collisions at a centre-of-mass energy \sqrts collected with the CMS detector at the LHC from 2016 to 2018, 
corresponding to an integrated luminosity of 137\fbinv. 
Analysis categories enriched in events produced via gluon fusion, vector boson fusion, 
vector boson associated production, production associated with two top quarks,  
and production associated with one top quark are constructed.

A range of production and coupling properties of the Higgs boson are measured.
The total Higgs boson signal strength, relative to the standard model (SM) prediction, 
is measured to be $1.12\pm0.09$.
A simultaneous measurement of the signal strengths of the four principal Higgs boson 
production mechanisms is performed and found to be compatible with the SM prediction 
with a $p$-value of 50\%.
Two different measurements are performed within the simplified template cross section framework, 
in which 17 and 27 independent kinematic regions are measured simultaneously, 
with corresponding $p$-values with respect to the SM of 31 and 70\%, respectively.
Many of these kinematic regions are measured for the first time, 
including a simultaneous measurement of Higgs boson production in association with two top quarks
in five different regions of the Higgs boson transverse momentum $\ptH$.
Furthermore, several additional measurements are the most precise made in a single channel to date.
These include cross sections of vector boson fusion in different kinematic regions, 
gluon fusion in association with jets, 
and the region of gluon fusion production with $\ptH>200\GeV$, 
which is particularly sensitive to physics beyond the SM.
The gluon fusion cross section with $\ptH>200\GeV$ is found to be consistent with the SM, 
with a measured value of $0.9_{-0.3}^{+0.4}$ relative to the SM prediction.
An upper limit on the rate of Higgs boson production in association with a single top quark is also presented. 
The observed (expected) limit at 95\% confidence level is found to be $14$ ($8$) times the SM prediction.
All other results, such as measurements of the Higgs boson's couplings to vector bosons and to fermions, 
are also in agreement with the SM expectations.

\begin{acknowledgments}
  We congratulate our colleagues in the CERN accelerator departments for the excellent performance of the LHC and thank the technical and administrative staffs at CERN and at other CMS institutes for their contributions to the success of the CMS effort. In addition, we gratefully acknowledge the computing centres and personnel of the Worldwide LHC Computing Grid and other centres for delivering so effectively the computing infrastructure essential to our analyses. We also gratefully acknowledge the LHC Higgs Working Group for its role in developing stage 1.2 of the simplified template cross section framework. Finally, we acknowledge the enduring support for the construction and operation of the LHC, the CMS detector, and the supporting computing infrastructure provided by the following funding agencies: BMBWF and FWF (Austria); FNRS and FWO (Belgium); CNPq, CAPES, FAPERJ, FAPERGS, and FAPESP (Brazil); MES (Bulgaria); CERN; CAS, MoST, and NSFC (China); COLCIENCIAS (Colombia); MSES and CSF (Croatia); RIF (Cyprus); SENESCYT (Ecuador); MoER, ERC PUT and ERDF (Estonia); Academy of Finland, MEC, and HIP (Finland); CEA and CNRS/IN2P3 (France); BMBF, DFG, and HGF (Germany); GSRT (Greece); NKFIA (Hungary); DAE and DST (India); IPM (Iran); SFI (Ireland); INFN (Italy); MSIP and NRF (Republic of Korea); MES (Latvia); LAS (Lithuania); MOE and UM (Malaysia); BUAP, CINVESTAV, CONACYT, LNS, SEP, and UASLP-FAI (Mexico); MOS (Montenegro); MBIE (New Zealand); PAEC (Pakistan); MSHE and NSC (Poland); FCT (Portugal); JINR (Dubna); MON, RosAtom, RAS, RFBR, and NRC KI (Russia); MESTD (Serbia); SEIDI, CPAN, PCTI, and FEDER (Spain); MOSTR (Sri Lanka); Swiss Funding Agencies (Switzerland); MST (Taipei); ThEPCenter, IPST, STAR, and NSTDA (Thailand); TUBITAK and TAEK (Turkey); NASU (Ukraine); STFC (United Kingdom); DOE and NSF (USA).
  
  \hyphenation{Rachada-pisek} Individuals have received support from the Marie-Curie programme and the European Research Council and Horizon 2020 Grant, contract Nos.\ 675440, 724704, 752730, and 765710 (European Union); the Leventis Foundation; the Alfred P.\ Sloan Foundation; the Alexander von Humboldt Foundation; the Belgian Federal Science Policy Office; the Fonds pour la Formation \`a la Recherche dans l'Industrie et dans l'Agriculture (FRIA-Belgium); the Agentschap voor Innovatie door Wetenschap en Technologie (IWT-Belgium); the F.R.S.-FNRS and FWO (Belgium) under the ``Excellence of Science -- EOS" -- be.h project n.\ 30820817; the Beijing Municipal Science \& Technology Commission, No. Z191100007219010; the Ministry of Education, Youth and Sports (MEYS) of the Czech Republic; the Deutsche Forschungsgemeinschaft (DFG), under Germany's Excellence Strategy -- EXC 2121 ``Quantum Universe" -- 390833306, and under project number 400140256 - GRK2497; the Lend\"ulet (``Momentum") Programme and the J\'anos Bolyai Research Scholarship of the Hungarian Academy of Sciences, the New National Excellence Program \'UNKP, the NKFIA research grants 123842, 123959, 124845, 124850, 125105, 128713, 128786, and 129058 (Hungary); the Council of Science and Industrial Research, India; the Ministry of Science and Higher Education and the National Science Center, contracts Opus 2014/15/B/ST2/03998 and 2015/19/B/ST2/02861 (Poland); the National Priorities Research Program by Qatar National Research Fund; the Ministry of Science and Higher Education, project no. 0723-2020-0041 (Russia); the Programa Estatal de Fomento de la Investigaci{\'o}n Cient{\'i}fica y T{\'e}cnica de Excelencia Mar\'{\i}a de Maeztu, grant MDM-2015-0509 and the Programa Severo Ochoa del Principado de Asturias; the Thalis and Aristeia programmes cofinanced by EU-ESF and the Greek NSRF; the Rachadapisek Sompot Fund for Postdoctoral Fellowship, Chulalongkorn University and the Chulalongkorn Academic into Its 2nd Century Project Advancement Project (Thailand); the Kavli Foundation; the Nvidia Corporation; the SuperMicro Corporation; the Welch Foundation, contract C-1845; and the Weston Havens Foundation (USA).
\end{acknowledgments}
\bibliography{auto_generated}
\cleardoublepage \appendix\section{The CMS Collaboration \label{app:collab}}\begin{sloppypar}\hyphenpenalty=5000\widowpenalty=500\clubpenalty=5000\vskip\cmsinstskip
\textbf{Yerevan Physics Institute, Yerevan, Armenia}\\*[0pt]
A.M.~Sirunyan$^{\textrm{\dag}}$, A.~Tumasyan
\vskip\cmsinstskip
\textbf{Institut f\"{u}r Hochenergiephysik, Wien, Austria}\\*[0pt]
W.~Adam, J.W.~Andrejkovic, T.~Bergauer, S.~Chatterjee, M.~Dragicevic, A.~Escalante~Del~Valle, R.~Fr\"{u}hwirth\cmsAuthorMark{1}, M.~Jeitler\cmsAuthorMark{1}, N.~Krammer, L.~Lechner, D.~Liko, I.~Mikulec, P.~Paulitsch, F.M.~Pitters, J.~Schieck\cmsAuthorMark{1}, R.~Sch\"{o}fbeck, M.~Spanring, S.~Templ, W.~Waltenberger, C.-E.~Wulz\cmsAuthorMark{1}
% ...
\vskip\cmsinstskip
\dag: Deceased\\
1:  Also at Vienna University of Technology, Vienna, Austria\\
% ...
\end{sloppypar}
%%% END EDITABLE REGION %%%
% skeleton_end
\end{document}